\documentclass[12pt,  a4paper]{article}
\usepackage{eurosym}
\usepackage[top=1in, bottom=1.5in,left=1in,right=1in]{geometry}
\usepackage{amsmath}
\usepackage{amssymb}
\usepackage{amsfonts}
\usepackage{float}
\usepackage{bm}
\usepackage{tabularx}
\usepackage{diagbox}
\usepackage{booktabs}
\usepackage{enumitem}
\usepackage[usenames,dvipsnames]{xcolor}
\usepackage{tikz}
\usepackage[ruled]{algorithm2e}
\usepackage[round]{natbib}
\usepackage[utf8]{inputenc}
\usepackage[toc]{appendix}

\usetikzlibrary{calc, arrows.meta, intersections, patterns, positioning, shapes.misc, fadings, through,decorations.pathreplacing}
\definecolor{ColorOne}{named}{MidnightBlue}
\definecolor{ColorTwo}{named}{Dandelion}
\definecolor{ColorThree}{named}{Plum}

\newtheorem{theorem}{Theorem}[section]

\newtheorem{lemma}{Lemma}[section]

\newtheorem{corollary}{Corollary}[section]
\newtheorem{assumption}{Assumption}[section]

\renewcommand{\epsilon}{\varepsilon}

\newenvironment{proof}[1][Proof]{\noindent \textbf{#1.} }{\  \rule{0.5em}{0.5em}}
\newcommand{\func}[1]{\operatorname{#1}}
\numberwithin{equation}{section}

\begin{document}

\author{Dennis Kristensen\thanks{%
Department of Economics, UCL. E-mail: d.kristensen@ucl.ac.uk} \and Ruolei Zhu%
\thanks{%
Department of Economics, UCL. E-mail: ruolei.zhu.19@ucl.ac.uk}}
\title{Identification of Rational Inattention--Additive Random Utility Models%
}
\date{\today }
\maketitle

\begin{abstract}
This paper introduces a novel class of econometric models, called Rational
Inattention-Additive Random Utility Models. It is based on the discrete
choice rational inattention model of \citet{matvejka2015rational} with
additive unobserved heterogeneity incorporated. We show that some of the
systematic components of the primitives of the model are nonparametrically
identified even if the distributions of unobserved variables and the prior
of decision makers are both unknown to the Econometrician. If additional
information about the prior is known to the Econometrician, even stronger
identification results become available.
\end{abstract}

\section{Introduction}

Many economic decision problems are discrete in nature. Additive Random
Utility Models (ARUMs) \citep{mcfadden1973conditional} is one of the primary
tools for the analysis of such problems. ARUMs assume that the problem is
deterministic from the decision makers' (DMs') perspective in the sense that
the DMs\ have full knowledge of their payoffs of each of the discrete
alternatives. At the same time, ARUMs generate stochastic choice behaviour
from the viewpoint of the researchers through the assumption that a
component of each payoff is unobserved by them. This feature is important in
empirical applications where unobserved heterogeneity is ubiquitous.

The assumption of DMs having full knowledge of the payoffs is very strong
though, and is most likely violated in many economic applications:
individuals rarely know the exact utility they will earn from a given
decision. To relax this assumption, researchers have developed theories and
models considering internal behavioural biases and external imperfect
information of DMs. The focus of this paper is on internal behavioural
biases, which can be explored using the discrete choice rational inattention
model first developed by \citet{matvejka2015rational}; see also %
\citet{caplin2019rational,caplin2022rationally}. Under rational inattention,
DMs are uncertain about the payoffs of the available alternatives and
gathering information about these is costly. DMs then have to decide on the
optimal trade-off between expected utility and information cost, resulting
in stochastic choice behaviour also from the DMs' perspective. These models
have attracted recent attention in applied work, with applications in
marketing and economics including \citet{bertoli2020rational}, %
\citet{caplin2020competitive}, \citet{brown2024endogenous}, %
\citet{joo2023rational}, \citet{natan2025choice}, and %
\citet{turlo2025discrete}. However, there is precious little theoretical
work on the econometrics of rational inattention discrete choice models.
This paper aims at filling this gap.

First, we propose a general class of Rational Inattention-Additive Random
Utility Models (RI-ARUMs)\footnote{%
Although our class of models shares the same name as the one proposed by %
\citet{fosgerau2020discrete}, ours is fundamentally different from theirs.}
as an econometric toolkit to be used in the analysis of empirical
behavioural biases in decision-making processes. The class of models takes
as starting point the classic representative agent model of %
\citet{matvejka2015rational}, where it is assumed that the DM has a prior on
the distribution of the uncertain payoffs. We then extend it to allow for
the DMs to have knowledge about parts of the uncertain payoffs. This is
highly empirically relevant since in most applications, the DMs have
knowledge of some features of the problem at hand, such as the price of each
alternative. We model this by allowing the DM's prior to be a function of
her existing knowledge. That is, we replace the unconditional prior usually
considered in rational inattention models by a conditional one. The
resulting conditional rational inattention model is then characterised by
the following three key components: The conditional prior of the payoffs,
the cost of information, and the realised payoffs. We then propose to model
each of these three components as being additive in observed and unobserved
(from the Econometrician's perspective) heterogeneity. This model contains
as special case the classic full--information ARUM corresponding to RI-ARUM
with zero information cost, or a degenerate prior with all probability mass
at the truth.

Second, we provide theoretical new insights into the features and properties
of rational inattention discrete choice models that should be of independent
interest. We analyse the conditional choice probabilities (CCPs) that the DM
faces as functions of the underlying primitives using tools from convex
optimisation theory. As such this part of the paper shares ideas and
techniques with \citet{sorensen2022mcfadden} who use similar tools in their
analysis of CCPs in ARUM. We, however, face a more complex problem since the
DM's problem is more complicated; in particular, the optimal CCPs may in
fact be situated on the boundary of the probability simplex. These results
are important for the subsequent identification analysis since the CCPs
carry very little information content about the underlying primitives
outside what we call the regular part of the parameter set of primitives.

Third, we develop nonparametric identification results of the unknown
conditional prior, the cost of information and the realised payoffs. This is
done under the following four scenarios: We consider separately the two
scenarios of unobserved heterogeneity not being present and being present in
the population with the latter being the particularly challenging one from
an identification perspective. Within each of these two benchmarks, we
analyse the following two sub--scenarios: In the first one, the DM has
correct beliefs about the payoffs, in which case the prior is identical to
the physical law of the payoffs, and the Econometrician knows this is the
case. This is the most benevolent sub--scenario which leads to strong
identification results. In the second, more testing, environment, the
Econometrician has no knowledge of the DM's prior and it may be incorrect.
As a consequence, we are only able to identify certain features of the
model. As such, our identification results provide guidelines for
researchers to develop novel, flexible, yet well-identified RI-ARUMs for
empirical work; see, for example, \citet{Zhu2026MLE}

Fourth, our modelling approach and theoretical analysis provides a general
blueprint for taking complex structural representative agent models to data.
The ideas and techniques employed in this paper are sufficiently general
that they could in principle be applied in the construction of econometric
versions of other representative agent models that incorporate observed and
unobserved heterogeneity into the primitives of the structural model.
Moreover, as part of our identification analysis, we develop a general set
of identification results that can be applied to problems outside of the
RI--ARUM that is the focus of this paper.

Our identification results extend existing nonparametric identification
results for ARUMs to allow for rational inattention; see, e.g., %
\citet{matzkin1991monotone,matzkin1992binary,matzkin1993nonparametric,matzkin1994restrictions,matzkin2007nonparametric}
and references therein. Identification of random-coefficient distributions
in discrete choice models is considered by \citet{ichimuraThompson1998}, %
\citet{gautierKitamura2013}, \citet{dunkerHoderleinKaido2013}, %
\citet{foxGandhi2016}, and \citet{fox2021note}. Related results for discrete
choice with multidimensional unobserved heterogeneity include %
\citet{matzkin2007heterogeneous}, \citet{berryHaile2009}, %
\citet{brieschChintaguntaMatzkin2010}, and \citet{matzkin2012limited} and %
\citet{matzkin2019constructive}. Finally, \citet{allen2019identification}
provides identification results for an extended version of ARUM that allows
for perturbations of the random utilities, while %
\citet{allenRehbeck2025latent} identify random-coefficient distributions in
latent utility models using restrictions on the mean demand mapping.

Our model and techniques share features with this literature, but rational
inattention introduces additional challenges that require new tools and
results. Similar to random--coefficient ARUM, the CCPs that the DM faces in
our RI-ARUM model have a mixed--logit form due to the uncertainty regarding
payoffs. However, there are two important differences: First, the mixture
weights are endogenous and depend on the underlying prior in a highly
complex manner. In contrast, the distribution of random coefficients is
primitive and the conditional logit specification does not contain these
endogenous unconditional choice probabilities. Second, the prior represents
subjective uncertainty rather than necessarily population heterogeneity. It
describes the DM's beliefs about payoffs. When these beliefs coincide with
the true distribution, the prior also describes heterogeneity across
realised states. Otherwise, it need not coincide with the objective
distribution generating observed choices.

There is a small existing literature on the econometric analysis of rational
inattention discrete choice models. The most closely related paper to ours
is \citet{wang2026semiparametric}, who studies a special case of our general
class of models where the Econometrician knows the DMs have correct prior
beliefs and no additional unobserved heterogeneity enters the prior or the
information cost. His identification strategy is different from ours in that
he relies on information costs approaching zero at which the model collapses
to full-information ARUM limit, where existing identification results can
then be employed. In contrast, our results do not require any such
restrictions. \citet{liao2024identification} considers partial
identification of a Berry--Levinsohn--Pakes (BLP) demand model, where
consumers are rationally inattentive. For related nonparametric
identification results in differentiated-products demand models using
market-level and individual-level data, respectively, see %
\citet{Berry2014identification} and \citet{berryHaile2024micro}. The
framework of these papers is different from ours since in theirs the
Econometrician observes aggregate market shares across different markets,
while we consider the case of individual choices in a single population. As
such, our identification results are not directly comparable. However, our
fundamental analysis of the CCPs of our RI-ARUM model can be used to develop
and analyse RI--ARUM BLP models as done in the follow--up paper %
\citet{Zhu2026BLP} that extends the model of \citet{liao2024identification}
to allow for more flexibility and richer heterogeneity.

Our paper also speaks to an existing literature on modelling,
identification, and estimation of models, where agents have subjective
beliefs. \citet{chenhansenhansen2024robust} consider models where agents'
subjective probabilities differ from the DGP and study how this affects
identification and construct robust confidence sets for set-identified
functionals. Similarly, \citet{anhuxiao2021dynamic} study dynamic
discrete-choice models where agents have subjective and potentially
incorrect transition beliefs. Under exclusion restrictions and CCP variation
they can recover both preferences and subjective beliefs, in finite- and
infinite-horizon models. Our analysis complement theirs and provide a more
comprehensive set of results for static models with particular focus on the
interplay between modelling assumptions and identification strength. We show
that with sufficient observed variation in unknown primitives, we can point
identify the systematic components of these regardless of the precise form
of the prior. In particular, if there is no unobserved heterogeneity in the
model, then even the prior is nonparametrically identified. This final
result, however, breaks down if unobserved heterogeneity is present, unless
the Econometrician is willing to impose the restriction that the prior is
not too different from the physical law.

Finally, as part of our identification analysis, we develop a number of
master theorems that provide general conditions for identification of
composite mappings, where the "outer" mapping is one-to-one on a set that is
allowed to vary with other covariates. This result extends existing results
found in the literature such as \citet{allen2019identification} and %
\citet{Fosgerau2020}, and should be useful in showing identification in
other complex models of discrete choice as noted earlier. These general
results highlight the interaction between exclusion restrictions imposed on
the regression function of interest and the underlying structural model that
this function enters into. At one end, we show that without exclusion
restrictions and little known structure on the model, very little can be
learned about the model. At the other end, with exclusion restrictions and
with more benevolent structure of the model, the model is point identified.

The remainder of the paper is organized as follows. In Section \ref{sec: DMs
problem}, we introduce a novel class of conditional rational inattention
discrete choice models for a given population of DMs and analyse the
properties of the CCPs coming out of this class of models. In Section \ref%
{sec: Econometric Model}, we formulate our RI--ARUM and set up the general
identification problem that the Econometrician faces. We then develop a
number of general identification results in Section \ref{sec: ID composite
injectivity} that we then employ in the analysis of RI--ARUM in Section \ref%
{Sec: ID RI-ARUM}. All proofs have been relegated to the appendix.

Throughout the paper, we use the following notation: $\mathcal{J}%
=\{0,1,\ldots ,J\}$, $J\geq 1$, denotes the choice set that the DM faces,
where $j=0$ is the outside alternative. The DM will form CCP's over these
alternatives. For any vector of probabilities over $\mathcal{J}$, say, $%
\left( p_{0},p_{1},...,p_{J}\right) $, we let $p=\left(
p_{1},...,p_{J}\right) \in \Delta (\mathcal{J})$ denote the vector of the
probabilities of the inside alternatives, where $\Delta (\mathcal{J})=\{p\in 
\mathbb{R}^{J}:p_{j}\geq 0,\ j\in \{1,\ldots ,J\},\ \text{and}\
\sum_{j=1}^{J}p_{j}\leq 1\}$.

\section{A class of conditional Rational Inattention models\label{sec: DMs
problem}}

In this section, we develop a general class of conditional rational
inattention models to describe a given population of DMs. The theory for
this conditional class of models is a straightforward extension of the
existing theory of RI models as found in, e.g., \citet{matvejka2015rational}
and \citet{caplin2019rational}, but provides a novel framework within which
to build econometric models. Next, we introduce a particular additive model
within this general class and for this derive a number of novel results for
this model, including invertibility results w.r.t. the structural components
of the model. These results will form the basis of our subsequent
identification analysis. Proofs of the results of this section can be found
in Appendix \ref{appendix:DMCCP}.

We consider a rational inattentive DM in a given population who faces the
choice set $\mathcal{J}$. If she chooses alternative $j\in \mathcal{J}$, she
will earn an uncertain payoff $V_{j}$. We collect these in $V=\left\{
V_{j}\right\} _{j\in \mathcal{J}}\in \mathcal{V}\subseteq \mathbb{R}^{J+1}$.
The DM has obtained partial information regarding the distribution of $V$ in
terms of a known vector $v_{0}$. One can think of this as an observed
outcome of a random variable $V_{0}$. Moreover, the DM\ has perfect
information regarding her unit cost of information $\Lambda >0$. Conditional
on $V_{0}=v_{0}$ and $\Lambda =\lambda $, the DM forms a prior $\pi
_{V|V_{0}}\left( v|v_{0}\right) $ on the distribution of $V|V_{0}=v_{0}$.
Importantly, we assume that her prior does not depend on the information
cost.

The DM is able to obtain a signal $S=(S_{0},\ldots ,S_{J})\in \mathcal{S}$,
where $\mathcal{S}$ is a measurable space, that informs her about $V$. The
signal is characterised by the conditional distribution of $S$ given $V=v$
and $V_{0}=v_{0}$ denoted $P_{S|V,V_{0}}(\cdot |v,v_{0})$, reflecting the
signal's informativeness about the state. We refer to $P_{S|V,V_{0}}$ as the
signal rule and let $\mathcal{P}_{S|V,V_{0}}$ denote the set of admissible
signal rules, i.e., measurable mappings $v\mapsto P_{S|V,V_{0}}(\cdot
|v,v_{0})$ with $P_{S|V,V_{0}}(\cdot |v,v_{0})$ being a probability measure
for all $v\in \mathcal{V}$.

Upon receiving signal $S=s$ from a given signal rule, the DM updates her
beliefs and forms a posterior of $V$ given $S=s$ and $V_{0}=v_{0}$ denoted $%
P_{V|S,V_{0}}(\cdot |s,v_{0})$, which enables her to compute expected
payoffs $\mathbb{E}_{P_{V|S,V_{0}}}[V_{j}|S=s,V_{0}=v_{0}]$ for $j\in 
\mathcal{J}$. However, there is an information cost $\lambda \cdot
I(P_{S|V,V_{0}})$ of receiving a given type of signal, where $%
I(P_{S|V,V_{0}})$ quantifies the amount of information she receives, as
measured by Shannon mutual information,%
\begin{equation*}
I(P_{S|V,V_{0}}|V_{0}=v_{0})=H(\pi _{V|V_{0}}(\cdot |v_{0}))-\mathbb{E}%
_{P_{S|V_{0}}}\left[ H(P_{V|S,V_{0}}(\cdot |S,v_{0}))\right] ,
\end{equation*}%
with $H(\cdot )$ denoting the Shannon entropy and $P_{S|V_{0}}(s|v_{0}):=%
\int_{\mathcal{V}}P_{S|V,V_{0}}(s|v,v_{0})d\pi _{V|V_{0}}(v|v_{0})$ is the
marginal distribution of $S|V_{0}=v_{0}$. Using the symmetry of Shannon
mutual information\footnote{%
The symmetry property follows from Bayes' rule through the identity $\Pr
(X,Y)=\Pr (X|Y)\Pr (Y)=\Pr (Y|X)\Pr (X)$. We also apply this rule to
formally derive Equation~\eqref{eq:shannonobjective}.}, this can be
rewritten as%
\begin{equation*}
I(P_{S|V,V_{0}}|V_{0}=v_{0})=H(P_{S|V_{0}}(\cdot |v_{0}))-\mathbb{E}_{\pi
_{V|V_{0}}}\left[ H(P_{S|V,V_{0}}(\cdot |V,v_{0}))\right] .
\end{equation*}%
The DM then seeks the optimal signal rule by solving the following problem:%
\begin{equation*}
\max_{P_{S|V,V_{0}}\in \mathcal{P}_{S|V,V_{0}}}\int_{\mathcal{V}}\int_{%
\mathcal{S}}\max_{j\in \mathcal{J}}\left[ \mathbb{E}_{P_{V|S,V_{0}}}[V_{j}]%
\right] dP_{S|V,V_{0}}(s|v,v_{0})d\pi _{V|V_{0}}(v|v_{0})-\lambda \cdot
I(P_{S|V,V_{0}}|V_{0}=v_{0}).
\end{equation*}%
This produces an optimal signal rule $P_{S|V,V_{0}}^{\ast }$ which in turn
the DM uses to obtain $P_{V|S,V_{0}}^{\ast }$. Based on this, she chooses 
\begin{equation*}
Y^{\ast }\left( s,v_{0}\right) =\arg \max_{j\in \mathcal{J}}\left\{ \mathbb{E%
}_{P_{V|S,V_{0}}^{\ast }}\left[ V_{j}|S=s,V_{0}=v_{0}\right] \right\} .
\end{equation*}

We now wish to characterise the CCPs implied by this action once the full
payoff vector $V$ has been realised. With $\mathcal{S}_{j}\left(
v_{0}\right) =\left\{ s:Y^{\ast }\left( s,v_{0}\right) =j\right\} $, the CCP
of alternative $j$ is given by $P_{j}^{\ast }\left( v_{0},v\right) =\Pr
\left( S\in \mathcal{S}_{j}\left( v_{0}\right) |V_{0}=v_{0},V=v\right) $. To
characterise these, we now apply the same logic as %
\citet{matvejka2015rational}: Under an optimal information structure, the DM
never wastes costly information by assigning distinct signals to the same
action. We can therefore index the relevant signals by the actions they
induce, $S=j$. Thus, the optimal CCPs that the DM acts upon are identical to
the optimal signal rule after aggregating all signals that induce the same
action, 
\begin{equation*}
P_{j}^{\ast }\left( v_{0},v\right) =\Pr \left( S=j|V_{0}=v_{0},V=v\right)
=P_{S|V,V_{0}}^{\ast }\left( j|V_{0}=v_{0},V=v\right) .
\end{equation*}%
For a given signal $P\left( v_{0},v\right) =\left( P_{0}\left(
v_{0},v\right) ,...,P_{J}\left( v_{0},v\right) \right) $, we can now
reformulate the DM's objective in terms of $P\left( v_{0},v\right) $. First, 
\begin{equation}
\int_{\mathcal{V}}\int_{\mathcal{S}}\max_{j\in \mathcal{J}}\left[ \mathbb{E}%
_{P_{V|S,V_{0}}(\cdot |s,v_{0})}[V_{j}]\right] dP_{S|V,V_{0}}(s|v,v_{0})d\pi
_{V|V_{0}}(v|v_{0})=\int_{\mathcal{V}}\sum_{j\in \mathcal{J}%
}v_{j}P_{j}\left( v_{0},v\right) d\pi _{V|V_{0}}(v|v_{0}).
\label{eq:shannonobjective}
\end{equation}%
Similarly, $I(P_{S|V,V_{0}}|V_{0}=v_{0})=I\left( P\left( v_{0},\cdot \right)
\right) $, where 
\begin{equation*}
I\left( P\left( v_{0},\cdot \right) \right) :=-\sum_{j\in \mathcal{J}%
}P_{j}^{0}(v_{0})\log P_{j}^{0}(v_{0})+\int_{\mathcal{V}}\left( \sum_{j\in 
\mathcal{J}}P_{j}\left( v_{0},\cdot \right) \log P_{j}\left( v_{0},\cdot
\right) \right) d\pi _{V|V_{0}}(v|v_{0}),
\end{equation*}%
and $P_{j}^{0}(v_{0})=\int_{\mathcal{V}}P_{j}\left( v_{0},v\right) d\pi
_{V|V_{0}}(v|v_{0})$ is the integrated CCP of alternative $j\in \mathcal{J}$%
. In total, the optimal CCPs/signal rule $P^{\ast }\left( v_{0},\cdot
\right) $ solves $\max_{P\left( v_{0},\cdot \right) \in \mathcal{P}\left(
v_{0}\right) }Q(P\left( \cdot |v_{0}\right) ;v_{0})$, where $\mathcal{P}%
\left( v_{0}\right) =\{P\left( v_{0},v\right) :P\left( v_{0},v\right) \in
\Delta (\mathcal{J})\ $for all $v\in \mathcal{V}\}$ and 
\begin{equation}
Q(P\left( v_{0},\cdot \right) ;v_{0}):=\int_{\mathcal{V}}\sum_{j\in \mathcal{%
J}}v_{j}P_{j}\left( v_{0},v\right) d\pi _{V|V_{0}}(v|v_{0})-\lambda \cdot
I(P\left( v_{0},\cdot \right) ;v_{0}).  \label{eq:DMproblemobjective}
\end{equation}

The following lemma provides a partial characterisation of the optimal CCPS $%
P^{\ast }\left( v_{0},v\right) =\arg \max_{P\left( v_{0},\cdot \right)
}Q(P\left( v_{0},\cdot \right) ;v_{0})$

\begin{lemma}
\label{lemma:DMCCP}If an optimal strategy $P^{\ast }\left( v_{0},v\right) $
exists, then it takes the form 
\begin{equation}
P_{j}^{\ast }\left( v_{0},v\right) =\frac{P_{j}^{0\ast }(v_{0})\exp \left(
v_{j}/\lambda \right) }{\sum_{\ell \in \mathcal{J}}P_{\ell }^{0\ast
}(v_{0})\exp \left( v_{\ell }/\lambda \right) },\qquad P_{j}^{0\ast
}(v_{0})=\int_{\mathcal{V}}P_{j}^{\ast }\left( v_{0},v\right) d\pi
_{V|V_{0}}(v|v_{0})  \label{eq:DMCCP}
\end{equation}%
for $j\in \{1,\ldots ,J\}$.
\end{lemma}

Equation\ \eqref{eq:DMCCP} takes the form of a multinomial mixed--logit
model with the optimal integrated CCPs $P_{j}^{0\ast }(v_{0})$ entering as
alternative-specific weights. In particular, the DM will assign zero
probability to a given inside alternative $j$ if and only if $P_{j}^{0\ast
}(v_{0})=0$. Thus, in contrast to most ARUMs, the optimal CCPs of the RI\
model may lie on the boundary of the simplex. This property makes the model
attractive for applications where certain alternatives are never chosen by
the DMs, but it also makes the theoretical analysis and numerical
computation of the optimal solution(s) more difficult. In particular, given
the model allows for corner solutions, we cannot characterise $P_{j}^{\ast
}(v_{0},v_{1})$ as a solution to a simple set of first-order conditions.

Instead, we first show that $P^{0\ast }(v_{0})$ can be expressed as the
solution to a maximisation problem implied by the original one: Replacing $%
P_{j}^{0\ast }(v_{0})$ by a fixed candidate $P^{0}(v_{0})$ in Equation %
\eqref{eq:DMCCP} and then substituting the resulting expression %
\eqref{eq:DMCCP} into Equation \eqref{eq:DMproblemobjective}, we obtain the
following characterisation of $P^{0\ast }(v_{0})$ as the maximiser of an
integrated version $Q^{0}$ of the original objective function $Q$:

\begin{lemma}
\label{lemma:optimalp0} The optimal vector of integrated CCPs satisfies 
\begin{equation}
P^{0\ast }\left( v_{0}\right) \in \arg \max_{P^{0}\in \Delta (\mathcal{J}%
)}Q^{0}(P^{0};v_{0}),  \label{eq: P0 charac}
\end{equation}%
where 
\begin{equation}
Q^{0}(P^{0};v_{0}):=\lambda \int_{\mathcal{V}}\log \left( \sum_{j\in 
\mathcal{J}}P_{j}^{0}(v_{0})\exp \left( v_{j}/\lambda \right) \right) d\pi
_{V|V_{0}}(v|v_{0}),  \label{eq: Q0 def}
\end{equation}

If $\left\{ \exp \left( V_{j}/\lambda \right) \right\} _{j\in \mathcal{J}}$
is affinely independent $\pi _{V|V_{0}}(\cdot |v_{0})$-almost surely, then
there exists a unique solution to (\ref{eq: P0 charac}).
\end{lemma}

The first part of the lemma provides a characterisation of (the set of)
optimal inctegrated CCPs as the maximisers of $P^{0}\mapsto
Q^{0}(P^{0};v_{0})$. This is objcective function, however, is not strictly
concave on $\Delta (\mathcal{J})$ and so a unique solution is not
guaranteed. If the prior $\pi _{V|V_{0}}$ has degeneracies, multiple
solutions may exist. The second part requires the prior to exhibit
sufficient variation to make the integrand appearing in the definition of $%
Q^{0}(P^{0};v_{0})$ in (\ref{eq: Q0 def}) strictly concave on the relevant
simplex, cf. \citet{matvejka2015rational} and \citet{caplin2019rational}.

We will now analyse the maximisation problem (\ref{eq: P0 charac}) with
particular focus on the behaviour of the mapping $v_{0}\mapsto P^{0\ast
}\left( v_{0}\right) $. For this analysis, we restrict ourselves to a class
of additive conditional RI models.

\subsection{A class of additive models\label{sec: additive RI}}

We here impose more structure on the conditional model to ensure that the
mapping $v_{0}\mapsto P^{0\ast }\left( v_{0}\right) $ is sufficiently
regular. This involves imposing sufficient conditions on the prior $\pi
_{V|V_{0}}\left( \cdot |v_{0}\right) $ as a function of $v_{0}$. We will
here restrict ourselves to an additive class of priors for which $\pi
_{V|V_{0}}\left( \cdot |v_{0}\right) $ takes a simple form thereby
facilitating our analysis. We conjecture that many of our results can be
generalised to non--additive versions, where $v_{0}$ enters the prior $\pi
_{V|V_{0}}\left( \cdot |v_{0}\right) $ in a more complex manner, under
suitable regularity conditions but leave this for future research.

Specifically, we assume that the total payoff $V_{j}$ can be decomposed into

\begin{equation*}
V=V_{0}+V_{1},
\end{equation*}%
where $V_{0}=\left\{ V_{0,j}\right\} _{j\in \mathcal{J}}\in \mathcal{V}%
_{0}\subseteq \mathbb{R}^{J+1}$ is the known component, while $V_{1}$ is
uncertain to the DM. The DM holds prior beliefs about $V_{1}|V_{0}$
represented by a probability measure $\pi _{V_{1}}\left( v_{1}\right) $ on $%
\mathcal{V}_{1}$. This induces the following conditional prior for $%
V|V_{0}=v_{0}$, 
\begin{equation}
\pi _{V|V_{0}}\left( v|v_{0}\right) =\pi _{V_{1}}\left( v-v_{0}\right) .
\label{eq: additive prior}
\end{equation}%
A strong restriction of this specification is that the prior $\pi $ on $%
V_{1}|V_{0}=v_{0}$ does not depend on $v_{0}$. This assumes that $V_{0}$ and 
$V_{1}$ are treated as independent by the DM: Knowledge of $V_{0}$ only
affects the prior mean of $V$ -- not other parts of the prior distribution.
This feature is important since it ensures that $v_{0}$ only enters $\pi
_{V|V_{0}}\left( v|v_{0}\right) $ in above simple additive fashion. This
greatly facilitates the subsequent analysis; one could imagine more flexible
specifications, but then analysis would involve more complex conditions
ensuring that $v_{0}\mapsto \pi _{V|V_{0}}\left( v|v_{0}\right) $ is
well--behaved. At the same time, we will impose no regularity conditions on $%
\pi _{V_{1}}$ and allow it to be completely unrestricted besides requiring
it to deliver a unique solution $P^{\ast }$. As such, the following results
are fully nonparametric in nature. Under (\ref{eq: additive prior}) and with
a slight abuse of notation, we will work with the following reparameterised
version of the conditional CCPs, 
\begin{equation*}
P_{j}^{\ast }\left( v_{0},v_{1},\lambda \right) =\Pr \left(
Y=j|V_{0}=v_{0},V_{1}=v_{1}\right) =\Pr \left(
Y=j|V_{0}=v_{0},V=v_{1}+v_{0}\right) ,
\end{equation*}%
where we now are explicit about its dependence on the underlying cost of
information $\lambda $, which was suppressed in the general conditional
model. The corresponding vector of integrated CCPs are given by%
\begin{equation*}
P^{0\ast }\left( v_{0},\lambda \right) =\int_{\mathcal{V}_{1}}P^{\ast
}\left( v_{0},v_{1}\right) d\pi _{V_{1}}(v_{1}).
\end{equation*}%
Similarly, we also now are explicit about the dependence of $%
Q^{0}(P^{0};v_{0},\lambda )$ on $\lambda $.

Above additive model for the conditional prior is a special case of the
general framework of the previous section. In particular, the following
version of the affine independence assumption of Lemma \ref{lemma:optimalp0}
ensures that a unique solution exists to the class of additive models:

\begin{assumption}
\label{ass: Affine ind}$\left\{ \exp \left( V_{1,j}/\lambda \right) \right\}
_{j\in \mathcal{J}}$ is affinely independent $\pi _{V_{1}}$-almost surely
and $E_{\pi _{V_{1}}}\left[ \left\Vert V_{1}\right\Vert \right] <\infty $.
\end{assumption}

Unlike some existing treatments of rational inattention models, our analysis
does not require exponential moments of the uncertain payoffs. The
first-moment restriction suffices to ensure finiteness of $Q^{0}$, while
almost-sure affine independence of the exponential transformations suffices
for strict concavity. Thus, our results accommodate substantially
heavier-tailed payoff distributions.

The first-moment restriction is imposed for convenience and can be relaxed.
Following the normalisation approach of \citet{mcfadden1981econometric} and %
\citet{sorensen2022mcfadden}, one could instead work with a centred version
of $Q^{0}$, obtained by subtracting its value at a fixed interior
probability vector. This would allow us to dispense with moment restrictions
on $V_{1}$, at the cost of some additional technical arguments.

We now analyse the mapping $v_{0}\mapsto P^{0\ast }\left( v_{0},\lambda
\right) $ under the additive specification. To this end, we first enforce
normalisations on the payoff vector that are well-known in the literature on
ARUM:

\begin{lemma}
\label{lemma:outsidenormalization} Any additive transformation of the prior $%
\pi _{V_{1}}$ on $V_{1}$ to a new prior $\tilde{\pi}_{V_{1}}$ corresponding
to a common location shift $V+\mathbf{1}_{J+1}C$, where $C\in \mathbb{R}$ is
a random variable with $\mathbb{E}\left[ \left\vert C\right\vert \right]
<\infty $, does not affect the DM's strategy:%
\begin{equation*}
P^{0\ast }(v_{0},\lambda )\in \arg \max_{P^{0}\in \Delta (\mathcal{J}%
)}Q^{0}(P^{0};v_{0},\lambda )=\arg \max_{P^{0}\in \Delta (\mathcal{J})}%
\tilde{Q}^{0}(P^{0};v_{0},\lambda ).
\end{equation*}%
where $Q^{0}(P^{0};v_{0},\lambda )$ and $\tilde{Q}^{0}(P^{0};v_{0},\lambda )$
are the objective functions corresponding to $\pi _{V_{1}}$ and $\tilde{\pi}%
_{V_{1}}$, respectively.
\end{lemma}

Hence, only relative differences in $V$ affect the optimal probabilities,
and the outside option's payoff $V_{0}=V_{0,0}+V_{1,0}$ can be normalised to
an arbitrary constant without loss of generality. From this point on, we
impose the normalisation $V_{0,0}=0$ and $V_{1,0}=0$ so that the payoff
vector can be represented by the inside components alone and so redefine the
payoff vectors as $v_{0}=(v_{0,1},\ldots ,v_{0,J})\in \mathbb{R}^{J}$ and $%
V_{1}=\left( V_{1,1},...,V_{1,J}\right) \in \mathbb{R}^{J}$. Similarly, we
redefine all above functions so that they are now functions of these
re-defined variables. For example, $P^{0\ast }(v_{0},\lambda )$ denotes the
optimal integrated CCP's corresponding to a given $v_{0}\in \mathbb{R}^{J}$.

In what follows, we define $P^{0\ast }(v_{0},\lambda )$ for arbitrary $%
v_{0}\in \mathbb{R}^{J}$, and $P^{\ast }(v_{0},v_{1},\lambda )$ for
arbitrary $\left( v_{0},v_{1}\right) \in \mathbb{R}^{2J}$, irrespective of
the supports of $V_{0}$ and $V_{1}$ in the DM's model, including her prior $%
\pi _{V_{1}}$, and our econometric model. These extensions are allowed since 
$P^{0\ast }(v_{0},\lambda )$ given by (\ref{eq: P0 charac}) is well-defined
for any $v_{0}\in \mathbb{R}^{J}$ which in turn implies that $P^{\ast
}(v_{0},v_{1},\lambda )$ given in (\ref{eq:DMCCP}) is well-defined. Note
here that this is one of the attractive features of working with the
additive model since this ensures that $\pi _{V|V_{0}}\left( \cdot
|v_{0}\right) $ in (\ref{eq: additive prior}) is well-defined for all $%
v_{0}\in \mathbb{R}^{J}$. This extension allows us to analyse the global
properties of the CCP mappings. In our econometric identification analysis
in the following section, however, $\left( v_{0},v_{1}\right) $ will have to
be restricted to the support of the data-generating processes $V_{0}$ and $%
V_{1}$. Also note that the additive specification allows us to transfer the
location parameter vector $v_{0}$ from the prior that we are integrating
over to the integrand in the defintiion of $Q^{0}(P^{0};v_{0},\lambda )$.
This allows us to treat $v_{0}\mapsto Q^{0}(P^{0};v_{0},\lambda )$ as a
smooth mapping even if $\pi _{V_{1}}$ is non--smooth, e.g., discrete.

We first record a useful concavity property of the integrated objective $%
Q^{0}$:

\begin{lemma}
\label{lemma:Q0_concavity} For every $v_{0}\in \mathbb{R}^{J}$ and $\lambda
>0$, the mapping $P^{0}\mapsto Q^{0}(P^{0};v_{0},\lambda )$ is concave on $%
\Delta (\mathcal{J})$. Under Assumption \ref{ass: Affine ind}, it is finite
and strictly concave on $\func{int}\Delta (\mathcal{J})$. Moreover, for
every $P^{0}\in \func{int}\Delta (\mathcal{J})$, the mapping $v_{0}\mapsto
Q^{0}(P^{0};v_{0},\lambda )$ is twice continuously differentiable, with 
\begin{equation*}
\nabla_{v_{0,j}}Q^{0}(P^{0};v_{0},\lambda )=\int_{\mathcal{V}_{1}}\frac{%
P_{j}^{0}\exp ((v_{0,j}+v_{1,j})/\lambda )}{1+\sum_{k=1}^{J}P_{k}^{0}\left\{
\exp ((v_{0,k}+v_{1,k})/\lambda )-1\right\} }d\pi _{V_{1}}(v_{1}).
\end{equation*}
\end{lemma}

Note that we here characterise $Q^{0}(P^{0};v_{0},\lambda )$ for any $%
v_{0}\in \mathbb{R}^{J}$. This is formally valid since $Q^{0}(P^{0};v_{0},%
\lambda )$ and $P^{0\ast }(v_{0},\lambda )$ are well-defined for all for $%
v_{0}\in \mathbb{R}^{J}$, and this even if the underlying random variable $%
V_{0}$ in fact has bounded support.

We next characterise $P^{0\ast }(v_{0},\lambda )$ and its derivative in
terms of the corresponding surplus function, 
\begin{equation*}
Q_{\max }^{0}(v_{0},\lambda ):=Q^{0}(P^{0\ast }(v_{0},\lambda
);v_{0},\lambda ).
\end{equation*}%
This is a generalisation of the classical surplus function known from ARUM.
Similar to the ARUM, we can characterise $P^{0\ast }$ as $P^{0\ast }\left(
v_{0},\lambda \right) =\nabla _{v_{0}}Q_{\max }^{0}(v_{0},\lambda )$. This
identity in turn allows us to analyse the mapping $\left( v_{0},\lambda
\right) \mapsto P^{0\ast }\left( v_{0},\lambda \right) $. In particularly,
we will show that $P^{0\ast }\left( \cdot ,\lambda \right) :\mathcal{V}_{0,%
\mathrm{reg}}\mapsto \func{int}\Delta (\mathcal{J)}$ is one-to-one, where 
\begin{eqnarray}
\mathcal{V}_{0,\mathrm{reg}}\left( \lambda \right) &=&\left\{ v_{0}\in 
\mathbb{R}^{J}:P^{0\ast }(v_{0},\lambda )\in \func{int}\Delta (\mathcal{J}%
)\right\}  \label{eq: V_0,reg def} \\
&=&\left\{ v_{0}\in \mathbb{R}^{J}:P^{0\ast }(v_{1}|v_{0},\lambda )\in \func{%
int}\Delta (\mathcal{J})\text{ for all }v_{1}\in \mathbb{R}^{J}\right\} 
\notag
\end{eqnarray}%
is the set of "regular" known payoffs, where the second equality uses Eq. (%
\ref{eq:DMCCP}). The following theorem first characterises $\mathcal{V}_{0,%
\mathrm{reg}}\left( \lambda \right) $, and then analyses the behaviour of $%
P^{0\ast }\left( v_{0},\lambda \right) $ on this set:

\begin{theorem}
\label{Thm: P0 one-to-one}Under Assumption \ref{ass: Affine ind} and for any 
$\lambda >0$, the following hold:

\begin{enumerate}
\item[(i)] The surplus function $Q_{\max }^{0}\left( \cdot ,\lambda \right) $
is convex on $\mathbb{R}^{J}$ and strictly convex along every non-degenerate
line segment contained in $\mathcal{V}_{0,\mathrm{reg}}\left( \lambda
\right) $.

\item[(ii)] For $\nu _{0}\in (0,\infty )^{J}$, define 
\begin{equation*}
p_{j}^{0}(\nu _{0},\lambda )=\int_{\mathcal{V}_{1}}\frac{\nu _{0,j}\exp
(v_{1,j}/\lambda )}{1+\sum_{k=1}^{J}\nu _{0,k}\exp (v_{1,k}/\lambda )}d\pi
_{V_{1}}(v_{1}),\quad j=1,\ldots ,J,
\end{equation*}%
and $p_{0}^{0}(\nu _{0},\lambda )=1-\sum_{j=1}^{J}p_{j}^{0}(\nu _{0},\lambda
)$. Then $\mathcal{V}_{0,\mathrm{reg}}(\lambda )$ is the open, non-empty and
connected set 
\begin{equation}
\mathcal{V}_{0,\mathrm{reg}}(\lambda )=\left\{ \left( \lambda \ln \frac{\nu
_{0,j}p_{0}^{0}(\nu _{0},\lambda )}{p_{j}^{0}(\nu _{0},\lambda )}\right)
_{j=1}^{J}:\nu _{0}\in (0,\infty )^{J}\right\} .  \label{eq: V0reg exact}
\end{equation}

\item[(iii)] For every $v_{0}\in \mathcal{V}_{0,\mathrm{reg}}\left( \lambda
\right) $, 
\begin{equation*}
P^{0\ast }\left( v_{0},\lambda \right) =\nabla _{v_{0}}Q_{\max
}^{0}(v_{0},\lambda ),\text{ \ \ }\nabla _{v_{0}}P^{0\ast }\left(
v_{0},\lambda \right) =\nabla _{v_{0}}^{2}Q_{\max }^{0}(v_{0},\lambda )\succ
0.
\end{equation*}%
In particular, the restricted integrated CCP map $P^{0\ast }(\cdot ,\lambda
):\mathcal{V}_{0,\mathrm{reg}}(\lambda )\mapsto \func{int}\Delta (\mathcal{J}%
)$ is a $C^{1}$ diffeomorphism.

\item[(iv)] At every $v_{0}\in \mathcal{V}_{0,\mathrm{reg}}(\lambda )$, $%
P^{0\ast }(v_{0},\lambda )$ is continuously differentiable in $\lambda $ and 
\begin{equation}
\nabla _{\lambda }P^{0\ast }(v_{0},\lambda )=\nabla _{v_{0},\lambda
}^{2}Q_{\max }^{0}(v_{0},\lambda )=-\left[ \nabla _{P^{0}P^{0}}^{2}Q^{0\ast }%
\right] ^{-1}\nabla _{P^{0}\lambda }^{2}Q^{0\ast },
\label{eq: P0 lambda derivative}
\end{equation}%
where $\nabla _{P^{0}\lambda }^{2}Q^{0\ast }=\nabla _{P^{0}\lambda
}^{2}Q^{0}(P^{0\ast }(v_{0},\lambda );v_{0},\lambda )$ and similar for $%
\left[ \nabla _{P^{0}P^{0}}^{2}Q^{0\ast }\right] $.
\end{enumerate}
\end{theorem}

The characterisation \eqref{eq: V0reg exact} makes the dependence of $%
\mathcal{V}_{0,\mathrm{reg}}(\lambda )$ on $\lambda $ explicit, although it
does not generally yield a closed-form expression for the set. For $J=1$, $%
\mathcal{V}_{0,\mathrm{reg}}(\lambda )$ admits a simple explicit
characterisation: Since $Q^{0}(P^{0};v_{0},\lambda )$ is strictly concave in 
$P^{0}\in \left( 0,1\right) $, its maximiser lies in $\left( 0,1\right) $ if
and only if 
\begin{eqnarray*}
\left. \nabla _{P^{0}}Q^{0}(P^{0};v_{0},\lambda )\right\vert _{P^{0}=0}
&=&\lambda \left\{ \exp (v_{0}/\lambda )E_{\pi _{V_{1}}}\left[ \exp
(V_{1}/\lambda )\right] -1\right\} >0, \\
\left. \nabla _{P^{0}}Q^{0}(P^{0};v_{0},\lambda )\right\vert _{P^{0}=1}
&=&\lambda \left\{ 1-\exp (-v_{0}/\lambda )E_{\pi _{V_{1}}}\left[ \exp
(-V_{1}/\lambda )\right] \right\} <0.
\end{eqnarray*}%
Solving these two inequalities w.r.t. $v_{0}$ yields 
\begin{equation*}
\mathcal{V}_{0,\mathrm{reg}}(\lambda )=\left( -\lambda \ln E_{\pi _{V_{1}}}%
\left[ \exp (V_{1}/\lambda )\right] ,\lambda \ln E_{\pi _{V_{1}}}\left[ \exp
(-V_{1}/\lambda )\right] \right) ,
\end{equation*}%
which shows that, for $J=1$, the regular region is an interval whose
endpoints depend on $\lambda $. The interval contracts as the information
cost increases. For $J\geq 2$, however, the parametrisation 
\eqref{eq: V0reg
exact} does not generally imply nested regular regions. Related to this, the
signs of the vector of derivatives in part (iv) are likewise not determined
in general and its components may change sign as $\lambda $ varies. This
complicates identifying $\lambda $ from the CCPs.

The characterisation $P^{0\ast }=\nabla _{v_{0}}Q_{\max }^{0}$ is a
generalisation of the Roy's Identity of ARUM models. The positive
definiteness of $\nabla _{v_{0}}P^{0\ast }$ is used to show that $P^{0\ast }$
is injective on $\mathcal{V}_{0,\mathrm{reg}}(\lambda )$. In contrast, the
sign of $\nabla _{\lambda }P^{0\ast }(v_{0},\lambda )$ has no global sign
and it can change depending on the values of $\left( v_{0},\lambda \right) $%
. Two limiting cases also clarify the effect of information costs. If the
solution to $\max_{j\in \mathcal{J}}\{v_{0,j}+V_{1,j}\}$ is unique $\pi
_{V_{1}}$-a.s., then, for each $j=0,\ldots ,J$, 
\begin{equation*}
\lim_{\lambda \downarrow 0}P_{j}^{0\ast }(v_{0},\lambda )=\pi _{V_{1}}\left(
j=\arg \max_{k\in \mathcal{J}}\{v_{0,k}+V_{1,k}\}\right) ,\text{ \ \ }%
\lim_{\lambda \rightarrow \infty }P_{j}^{0\ast }(v_{0},\lambda )=\mathbf{1}%
\{j=j^{\ast }\}.
\end{equation*}%
The first result follows from the cost of full information being bounded by $%
\lambda \ln (J+1)$, which vanishes as $\lambda \downarrow 0$, while for the
second, comparison with the best constant choice implies that optimal mutual
information tends to zero as $\lambda \rightarrow \infty $. In the first
case, the integrated CCPs converge to the full-information ARUM choice
probabilities, while in the second the DM acquires no information and
selects the alternative with the highest expected payoff.

Next, we combine Lemma \ref{lemma:DMCCP} and Theorem \ref{Thm: P0 one-to-one}
to analyse the derivatives and injectivity of $P^{\ast }$, where here and in
the following we use $\Sigma \left( p\right) :=\left\{ \func{diag}%
(p)-pp^{\prime }\right\} $ to denote the variance of any multinomial choice
variable with choice probability vector $p$.

\begin{theorem}
\label{Thm: P one-to-one}Under Assumption \ref{ass: Affine ind}, for every $%
\lambda>0$ the following hold:

\begin{enumerate}
\item[(i)] For every $v_{0}\in \mathbb{R}^{J}$, $P^{\ast
}(v_{0},v_{1},\lambda ):\mathbb{R}^{J}\mapsto \Delta (\mathcal{J})$ is
continuously differentiable with symmetric and positive semidefinite
Jacobian 
\begin{equation}
\nabla _{v_{1}}P^{\ast }=\frac{1}{\lambda }\Sigma \left( P^{\ast }\right)
\label{eq: P v1 Jacobian}
\end{equation}

If $v_{0}\in \mathcal{V}_{0,\mathrm{reg}}(\lambda )$, then $\nabla
_{v_{1}}P^{\ast }(v_{0},v_{1},\lambda )\succ 0$ for every $v_{1}\in \mathbb{R%
}^{J}$. Consequently, $P^{\ast }(v_{0},\cdot ,\lambda ):\mathbb{R}%
^{J}\mapsto \func{int}\Delta (\mathcal{J})$ is one-to-one. Moreover, for $%
j\neq k$, 
\begin{equation}
\nabla _{v_{1,k}}P_{j}^{\ast }=-\frac{P_{j}^{\ast }P_{k}^{\ast }}{\lambda }%
<0.  \label{eq: P v1 cross derivative}
\end{equation}

\item[(ii)] For every $v_{1}\in \mathbb{R}^{J}$, $P^{\ast }(\cdot
,v_{1},\lambda ):\mathcal{V}_{0,\mathrm{reg}}(\lambda )\mapsto \func{int}%
\Delta (\mathcal{J})$ is one-to-one with%
\begin{equation}
\nabla _{v_{0}}P^{\ast }=\Sigma (P^{\ast })\left[ \frac{1}{\lambda }%
I_{J}+\Sigma (P^{0\ast })^{-1}\nabla _{v_{0}}P^{0\ast }\right] ,
\label{eq: P v0 Jacobian}
\end{equation}%
This Jacobian has full rank $J$, but is not symmetric in general.

\item[(iii)] For every $v_{0}\in \mathcal{V}_{0,\mathrm{reg}}(\lambda )$ and 
$v_{1}\in \mathbb{R}^{J}$, the information cost satisfies 
\begin{equation}
\lambda =\frac{\func{tr}\Sigma (P^{\ast }(v_{0},v_{1},\lambda ))}{\func{tr}%
\nabla _{v_{1}}P^{\ast }(v_{0},v_{1},\lambda )}.
\label{eq: P stacked lambda identification}
\end{equation}
\end{enumerate}
\end{theorem}

Similar to Lemma \ref{lemma:Q0_concavity}, above analysis is not restricted
to $v_{0}$ and $v_{1}$ lying within the potentially bounded supports of $%
V_{0}$ and $V_{1}$ in the population of interest. It is here also important
to stress that while $v_{0}\mapsto P^{\ast }(v_{0},v_{1},\lambda )$ is
one--to--one on $\mathcal{V}_{0,\mathrm{reg}}\left( \lambda \right) $, it is
not monotonic, which is in contrast to $v_{0}\mapsto P^{0\ast
}(v_{0},\lambda )$. This weaker injectivity property of $P^{\ast }$ becomes
a critical obstacle for showing identification of the econometric model
presented in the next section when unobserved heterogeneity is present in
it. Part (iii) provides an explicit formula for $\lambda $ in terms of the
conditional choice probabilities and their $v_{1}$-derivatives. An
alternative version not involving derivatives is, for any $v_{1},\tilde{v}%
_{1}\in \mathbb{R}^{J}$ with $v_{1,j}\neq \tilde{v}_{1,j}$, 
\begin{equation}
\lambda =\frac{v_{1,j}-\tilde{v}_{1,j}}{\log \frac{P_{j}^{\ast
}(v_{0},v_{1},\lambda )}{P_{0}^{\ast }(v_{0},v_{1},\lambda )}-\log \frac{%
P_{j}^{\ast }(v_{0},\tilde{v}_{1},\lambda )}{P_{0}^{\ast }(v_{0},\tilde{v}%
_{1},\lambda )}}.  \label{eq: P lambda log odds difference}
\end{equation}%
Thus $\lambda $ is determined either by the conditional choice probabilities
and their $v_{1}$-derivatives, or by conditional choice probabilities at two
values of $v_{1}$.

\section{The Econometric Model\label{sec: Econometric Model}}

We now build an econometric model based on the additive conditional RI model
of the previous sectin. We specify the payoff vectors of the inside
alternatives, $V_{0},V_{1}\in \mathbb{R}^{J}$, and the unit cost of
information $\Lambda >0$ as follows: 
\begin{equation}
V_{0}=u_{0}(X_{0})+\varepsilon _{0},\text{ \ \ }V_{1}=u_{1}(X_{1})+%
\varepsilon _{1},\text{ \ \ }\Lambda =\lambda (X_{2})+\varepsilon _{2},
\label{eq:Vdef}
\end{equation}%
where $u_{0}(\cdot )$, $u_{1}(\cdot )$, and $\lambda (\cdot )\geq 0$ are
unknown functions to be identified, while $X=(X_{0},X_{1},X_{2})$ and $%
\varepsilon =(\varepsilon _{0},\varepsilon _{1},\varepsilon _{2})$ are
random variables from the Econometrician's perspective. The terms $%
\varepsilon _{0}=(\varepsilon _{0,1},\ldots ,\varepsilon _{0,J})\in \mathcal{%
E}_{0}\subseteq \mathbb{R}^{J}$ and $\varepsilon _{1}=(\varepsilon
_{1,1},\ldots ,\varepsilon _{1,J})\in \mathcal{E}_{1}\subseteq \mathbb{R}%
^{J} $ are vectors over the inside alternatives while $\varepsilon _{2}\geq
0 $ is a scalar. For\ $t\in \{0,1\}$, suppose furthermore that $X_{t}=\left(
X_{t,1},X_{t,2}\right) $, where $X_{t,1}=(X_{t,1,1},\ldots
,X_{t,1,J})^{\prime }\in \mathcal{X}_{t,1}\subseteq \mathbb{R}^{J}$ contains
alternative specific covariates, while $X_{t,2}\in \mathcal{X}_{t,2}$
contains variables that enter all payoffs, 
\begin{equation*}
u_{t}(X_{t})=(u_{t,1}(X_{t,1,1},X_{t,2}),\ldots
,u_{t,J}(X_{t,1,J},X_{t,2}))\in {\mathbb{R}}^{J}.
\end{equation*}%
To avoid overly complicated notation and arguments in our identification
results, we restrict ourselves to the case of scalar alternative specific
covariates, $X_{t,1,j}\in \mathbb{R}$. Our results can be extended to the
multivariate case at the price of heavier notation and more complicated
assumptions. Finally, $X_{2}\in \mathcal{X}_{2}$ contains variables that
affect information--costs. These are individual--specific variables that
(partially) explain DM's ability to process information.

Next, we introduce the Econometrician and describe her information set.
After $V_{1}$ has been realised, the Econometrician arrives and gathers data
from the population of heterogeneous DMs. For each DM, the Econometrician
observes the DM's choice $Y\in \mathcal{J}$, together with the covariates $X$
entering our model (\ref{eq:Vdef}). However, the Econometrician does not
observe $\varepsilon =\left( \varepsilon _{0},\varepsilon _{1},\varepsilon
_{2}\right) $. Thus, there is a potential information asymmetry between the
Econometrician and the DM as already discussed in the introduction: The DM
is uncertain about $V_{1}$ but knows $V_{0}$; the Econometrician observes
parts of each of these variables.

Since the Econometrician observes the entire population, the joint
distribution of the observables $(Y,X)$ is known to her. It follows that the
following CCPs,%
\begin{equation}
\Phi _{j}(x):=\Pr (Y=j\mid X=x),\qquad j\in \{1,\ldots ,J\},
\label{eq: Phi def}
\end{equation}%
are identified from the population distribution of $\left( Y,X\right) $. It
is here important to distinguish between the DM's prior distribution of $%
V_{1}$, $\pi _{V_{1}}$, and the true/physical distribution of $V_{1}$, $%
F_{V_{1}}$. We allow for the DM to be irrational and having incorrect prior
on this, $\pi _{V_{1}}\neq F_{V_{1}}$. Fortunately, this does not affect the
physical distribution of $Y|\left( V_{0},V_{1}\right) $ since the incorrect
prior only affects the distribution with which states occur, not the
conditional action rule given the state. So, if the econometrician should
observe $\left( Y,V_{0},V_{1}\right) $, with $Y|\left( V_{0},V_{1}\right)
\sim P^{\ast }$, then, under the physical law $F_{V_{1}}$, $\Pr_{F}\left(
Y|V_{0}=v_{0},V_{1}=v_{1}\right) =P^{\ast }(v_{0},v_{1},\lambda )$.

The aim of the subsequent analysis is to show that knowledge of $\Phi
=\left( \Phi _{1},...,\Phi _{J}\right) $ allows us to nonparametrically
identify (up to normalisations) the functions $u_{0}(\cdot )$, $u_{1}(\cdot
) $ and $\lambda (\cdot )$ and potentially also the prior $\pi _{V_{1}}$. In
order to do so, we first establish a mapping taking the DM's CCPs $P^{\ast }$
into $\Phi $. This will involve integrating out the unobserved components in 
$\varepsilon $. To ensure that variation in $X$ only generates shifts in $%
u_{0}(X_{0})$, $u_{1}(X_{1})$ and $\lambda \left( X_{2}\right) $ and not the
distribution of $\varepsilon $, we impose the following exogeneity
assumption:

\begin{assumption}
\label{ass:idependentXWZande}$X=\left( X_{0},X_{1},X_{2}\right) $ and $%
\varepsilon =\left( \varepsilon _{0},\varepsilon _{1},\varepsilon
_{2}\right) \sim F_{\varepsilon }$ are mutually independent with $%
\varepsilon _{2}\geq 0$ a.s.
\end{assumption}

The assumption that $X$ is independent of the unobserved components is
standard in most non-linear structural econometric models, including ARUM
and is essential for our identification argument to work. If this is
violated, then we would struggle to identify $u_{0}(\cdot )$, $u_{1}(\cdot )$
and $\lambda (\cdot )$ since we will rely on the observed variation in $X$
to trace out their values.

In the analysis of the mapping $\left( u_{0}(\cdot ),u_{1}(\cdot ),\lambda
(\cdot )\right) \mapsto \Phi $, it proves convenient to introduce the
following vector-valued function, 
\begin{equation*}
\Pi (u_{0},u_{1},\lambda ):=\int_{\mathcal{E}}P^{\ast
}(u_{0}+e_{0},u_{1}+e_{1},\lambda +e_{2})dF_{\varepsilon
}(e_{0},e_{1},e_{2}),
\end{equation*}%
where $\mathcal{E}$ is the support of $\varepsilon $. The identified CCPs in
(\ref{eq: Phi def}) then satisfy 
\begin{equation}
\Phi (x_{0},x_{1},x_{2})=\Pi \left( u_{0}(x_{0}),u_{1}(x_{1}),\lambda
(x_{2})\right)  \label{eq: Phi = Pi}
\end{equation}%
under Assumption \ref{ass:idependentXWZande}. This identity will form the
basis of our identification analysis. It will allow us to use $\Phi (\cdot )$
to learn about the unknown structural functions $u_{0}(\cdot )$, $%
u_{1}(\cdot )$, and $\lambda (\cdot )$. This, however, requires $\Pi
(u_{0},u_{1},\lambda )$ to be injective in each of its three components $%
(u_{0},u_{1},\lambda )$. As should be clear from the injectivity results for 
$P^{\ast }$, this property will not hold globally and so we will generally
only be able to identify the relevant functions on a subset of the support
of $X$.

\section{Identification of composite functions under injectivity}

\label{sec: ID composite injectivity}

We here present a set of new identification results for the following
abstract class of models,%
\begin{equation}
\Phi (w,z)=\Psi (\nu (w),z),  \label{eq: composite model}
\end{equation}%
where $\left( w,z\right) \in \mathcal{D}\subseteq \mathbb{R}^{d_{w}+d_{z}}$.
The function $\Phi $ together with $\mathcal{D}$ is assumed to be identified
(known) together with the model/structure (), while $\Psi $ and $\nu $ are
unknown functions to be identified. Below, we provide sufficient conditions
for identification under an increasing number restriction on above model. We
will then apply these general results to the problem of identifying the
functions $u_{0}$, $u_{1}$ and $\lambda $ in the RI-ARUM model in the next
section. For example, for identification of $u_{1}\left( x_{1}\right) $ in
the RI--ARUM, we first set $w=x_{1}$, $z=\left( x_{0},x_{2}\right) $, $\nu
\left( w\right) =u_{1}\left( x_{1}\right) $ and then $\Psi \left(
u_{1},x_{0},x_{2}\right) =\Pi \left( u_{0}\left( x_{0}\right) ,u_{1},\lambda
\left( x_{2}\right) \right) $.

The master theorems should be of independent interest since they generalise
existing results found in the literature, including the ones found in %
\citet{allen2019identification} and \citet{Fosgerau2020}, and can
potentially be applied to other models, where the identification problem can
be expressed on the form of (\ref{eq: composite model}). Our results also
provide new insights into the interplay between model restrictions and
identification strength. Specifically, we first provide partial
identification results for $\nu $ under very weak restrictions on $\Psi $
and $\nu $. We then show how added restrictions on these allow us to
identify further features of $\nu $. Proofs of the results of this section
can be found in Appendix \ref{sec: Proofs ID composite}.

The domain $\mathcal{D}$ that $\left( w,z\right) $ is restricted to lie in
plays the role of the support of some underlying random variables $\left(
W,Z\right) $. Let $\mathcal{Z}=\left\{ z:\left( w,z\right) \in \mathcal{D}%
\right\} $ be the support of $Z$; for a given $z\in \mathcal{Z}$, define $%
\mathcal{W}\left( z\right) =\left\{ w:\left( w,z\right) \in \mathcal{D}%
\right\} $ and $\mathcal{V}\left( z\right) =\nu \left( \mathcal{W}\left(
z\right) \right) $. A key condition for the following identification result
is that $\nu \mapsto \Psi (\nu ,z)$ is injective. In the existing literature
concerned with identification of structural models on the form (\ref{eq:
composite model}), this property is assumed to hold globally; that is, for
any $z$, $\Psi (\cdot ,z):\mathcal{V}\left( z\right) \mapsto \mathbb{R}^{J}$
is injective. However, in our application to RI-ARUM, this assumption is
generally violated: The set on which the CCPs are injective may not coincide
with $\mathcal{V}\left( z\right) $, c.f. Theorems \ref{Thm: P0 one-to-one}
and \ref{Thm: P one-to-one}. To accommodate this feature, we impose the
following, weaker assumption on any $\Psi $ belonging to our class of models:

\begin{assumption}
\label{Ass: Pi injective}There exists $\mathcal{Z}_{0}\subseteq \mathcal{Z}$
so that for any $z\in \mathcal{Z}_{0}$, there exists $\mathcal{V}_{\mathrm{%
reg}}\left( z\right) \subseteq \mathbb{R}^{J}$ with $\Psi (\cdot ,z):%
\mathcal{V}_{\mathrm{reg}}\left( z\right) \mapsto \mathbb{R}^{J}$ injective.
\end{assumption}

This assumption allows for $\Psi $ to be non-injective w.r.t. $\nu$ outside
of $\mathcal{V}_{\mathrm{reg}}\left( z\right) $ in which case we are only
able to point identify functions on the part of the support of $\left(
w,z\right) $ for which $\nu\left( w\right) \in \mathcal{V}_{\mathrm{reg}%
}\left( z\right) $. If $\Psi $ has some curvature w.r.t. $\nu$ otuside of $%
\mathcal{V}_{\mathrm{reg}}\left( z\right) $, we may still be able to
partially identify (bound) the function $\nu$. We will not pursue this more
expansive identification strategy here and focus on point identification of $%
\nu$ on the part of its image that overlaps with $\mathcal{V}_{\mathrm{reg}%
}\left( z\right) $. Specifically, let%
\begin{equation}
\mathcal{V}_{0}(z)=\mathcal{V}_{\mathrm{reg}}\left( z\right) \cap \mathcal{V}%
\left( z\right) ,\text{ \ \ }\mathcal{W}_{0}\left( z\right) =\left\{ w\in 
\mathcal{W}(z):\nu(w)\in \mathcal{V}_{0}(z)\right\}  \label{eq: U_0 def}
\end{equation}%
denote the attained regular index region and its covariate pre-image,
respectively. The latter corresponds to the parts of the support of $w|Z=z$
for which we can hope to achieve point identification.

Focusing on $\left( w,z\right) $ for which $w\in \mathcal{W}_{0}\left(
z\right) $, we still face a formidable identification problem since there
exists a continuum of observationally equivalent representations. For any
one--to--one transformation $h:\mathbb{R}^{J}\mapsto \mathbb{R}$, 
\begin{equation*}
\Psi (\nu (w),z)=\tilde{\Psi}(\tilde{\nu}(w),z),\qquad \tilde{\Psi}(\nu
,z):=\Psi (h^{-1}(\nu ),z),,\qquad \tilde{\nu}(w):=h\left( \nu \left(
w\right) \right)
\end{equation*}%
Thus, without further restrictions on $\Psi $ and $\nu $, some cardinal
anchor on the latent index is required. Our first identification result
imposes only Assumption \ref{Ass: Pi injective}, in which case we can only
identify a single component of $\nu $ while having to assume known
(normalise) the remaining parts of it which functions as our anchor in this
case. We then introduce additional restrictions on $\Psi $ and $\nu $ under
which we can relax the degree of normalisation needed on $\nu $ to achieve
identification. We consider the following two types of restrictions : First,
we impose restrictions on the Jacobian of $\Psi $ while leaving $\nu $
unrestricted $\nu $; second, leaving $\Psi $ unrestricted, we introduce
exclusion restrictions on $\nu $. In each case, we show how these
restrictions help identifying the model. Finally, we combine the two sets of
restrictions to achieve our final identification result of this section.

Our fundamental identification strategy is imported from %
\citet{allen2019identification} and \citet{Fosgerau2020}:\ We will assume
that $w$ can be decomposed into $w=\left( w_{1},w_{2}\right) $, and then
require $w_{1}$ and $w_{1}\mapsto \nu \left( w_{1},w_{2}\right) $ to exhibit
sufficient variation in the sense that their support sets or sufficiently
large. This allows us to first identify a component of $\nu \left(
w_{1},w_{2}\right) $, $\nu _{0}(w_{1},w_{2})$, which then functions as an
anchor in the second step: If the anchor exhibits sufficient variation, then
we can identify the remaining components of $\nu $ through a matching
argument, which relies on the injectivity of $\Psi $ of Assumption \ref{Ass:
Pi injective}.

The first result extends the identification result of \citet{Fosgerau2020}
to allow for the known component of $\nu $ in their model to be a function
of both components, and to allow for $\Psi $ not be injective on the full
support of $\nu $. With $\nu _{0}(w_{1},w_{2})$ denoting the known
component, define the following regular parts of the conditional supports
for $W_{2}|Z=x$ and $W_{1}|Z=z$ and relevant images of these,%
\begin{eqnarray*}
\mathcal{W}_{0,2}(z) &=&\func{proj}_{w_{2}}\mathcal{W}_{0}(z),\text{ \ \ }%
\mathcal{W}_{0,1}(w_{2},z)=\{w_{1}:(w_{1},w_{2})\in \mathcal{W}_{0}(z)\}, \\
\mathcal{V}_{0}(w_{2},z) &=&\nu (\mathcal{W}_{0,1}(w_{2},z),w_{2}),\text{ \
\ }\mathcal{V}_{0,0}(w_{2},z)=\nu _{0}(\mathcal{W}_{0,1}(w_{2},z),w_{2})
\end{eqnarray*}

\begin{theorem}
\label{Th: Pi ID 1}Suppose that Assumption \ref{Ass: Pi injective} holds, 
\begin{equation}
\nu (w_{1},w_{2})=\nu _{0}(w_{1},w_{2})+\nu _{2}(w_{2}),
\label{eq: inversion model app}
\end{equation}%
where $\nu _{0}$ is known while $\nu _{2}$ is unknown, together with: (i) $%
\mathcal{V}_{0}(z)$ is connected for every $z\in \mathcal{Z}_{0}$ and $%
\mathcal{V}_{0,0}(w_{2},z)$ is nonempty and open for every $w_{2}\in 
\mathcal{W}_{0,2}(z)$; (ii) whenever $\mathcal{Z}_{1}\cup \mathcal{Z}_{2}=%
\mathcal{Z}_{0}$ with $\mathcal{Z}_{1},\mathcal{Z}_{2}\neq \varnothing $,
there exist $z_{1}\in \mathcal{Z}_{1}$ and $z_{2}\in \mathcal{Z}_{2}$ such
that 
\begin{equation*}
\mathcal{W}_{0,2}(z_{1})\cap \mathcal{W}_{0,2}(z_{2})\neq \varnothing .
\end{equation*}%
Then $\nu _{2}(w_{2})$ is identified for $w_{2}\in \bigcup_{z\in \mathcal{Z}%
_{0}}\mathcal{W}_{0,2}(z)$ under the following normalisation: For some known 
$z_{0}\in \mathcal{Z}_{0}$ and $w_{0,2}\in \mathcal{W}_{0,2}(z_{0})$, $\nu
_{2}(w_{0,2})=0$. Moreover, $\Psi (\nu ,z)$ is identified for $\nu \in 
\mathcal{V}_{0}(z)$ and $z\in \mathcal{Z}_{0}$.
\end{theorem}

Assumptions (i)--(ii) ensure that the subsets of the support of $\left(
W,Z\right) $ on which we can potentially achieve identification are suitably
connected so that we can extend identification from $\mathcal{W}%
_{0,2}(z_{0}) $, where $\left( w_{0,2},z_{0}\right) $ is the point at which
we impose normalisation, to the set $\bigcup_{z\in \mathcal{Z}_{0}}\mathcal{W%
}_{0,2}(z) $.

The normalisation of $\nu$ to satisfy (\ref{eq: inversion model app}) is a
convenient sufficient condition for identification, but it should not be
interpreted as the weakest possible one. A weaker normalisation may be
possible. What is essential is not necessarily knowledge of an entire
additive component, but sufficiently rich known translations of the latent
index $\nu$.

We next investigate how additional restrictions on $\Psi $ and/or $\nu $ can
help strengthening above benchmark identification result. These results will
be developed under the following common set of support restrictions:

\begin{assumption}
\label{Ass: Pi reference support} There exist a known $z_{0}\in \mathcal{Z}%
_{0}$, nonempty open intervals $\mathcal{I}_{j}\subseteq \mathbb{R}$, $%
j=1,\ldots ,J$, and a nonempty open connected set $\mathcal{W}_{2}^{\ast
}\subseteq \mathbb{R}^{d_{w,2}}$ such that $\mathcal{W}^{\ast
}(z_{0}):=\prod_{j=1}^{J}\mathcal{I}_{j}\times \mathcal{W}_{2}^{\ast
}\subseteq \mathcal{W}_{0}(z_{0})$. Moreover, for every $z\in \mathcal{Z}%
_{0} $, $\mathcal{V}_{0}(z)=\nu (\mathcal{W}^{\ast }(z_{0})\cap\mathcal{W}%
(z))$.
\end{assumption}

Assumption \ref{Ass: Pi reference support} requires a single product
reference set $\mathcal{W}^{\ast }(z_{0})$ and that every index value in $%
\mathcal{V}_{0}(z)$ be attained on $\mathcal{W}^{\ast }(z_{0})\cap \mathcal{W%
}(z)$ for every $z\in \mathcal{Z}_{0}$. The latter is a joint restriction on
the size of the conditional reference support and the variation of $\nu $
over it: a relatively small reference set $\mathcal{W}^{\ast }(z_{0})$ may
suffice if its image covers $\mathcal{V}_{0}(z)$, whereas a larger set need
not suffice if its image w.r.t. $\nu $ does not fully overlap with that of $%
\mathcal{V}_{0}(z)$. Thus, the substantive requirement is coverage in index
space rather than the size of the reference set in covariate space. These
restrictions can be replaced by the alternative support conditions stated in
Appendix \ref{sec: support conds} for Theorems \ref{Th: Pi ID joint rank}
and \ref{Th: Pi ID 3}. The alternative conditions permit reference regions
to vary with $w_{2}$ and $z$, but require their index images and covariate
supports to overlap sufficiently to link the local identification results.
Neither set of conditions is generally weaker than the other.

In many applications, including RI--ARUM, the Jacobian of the structural
mapping is symmetric. We here investigate the identifying contents of the
following symmetry and rank condition on the Jacobian $\nabla _{\nu }\Psi $:

\begin{theorem}
\label{Th: Pi ID joint rank} Suppose Assumptions \ref{Ass: Pi injective}--%
\ref{Ass: Pi reference support} hold, $J\geq 2$ and: (i) there exists a
nonempty $\mathcal{Z}^{\ast }\subseteq \mathcal{Z}_{0}$ such that, for every 
$\nu \in \bigcap_{z\in \mathcal{Z}^{\ast }}\mathcal{V}_{0}(z)$, $\Psi (\cdot
,z)$ is continuously differentiable in a neighbourhood of $\nu $ with
symmetric Jacobian $\nabla _{\nu }\Psi (\nu ,z)$ for every $z\in \mathcal{Z}%
^{\ast }$; (ii) there exist $J(J+1)/2$ points $z_{1}^{\ast },\ldots
,z_{J(J+1)/2}^{\ast }\in \mathcal{Z}^{\ast }$ such that 
\begin{equation}
\nabla _{\nu }\Psi (\nu ,z_{1}^{\ast }),\ldots ,\nabla _{\nu }\Psi (\nu
,z_{J(J+1)/2}^{\ast })\text{ are linearly independent}
\label{eq: Pi joint rank sufficient condition}
\end{equation}%
for every such $\nu $, and for every such $\nu $ there exists $z\in \mathcal{%
Z}^{\ast }$ such that $\nabla _{\nu }\Psi (\nu ,z)$ is nonsingular; (iii) $%
\mathcal{W}^{\ast }(z_{0})\subseteq \bigcap_{z\in \mathcal{Z}^{\ast }}%
\mathcal{W}_{0}(z)$ and $\nu (\mathcal{W}^{\ast }(z_{0}))$ is open and
connected. Then $\nu $ is identified on $\bigcup_{z\in \mathcal{Z}_{0}}%
\mathcal{W}_{0}(z)$ under the following normalisations: for known $w_{0},%
\bar{w}_{0}\in \mathcal{W}^{\ast }(z_{0})$ with $\nu _{1}(\bar{w}_{0})\neq
\nu _{1}(w_{0})$, 
\begin{equation}
\nu (w_{0})=0,\qquad \nu _{1}(\bar{w}_{0})=1.
\label{eq: Pi joint rank normalizations}
\end{equation}%
Moreover, $\Psi (\cdot ,z)$ is identified on $\mathcal{V}_{0}(z)$ for all $%
z\in \mathcal{Z}_{0}$.
\end{theorem}

The linear independence condition in (\ref{eq: Pi joint rank sufficient
condition}) can be replaced by the weaker requirement of 
\begin{equation*}
\left\{ H\in \mathbb{R}^{J\times J}:\nabla _{\nu }\Psi (\nu ,z)H\text{ is
symmetric for all }z\in \mathcal{Z}^{\ast }\right\} =\func{span}\{I_{J}\},
\end{equation*}%
without changing the conclusions of Theorem \ref{Th: Pi ID joint rank}. We
maintain (\ref{eq: Pi joint rank sufficient condition}) because it is easier
to interpret and verify.

We next examine the identifying power of additional restrictions on $\nu $
while only requiring $\Psi $ to satisfy Assumption \ref{Ass: Pi injective}.
Specifically, we impose exclusion restrictions on $\nu $ in the form of the
presence of alternative-specific regressors. We assume that $%
w_{1}=(w_{1,1},\ldots ,w_{1,J})$ contains these so that $\nu $ satisfies the
following $J$ exclusion restrictions: 
\begin{equation}
\nu (w_{1},w_{2})=(\nu _{1}(w_{1,1},w_{2}),\ldots ,\nu
_{J}(w_{1,J},w_{2}))^{\prime },  \label{eq: ASV common structure}
\end{equation}%
As before, this is a restriction on the model and so any observationally
equivalent representation should also satisfy these. This restriction again
allows us to strengthen the conclusions of Theorem \ref{Th: Pi ID 1}. To
state this result, for $j=1,\ldots ,J$, define 
\begin{equation*}
\mathcal{V}_{j}(w_{2},z):=\func{proj}_{j}\mathcal{V}_{0}(w_{2},z),\qquad 
\mathcal{V}_{j}:=\bigcup_{z\in \mathcal{Z}_{0}}\bigcup_{w_{2}\in \mathcal{W}%
_{0,2}(z)}\mathcal{V}_{j}(w_{2},z).
\end{equation*}

\begin{theorem}
\label{Th: Pi ID 2} Suppose Assumptions \ref{Ass: Pi injective}--\ref{Ass:
Pi reference support} hold together with: (i) 
\begin{equation*}
\nu _{j}(w_{1,j},w_{2})=\nu _{0j}(w_{1,j},w_{2})+\nu _{1j}(w_{1,j})+\nu
_{2j}(w_{2}),\qquad j=1,\ldots ,J,
\end{equation*}%
where $\nu _{0j}$ is known, while $\nu _{1j}$ and $\nu _{2j}$ are unknown
with three components continuous on their relevant domains; (ii) for every $%
j $ and $w_{1,j}\in \mathcal{I}_{j}$, $w_{2}\mapsto \nu _{j}(w_{1,j},w_{2})$
is nonconstant on $\mathcal{W}_{2}^{\ast }$; (iii) for every $j$, $\nu _{1j}$
is differentiable at $w_{0,1,j}$ and $\nu _{0j}(\cdot ,w_{2})$ is
differentiable there for every $w_{2}\in \mathcal{W}_{2}^{\ast }$, with 
\begin{equation*}
1+\nabla _{w_{1,j}}\nu _{0j}(w_{0,1,j},w_{2})-\nabla _{w_{1,j}}\nu
_{0j}(w_{0,1,j},w_{0,2})\neq 0,\qquad w_{2}\in \mathcal{W}_{2}^{\ast }.
\end{equation*}%
Then $\nu $ is identified on $\bigcup_{z\in \mathcal{Z}_{0}}\mathcal{W}%
_{0}(z)$ under the following normalisations: for some known $%
w_{0}=(w_{0,1},w_{0,2})\in \mathcal{W}^{\ast }\left( z_{0}\right) $, 
\begin{equation*}
\nu (w_{0})=0,\qquad \nabla _{w_{1,j}}\nu _{j}(w_{0})=1,\quad j=1,\ldots ,J.
\end{equation*}%
Moreover, $\Psi (\cdot ,z)$ is identified on $\mathcal{V}_{0}(z)$ for every $%
z\in \mathcal{Z}_{0}$.
\end{theorem}

In comparison with Theorem \ref{Th: Pi ID 1}, we are not able to relax the
normalisations imposed on $\nu$ and can now allow for two unknown additive
components of $\nu$. Separate identification of $\nu_{1j}$ and $\nu_{2j}$
still requires component location normalisations, since opposite constant
shifts leave their sum unchanged.

Finally, we combine alternative-specific restrictions on $\nu $ with
symmetry of $\nabla _{\nu }\Psi $. These two restrictions combined allow us
to weaken the rank condition imposed in Theorem \ref{Th: Pi ID joint rank}
substantially: For each $w_{2}\in \mathcal{W}_{2}^{\ast }$, let $\mathcal{G}%
_{\Psi }(w_{2})$ be the undirected graph with vertices $1,\ldots ,J$ and an
edge $\{j,k\}$ whenever 
\begin{equation*}
\nabla _{\nu _{k}}\Psi _{j}(\nu (w_{1},w_{2}),z_{0})\neq 0\qquad \text{for
every }w_{1}\in \prod_{l=1}^{J}\mathcal{I}_{l}.
\end{equation*}%
The following result only requires this graph to be connected.

\begin{theorem}
\label{Th: Pi ID 3} Suppose Assumptions \ref{Ass: Pi injective}--\ref{Ass:
Pi reference support} and \eqref{eq: ASV common
structure} hold, $J\geq 2$ and: (i) $\nu $ is continuous on $\mathcal{W}%
^{\ast }$ and each $\nu _{j}$ is continuously differentiable in $w_{1,j}$ on
neighbourhoods of $\mathcal{W}^{\ast }\left( z_{0}\right) $, and $\Psi
(\cdot ,z_{0})$ is continuously differentiable with symmetric Jacobian $%
\nabla _{\nu }\Psi (\nu ,z_{0})$ on $\nu (\mathcal{W}^{\ast }\left(
z_{0}\right) )$. Suppose further that, for every $w_{2}\in \mathcal{W}%
_{2}^{\ast }$: (ii) each $\nu _{j}(\cdot ,w_{2})$ is nonconstant on $%
\mathcal{I}_{j}$; (iii) the graph $\mathcal{G}_{\Psi }(w_{2})$ is connected.
Then $\nu $ is identified on $\bigcup_{z\in \mathcal{Z}_{0}}\mathcal{W}%
_{0}(z)$ under the following normalisations: for a known $%
w_{0}=(w_{0,1},w_{0,2})\in \mathcal{W}^{\ast }$ with $\nabla _{w_{1,1}}\nu
_{1}(w_{0})\neq 0$, 
\begin{equation}
\nu (w_{0})=0,\qquad \nabla _{w_{1,1}}\nu _{1}(w_{0})=1.
\label{eq: Pi3 matching normalization}
\end{equation}%
Moreover, $\Psi (\cdot ,z)$ is identified on $\mathcal{V}_{0}(z)$ for every $%
z\in \mathcal{Z}_{0}$.
\end{theorem}

Both Theorem \ref{Th: Pi ID joint rank} and \ref{Th: Pi ID 3} exploit
symmetry of $\nabla _{\nu }\Psi $, but differ in how they restrict
observationally equivalent transformations: The former theorem uses
variation in the Jacobian across $z$, whereas the latter one combines graph
connectedness with alternative-specific restrictions on $\nu $.

\section{Identification of RI-ARUM\label{Sec: ID RI-ARUM}}

We here apply the general results for composite mappings developed in the
previous section to show identification in the RI-ARUM model. We proceed in
two steps: First, we establish identification of $u_{1}$, $\lambda $, $u_{0}$
and the subjective prior $\pi _{V_{1}}$ in the benchmark case without
unobserved heterogeneity ($\varepsilon =0$). We then introduce unobserved
heterogeneity ($\varepsilon \neq 0$) and examine which of the benchmark
identification results survive in this more difficult environment and which
require additional conditions.

We will throughout assume that alternative--specific covariates are
available. Results for the case of only common variables being available can
be obtained by applying the relevant theorems of the previous section that
cover this case. Similarly, we will rely on the results obtained under the
support conditions stated in Assumption \ref{Ass: Pi reference support}. One
could alternative use the general results obtained under the support
conditions stated in Appendix \ref{sec: support conds}; for completeness, we
have stated the relevant applications to RI--ARUM of this alternative theory
in Appendix \ref{sec: support conds}. Proofs of the results of this section
can be found in Appendix \ref{sec: Proofs ID RI-ARUM}.

We will also throughout work under the following assumption on the
information cost:

\begin{assumption}
\label{Ass: RI ARUM positive lambda} $\lambda (x_{2})>0$ for all $x_{2}\in 
\mathcal{X}_{2}$.
\end{assumption}

We impose this assumption so that the results of Section \ref{sec: DMs
problem}, which require $\lambda >0$, can be applied without further
modifications. In principle, we could allow $\lambda (x_{2})=0$, but this
would require a more careful choice of the support regions on which we work
and/or additional restrictions on $\varepsilon _{2}$. This would complicate
assumptions and arguments unnecessarily, we find, and so refrain from doing
so.

Finally, it is important to emphasise that we will maintain the following
strong assumption: Not only does the Econometrician observe all relevant
covariates, as contained in $X$, but she also knows which of these
covariates go into the three indices $X_{0}$, $X_{1}$ and $X_{2}$. Without
this assumption, we face an additional identification (model selection)
problem, namely the identification of classification of the covariates in $X$
into these three indices. This is a novel problem that we are not aware of
any results on in the literature; we leave this interesting open question
for future research.

\subsection{Identification without unobserved heterogeneity ($\protect%
\varepsilon =0$)}

As a benchmark, we first develop identification result under the assumption
of $\varepsilon =0$. In this case, the identified CCPs $\Phi \left(
x_{0},x_{1},x_{2}\right) $ defined in (\ref{eq: Phi def}) satisfies%
\begin{equation}
\Phi \left( x_{0},x_{1},x_{2}\right) =P^{\ast }(u_{0}(x_{0}),u_{1}\left(
x_{1}\right) ,\lambda (x_{2})),  \label{eq: Phi def eps=0}
\end{equation}%
so that $\Pi =P^{\ast }$. We then proceed as follows: We first show that $%
u_{1}$ and $\lambda $ are identified under very weak assumptions and
normalisations regardless of whether the precise form of the prior. In
particular, the DMs beliefs could be incorrect and unknown to the
Econometrician. Next, we analyse the identifying contents of above identity
w.r.t. $u_{0}$ and find that this depends on whether the Econometrician has
any knowledge of the underlying prior $\pi _{V_{1}}$ or not. The latter case
proves to be particularly challenging, and we are only able to identify
certain features of $u_{0}$. Finally, we show that once $u_{1}$, $\lambda $
and $u_{0}$ have been identified, then the prior is also identified. That
is, we can in principle learn about the DMs' (potentially incorrect) beliefs
from their decisions.

\subsubsection{Identification of $u_{1}$ and $\protect\lambda$}

To identify $u_{1}\left( x_{1}\right) $, observe that the identity (\ref{eq:
Phi def eps=0}) can be rewritten as $\Phi (x_{1},z)=\Psi (u_{1}(x_{1}),z)$,
where $z:=(x_{0},x_{2})$ and $\Psi (v_{1},z):=P^{\ast
}(u_{0}(x_{0}),v_{1},\lambda (x_{2}))$. On the regular region, $\nabla
_{v_{1}}\Psi $ is symmetric and positive definite, with strictly negative
off-diagonal entries. Thus, Theorem \ref{Th: Pi ID 3} can be applied to
identify $u_{1}$. With $u_{1}\left( x_{1}\right) $ identified, $\lambda
(x_{2})$ could in principle be recovered from the derivative formula in
Theorem \ref{Thm: P one-to-one}(iii). However, this would require working
with derivatives of the identified choice probabilities. Instead, we use the
finite-difference log-odds formula \eqref{eq: P lambda log odds difference}
in the same theorem, which only requires choice probabilities at two
covariate values.

We summarise this argument formally for the case of alternative-specific
variables. For $(x_{0},x_{2})$ in the support of $(X_{0},X_{2})$, let $%
\mathcal{X}_{1}(x_{0},x_{2})=\func{int}\func{supp}(X_{1}\mid
X_{0}=x_{0},X_{2}=x_{2})$ and define 
\begin{equation*}
\mathcal{Z}_{0}^{0}=\{(x_{0},x_{2})\in \func{supp}(X_{0},X_{2}):u_{0}(x_{0})%
\in \mathcal{V}_{0,\mathrm{reg}}(\lambda (x_{2})),\ \mathcal{X}%
_{1}(x_{0},x_{2})\neq \varnothing \}.
\end{equation*}%
\begin{equation*}
\mathcal{X}_{1,\mathrm{reg}}^{0}=\bigcup_{(x_{0},x_{2})\in \mathcal{Z}%
_{0}^{0}}\mathcal{X}_{1}(x_{0},x_{2}),
\end{equation*}%
where $\mathcal{V}_{0,\mathrm{reg}}(\lambda )$ is defined in 
\eqref{eq:
V_0,reg def}.

\begin{assumption}
\label{Ass: RI ARUM u1 lambda support} There exist a known $%
(x_{0,0},x_{2,0})\in \mathcal{Z}_{0}^{0}$, nonempty open intervals $\mathcal{%
I}_{1j}\subseteq \mathbb{R}$, $j=1,\ldots ,J$, and a nonempty open connected
set $\mathcal{X}_{1,2}^{\ast }$ such that $\mathcal{X}_{1}^{\ast
}:=\prod_{j=1}^{J}\mathcal{I}_{1j}\times \mathcal{X}_{1,2}^{\ast }\subseteq 
\mathcal{X}_{1}(x_{0,0},x_{2,0})$. Moreover, (i) for every $x_{1,2}\in 
\mathcal{X}_{1,2}^{\ast }$ and $j=1,\ldots ,J$, $x_{1,1,j}\mapsto
u_{1,j}(x_{1,1,j},x_{1,2})$ is nonconstant on $\mathcal{I}_{1j}$; (ii) for
every $(x_{0},x_{2})\in \mathcal{Z}_{0}^{0}$, $u_{1}(\mathcal{X}%
_{1}(x_{0},x_{2}))=u_{1}(\mathcal{X}_{1}^{\ast }\cap \mathcal{X}%
_{1}(x_{0},x_{2}))$; (iii) for every $x_{2}\in \func{proj}_{x_{2}}\mathcal{Z}%
_{0}^{0}$ there exist $x_{0}$ with $(x_{0},x_{2})\in \mathcal{Z}_{0}^{0}$, $%
j\in \{1,\ldots ,J\}$ and $x_{1},\bar{x}_{1}\in \mathcal{X}_{1}(x_{0},x_{2})$
such that $u_{1,j}(x_{1})\neq u_{1,j}(\bar{x}_{1})$.
\end{assumption}

\begin{theorem}
\label{Th: RI ARUM u1 lambda eps = 0} Suppose $J\geq 2$, $\varepsilon =0$, $%
\lambda (x_{2})>0$, and write $x_{1}=(x_{1,1},x_{1,2})$, where $%
x_{1,1}=(x_{1,1,1},\ldots ,x_{1,1,J})$ and $x_{1,2}$ contain
alternative-specific and common covariates, respectively, so that 
\begin{equation*}
u_{1,j}(x_{1})=u_{1,j}(x_{1,1,j},x_{1,2}),\qquad j=1,\ldots ,J.
\end{equation*}%
Suppose furthermore that each $u_{1,j}$ is continuously differentiable in $%
x_{1,1,j}$ on neighbourhoods of the relevant supports, and Assumption~ \ref%
{Ass: RI ARUM u1 lambda support} holds. Then $u_{1}$ is identified on $%
\mathcal{X}_{1,\mathrm{reg}}^{0}$, and $\lambda $ is identified on $\func{%
proj}_{x_{2}}\mathcal{Z}_{0}^{0}$ under the following normalisations: For
some known $x_{1,0}=(x_{1,1,0},x_{1,2,0})$ in $\mathcal{X}_{1}^{\ast }$, 
\begin{equation*}
u_{1}(x_{1,0})=0,\qquad \nabla _{x_{1,1,1}}u_{1,1}(x_{1,1,1,0},x_{1,2,0})=1.
\end{equation*}
\end{theorem}

Alternative support conditions for this result are given in Appendix \ref%
{sec: support conds}.

\subsubsection{Identification of $u_{0}$}

We next consider identification of $u_{0}$ under the following two
scenarios: In the first one, the Econometrician knows the DM's prior in
which case the integrated CCP mapping $P^{0\ast }$ is known to us and $u_{0}$
can be inverted out from this and so is fully identified. In the second
case, the prior is unrestricted and unknown to us and so we cannot obtain $%
P^{0\ast }$. As a consequence, much less can be learned about $u_{0}$ from
data. Specifically, we have to impose much stronger identifiying
restrictions/normalisations on $u_{0}$ in order to identify the remaining
unknown parts of it.

\paragraph{Prior is known}

Suppose that $\pi _{V_{1}}$ is known, but not necessarily correctly
specified. In this case, for any given value of $\left( v_{0},\lambda
\right) $, we can in principle compute $P^{0\ast }\left( v_{0},\lambda
\right) \in \arg \max_{P^{0}\in \Delta (\mathcal{J})}Q^{0}(P^{0};v_{0},%
\lambda )$, where 
\begin{equation*}
Q^{0}(P^{0};v_{0},\lambda ):=\lambda \int_{\mathcal{V}}\log \left(
\sum_{j\in \mathcal{J}}P_{j}^{0}(v_{0})\exp \left( v_{j}/\lambda \right)
\right) d\pi _{V_{1}}(v-v_{0}).
\end{equation*}%
That is, for any given $\lambda >0$, the mapping $v_{0}\mapsto P^{0\ast
}\left( v_{0},\lambda \right) $ is known to us.

Recall the support sets introduced before Assumption \ref{Ass: RI ARUM u1
lambda support}. With $u_{1}$ identified on $\mathcal{X}_{1,\mathrm{reg}%
}^{0} $, cf. the first part of Theorem \ref{Th: RI ARUM u1 lambda eps = 0},
we can treat $V_{1}=u_{1}\left( X_{1}\right) $ as observed for $X_{1}\in 
\mathcal{X}_{1}\left( X_{0},X_{2}\right) $ and $\left( X_{0},X_{2}\right)
\in \mathcal{Z}^{0}$. In particular, 
\begin{equation*}
\Phi _{\pi ,j}(x_{0},v_{1},x_{2}):=\Pr \left(
Y=j|X_{0}=x_{0},V_{1}=v_{1},X_{2}=x_{2}\right) ,\text{ \ \ }j=1,...,J,
\end{equation*}%
are identified for all $v_{1}\in u_{1}\left( \mathcal{X}_{1}(x_{0},x_{2})%
\right) $ and $\left( x_{0},x_{2}\right) \in \mathcal{Z}_{0}^{0}$. Define 
\begin{equation*}
\mathcal{Z}_{\pi }^{0}=\left\{ \left( x_{0},x_{2}\right) \in \mathcal{Z}%
_{0}^{0}:\func{supp}(\pi _{V_{1}})\subseteq u_{1}(\mathcal{X}%
_{1}(x_{0},x_{2}))\right\} .
\end{equation*}%
For any $\left( x_{0},x_{2}\right) \in \mathcal{Z}_{\pi }^{0}$, we can then
compute/identify $\Phi _{\pi }^{0}(x_{0},x_{2})$ given by 
\begin{equation*}
\Phi _{\pi }^{0}(x_{0},x_{2})=\int_{\mathbb{R}^{J}}\Phi _{\pi
}(x_{0},v_{1},x_{2})d\pi _{V_{1}}(v_{1})=P^{0\ast }(u_{0}(x_{0}),\lambda
\left( x_{2}\right) ),
\end{equation*}%
where $\lambda \left( x_{2}\right) $ is identified by the second part of
Theorem \ref{Th: RI ARUM u1 lambda eps = 0}. Since $v_{0}\mapsto P^{0\ast
}\left( v_{0},\lambda \right) $ is known to us and it is one--to--one on $%
v_{0}\in \mathcal{V}_{0,\mathrm{reg}}(\lambda )$, we can solve $\Phi _{\pi
}^{0}(x_{0},x_{2})=P^{0\ast }(u_{0}(x_{0}),\lambda \left( x_{2}\right) )$
w.r.t. $u_{0}(x_{0})$ if this value lies in $\mathcal{V}_{0,\mathrm{reg}%
}(\lambda )$; but this is guaranteed by the definition of $\mathcal{Z}_{\pi
}^{0}$ together with $\left( x_{0},x_{2}\right) \in \mathcal{Z}_{\pi }^{0}$.
We conclude:

\begin{theorem}
\label{Th: RI ARUM u0 eps=0 pi known} Suppose $\varepsilon =0$, $\pi
_{V_{1}} $ is known and the assumptions of Theorem \ref{Th: RI ARUM u1
lambda eps = 0} hold. Then $u_{0}$ is identified on 
\begin{equation*}
\mathcal{X}_{0}^{0}=\func{proj}_{x_{0}}\mathcal{Z}_{\pi }^{0}=\left\{
x_{0}:\exists x_{2}\text{ such that }(x_{0},x_{2})\in \mathcal{Z}_{\pi
}^{0}\right\} .
\end{equation*}
\end{theorem}

\paragraph{Prior is unrestricted and unknown}

When $\pi _{V_{1}}$ is unknown, the mapping $P^{0\ast }(\cdot ,\lambda )$
cannot be computed from the known primitives, so the direct inversion
argument of Theorem \ref{Th: RI ARUM u0 eps=0 pi known} is unavailable.
Instead, we use the fact that $\Phi (x_{0},x_{1},x_{2})$ defined in (\ref%
{eq: Phi def}) satisfies $\Phi (x_{0},x_{1},x_{2})=P^{\ast
}(u_{0}(x_{0}),u_{1}\left( x_{1}\right) ,\lambda \left( x_{2}\right) )$ and
then wish to use the general results in Section \ref{sec: ID composite
injectivity} for identification. However, \ the Jacobian $\nabla
_{v_{0}}P^{\ast }(v_{0},v_{1},\lambda )$ is generally asymmetric and its
off--diaogonal elements are not necessarily non--zero. As such, we cannot
apply Theorem \ref{Th: Pi ID 3} and instead have to impose enough
restrictions on $u_{0}$ in order to apply Theorem \ref{Th: Pi ID 2}. With $%
x_{0}=(x_{0,1},x_{0,2})$, where $x_{0,1}=(x_{0,1,1},\ldots ,x_{0,1,J})$
contains alternative--specific variables, we will require that $u_{0,j}$
takes the form%
\begin{equation}
u_{0,j}(x_{0,1,j},x_{0,2})=u_{0j}^{\left( 0\right)
}(x_{0,1,j},x_{0,2})+u_{0j}^{\left( 1\right) }(x_{0,1,j})+u_{0j}^{\left(
2\right) }(x_{0,2}),\quad j=1,\ldots ,J,  \label{eq: u0 known interaction}
\end{equation}%
where $u_{0j}^{\left( 0\right) }$ is known and $u_{0j}^{\left( 1\right)
},u_{0j}^{\left( 2\right) }$ are unknown. This contrasts with Theorem \ref%
{Th: RI ARUM u0 eps=0 pi known}, where no such restrictions on $u_{0}$ are
required.

Under (\ref{eq: u0 known interaction}), we can apply Theorem \ref{Th: Pi ID
2}. To translate the conditions of the theorem into our setting, write 
\begin{equation*}
\Phi (x_{0},z)=\Psi (u_{0}(x_{0}),z),\qquad \Psi (v_{0},z)=P^{\ast
}(v_{0},u_{1}(x_{1}),\lambda (x_{2})),\qquad z=(x_{1},x_{2}).
\end{equation*}%
By Theorem \ref{Thm: P one-to-one}(ii), $v_{0}\mapsto \Psi (v_{0},z)$ is
injective on $\mathcal{V}_{0,\mathrm{reg}}(\lambda (x_{2}))$. For any $%
(x_{1},x_{2})$ in the joint support of $(X_{1},X_{2})$, let $\mathcal{X}%
_{0}(x_{1},x_{2})=\func{int}\func{supp}(X_{0}|(X_{1},X_{2})=(x_{1},x_{2}))$
be the conditional support of $X_{0}$ and%
\begin{equation*}
\mathcal{X}_{0,\mathrm{reg}}^{0}(x_{1},x_{2})=\left\{ x_{0}\in \mathcal{X}%
_{0}(x_{1},x_{2}):u_{0}(x_{0})\in \mathcal{V}_{0,\mathrm{reg}}(\lambda
(x_{2}))\right\}
\end{equation*}%
be the regular part of this and $\mathcal{U}_{0}(x_{1},x_{2}):=u_{0}(%
\mathcal{X}_{0,\mathrm{reg}}^{0}(x_{1},x_{2}))$ be its image w.r.t. $u_{0}$.
Finally,%
\begin{equation*}
\mathcal{Z}_{u_{0}}^{0}:=\left\{ (x_{1},x_{2})\in \func{supp}(X_{1},X_{2}):%
\mathcal{X}_{0,\mathrm{reg}}^{0}(x_{1},x_{2})\neq \varnothing \right\} ,
\end{equation*}%
is the subset of the support of $(X_{1},X_{2})$ for which this regular
region is non--empty. The following assumption translates the general
support conditions of Theorem \ref{Th: Pi ID 2} into our current application
of it:

\begin{assumption}
\label{Ass: RI ARUM u0 support}(i) For some $(x_{01},x_{02})\in \mathcal{Z}%
_{u_{0}}^{0}$, there exists non--empty open intervals $\mathcal{I}%
_{0j}\subseteq \mathbb{R}$, $j=1,\ldots ,J$, and a nonempty open connected
set $\mathcal{X}_{0,2}^{\ast }\subseteq \mathbb{R}^{d_{x_{0,2}}}$ such that $%
\mathcal{X}_{0}^{\ast }:=\prod_{j=1}^{J}\mathcal{I}_{0j}\times \mathcal{X}%
_{0,2}^{\ast }\subseteq \mathcal{X}_{0,\mathrm{reg}}^{0}(x_{01},x_{02}),$
and $x_{0,2}\mapsto u_{0,j}(x_{0,1,j},x_{0,2})$ is nonconstant on $\mathcal{X%
}_{0}^{\ast }$ for every $j$ and $x_{0,1,j}\in \mathcal{I}_{0j}$; (ii) for
every $j$, $u_{0j}^{\left( 1\right) }$ is differentiable at $x_{0,1,j,0}$,
and $u_{0j}^{\left( 0\right) }(\cdot ,x_{0,2})$ is differentiable at the
same point for every $x_{0,2}\in \mathcal{X}_{0,2}^{\ast }$, with 
\begin{equation*}
1+\nabla _{x_{0,1,j}}u_{0j}^{\left( 0\right) }(x_{0,1,j,0},x_{0,2})-\nabla
_{x_{0,1,j}}u_{0j}^{\left( 0\right) }(x_{0,1,j,0},x_{0,2,0})\neq 0,\qquad
x_{0,2}\in \mathcal{X}_{0,2}^{\ast };
\end{equation*}%
(iii) For every $(x_{1},x_{2})\in \mathcal{Z}_{u_{0}}^{0}$, $u_{0}\left( 
\mathcal{X}_{0,\mathrm{reg}}^{0}(x_{1},x_{2})\right) =u_{0}\bigl(\mathcal{X}%
_{0}^{\ast }\cap \mathcal{X}_{0}(x_{1},x_{2})\bigr)$.
\end{assumption}

\begin{theorem}
\label{Th: RI ARUM u0 eps=0 pi unknown} Suppose that the assumptions of
Theorem \ref{Th: RI ARUM u1 lambda eps = 0} are satisfied together with 
\eqref{eq: u0
known interaction} and Assumption~\ref{Ass: RI ARUM u0 support}. Then $u_{0}$
is identified on $\bigcup_{(x_{1},x_{2})\in \mathcal{Z}_{u_{0}}^{0}}\mathcal{%
X}_{0,\mathrm{reg}}^{0}(x_{1},x_{2})$, and $v_{0}\mapsto P^{\ast
}(v_{0},u_{1}(x_{1}),\lambda (x_{2}))$ is identified on $\mathcal{U}%
_{0}(x_{1},x_{2})$ for $(x_{1},x_{2})\in \mathcal{Z}_{u_{0}}^{0}$ under the
following normalisation: With $(x_{1,0},x_{2,0})\in \mathcal{Z}_{u_{0}}^{0}$
given in Assumption~\ref{Ass: RI ARUM u0 support}(i) and for some known $%
x_{0,0}=(x_{0,1,0},x_{0,2,0})\in \mathcal{X}_{0}^{\ast }$, 
\begin{equation*}
u_{0}(x_{0,0})=0,\qquad \nabla _{x_{0,1,j}}u_{0,j}(x_{0,0})=1,\quad
j=1,\ldots ,J;
\end{equation*}
\end{theorem}

This is a very weak identification result, where strong normalisations have
to be imposed on $u_{0}$. However, if we are willing to assume that the
prior is not too badly misspecified in the sense of the difference $\pi
_{V_{1}}-F_{V_{1}}$ not being "too large", then the much stronger result of
Theorem \ref{Th: RI ARUM u0 eps=0 pi known} will still hold. Since this
result holds regardless of whether $\varepsilon =0$ or not, we present this
finding under the scenario of $\varepsilon \neq 0$.

\subsubsection{Identification of prior $\protect\pi _{V_{1}}$}

Suppose the functions $u_{0}$, $u_{1}$ and $\lambda $ have been identified,
but the prior is unknown. This corresponds to the scenario considered in
Theorem \ref{Th: RI ARUM u0 eps=0 pi unknown}. We show that $\pi _{V_{1}}$
is then identified using arguments closely related to those for the mixed
multinomial logit models in \citet{foxkimryanbajari2012}.

By \eqref{eq:DMCCP}, at any $(x_{0},x_{1},x_{2})$ in the support of $X$
where $u_{0}(x_{0})$, $u_{1}(x_{1})$ and $\lambda (x_{2})$ are identified
and $u_{0}(x_{0})\in \mathcal{V}_{0,\mathrm{reg}}(\lambda (x_{2}))$, we have 
\begin{equation}
\frac{P_{j}^{0\ast }(u_{0}(x_{0}),\lambda (x_{2}))}{P_{0}^{0\ast
}(u_{0}(x_{0}),\lambda (x_{2}))}=\frac{\Phi _{j}(x_{0},x_{1},x_{2})}{\Phi
_{0}(x_{0},x_{1},x_{2})}\exp \left( -\frac{u_{0,j}(x_{0})+u_{1,j}(x_{1})}{%
\lambda (x_{2})}\right) ,\qquad j=1,\ldots ,J,  \label{eq: prior P0 odds}
\end{equation}%
so the integrated CCPs $P^{0\ast }\left( v_{0},\lambda \right) $ evaluated
at $\left( v_{0},\lambda \right) =(u_{0}(x_{0}),\lambda (x_{2}))$ are
identified. Now, the integrated CCPs satisfy the integral equation $%
P_{0}^{0\ast }(v_{0},\lambda )=\int P_{0}^{\ast }(v_{0},v_{1},\lambda )d\pi
_{V_{1}}\left( v_{1}\right) $, where $P_{0}^{\ast }(v_{0},v_{1},\lambda )$
satisfies the mixed--logit formula in (\ref{eq:DMCCP}). The components of
this formula is known to us and so we can identify $\pi _{V_{1}}$ as the
inverse of this integral equation:

\begin{theorem}
\label{Th: RI ARUM prior ID} Suppose that the assumptions of Theorems \ref%
{Th: RI ARUM u1 lambda eps = 0} and \ref{Th: RI ARUM u0 eps=0 pi unknown}
hold. Suppose furthermore that there exists $(x_{1,\pi },x_{2,\pi })\in 
\mathcal{Z}_{u_{0}}^{0}$ with $x_{1,\pi }\in \mathcal{X}_{1,\mathrm{reg}%
}^{0} $ and $x_{2,\pi }\in \func{proj}_{x_{2}}\mathcal{Z}_{u_{0}}^{0}$, and
a non--empty open set $\mathcal{X}_{0,\pi }\subseteq \mathcal{X}_{0,\mathrm{%
reg}}^{0}(x_{1,\pi },x_{2,\pi })$ such that $\func{int}u_{0}(\mathcal{X}%
_{0,\pi })\neq \varnothing $. Then $\pi _{V_{1}}$ is identified.
\end{theorem}

The assumed existence of $(x_{1,\pi },x_{2,\pi })$ and $\mathcal{X}_{0,\pi }$
ensures that we can identify $u_{0}(x_{0})$, $u_{1}(x_{1})$ and $\lambda
(x_{2})$ up to relevant normalisations via Theorems \ref{Th: RI ARUM u1
lambda eps = 0} and \ref{Th: RI ARUM u0 eps=0 pi unknown} on an open set.

\subsection{Identification with unobserved heterogeneity ($\protect%
\varepsilon \neq 0$)}

We now consider the more difficult scenario, where unobserved heterogeneity
is added to all three components, $u_{0}$, $u_{1}$ and $\lambda $. In this
case, we will have to rely on the identity (\ref{eq: Phi = Pi}), where the
composite mapping $\Pi $ is the integrated version of $P^{\ast }$ w.r.t. $%
F_{\varepsilon }$. Similar to ARUM, the integration w.r.t. $F_{\varepsilon }$
does not affect identification of $u_{1}$ since the Jacobian of the
integrand, $\nabla _{u_{1}}P^{\ast }$, is strictly positive definite and its
off--diagonal elements are negative on the regular parts of its domain. The
Jacobian of its integrated version w.r.t $\varepsilon $, $\nabla _{u_{0}}\Pi 
$, has the same attractive features and so $\Pi $ carries the same
identification strength as $P^{\ast }$ w.r.t. $u_{1}$. In contrast, the
mapping $u_{1}\mapsto P^{\ast }$ and $\lambda \mapsto P^{\ast }$ do not have
such nice features and so for these two components we arrive at much weaker
identification results in the sense that much stronger normalisations on
these two functions have to be imposed.

\subsubsection{Identification of $u_{1}$ and $\protect\lambda $ \label{sec:
ID u_1 eps neq 0}}

To show identification of $u_{1}$, we first establish that $u_{1}\mapsto \Pi 
$ inherits most of the attractive properties of $u_{1}\mapsto P^{\ast }$.
Recalling the definition of $\mathcal{V}_{0,\mathrm{reg}}$ in 
\eqref{eq:
V_0,reg def}, we define 
\begin{equation}
\mathcal{E}_{\mathrm{reg}}(u_{0},\lambda ):=\left\{ e\in \mathcal{E}%
:u_{0}+e_{0}\in \mathcal{V}_{0,\mathrm{reg}}(\lambda +e_{2})\right\} .
\label{eq: E reg Pi definition}
\end{equation}%
This is the regular part of the support of $\varepsilon $. For any $%
(u_{0},\lambda )$ for which this set has positive probability mass w.r.t. $%
F_{\varepsilon }$, $u_{1}\mapsto \Pi (u_{0},u_{1},\lambda )$ is
well--behaved in the following sense:

\begin{lemma}
\label{Lem: Pi u1 injectivity} Suppose Assumption \ref{ass: Affine ind}
holds. Then, for any $\lambda >0$ and any $u_{0}\in \mathbb{R}^{J}$, $%
u_{1}\mapsto \Pi (u_{0},u_{1},\lambda )$ is continuously differentiable on $%
\mathbb{R}^{J}$, with 
\begin{equation}
\nabla _{u_{1}}\Pi =\int_{\mathcal{E}}\frac{1}{\lambda +e_{2}}\Sigma
(P^{\ast })dF_{\varepsilon }(e)  \label{eq: Pi u1 Jacobian}
\end{equation}%
symmetric and positive semidefinite. For any $(u_{0},\lambda )$ satisfying 
\begin{equation}
F_{\varepsilon }\bigl(\mathcal{E}_{\mathrm{reg}}(u_{0},\lambda )\bigr)>0,
\label{eq: Pi u1 regular condition}
\end{equation}%
$\nabla _{u_{1}}\Pi (u_{0},u_{1},\lambda )$ is positive definite and, for $%
j\neq k$, $\nabla _{u_{1,k}}\Pi _{j}(u_{0},u_{1},\lambda )<0$ for all $%
u_{1}\in \mathbb{R}^{J}$. In particular, for any such $(u_{0},\lambda )$, $%
u_{1}\mapsto \Pi (u_{0},u_{1},\lambda )$ is globally injective.
\end{lemma}

We now use this result in conjunction with Theorem \ref{Th: Pi ID 3} to show
identification of $u_{1}$. Let $\mathcal{X}_{1}(x_{0},x_{2})$ denote the
interior of the conditional support of $X_{1}$ given $%
(X_{0},X_{2})=(x_{0},x_{2})$ and define 
\begin{equation}
\mathcal{Z}_{u_{1}}=\{(x_{0},x_{2})\in \func{supp}(X_{0},X_{2}):F_{%
\varepsilon }\bigl(\mathcal{E}_{\mathrm{reg}}(u_{0}(x_{0}),\lambda (x_{2}))%
\bigr)>0,\ \mathcal{X}_{1}(x_{0},x_{2})\neq \varnothing \}.
\label{eq: Pi u1 heterogeneity regular z}
\end{equation}%
This corresponds to the set $\mathcal{Z}_{0}^{0}$ used for identification in
the case of $\varepsilon =0$. The set $\mathcal{X}_{1,\mathrm{reg}%
}:=\bigcup_{(x_{0},x_{2})\in \mathcal{Z}_{u_{1}}}\mathcal{X}%
_{1}(x_{0},x_{2}) $ then corresponds to the region of the support of $X_{1}$
so that $\Pi (u_{0},u_{1},\lambda )$ is injective on $u_{1}\left( \mathcal{X}%
_{1,\mathrm{reg}}\right) $.

\begin{assumption}
\label{Ass: RI ARUM u1 support}$\mathcal{Z}_{u_{1}}$ defined in (\ref{eq: Pi
u1 heterogeneity regular z}) is non--empty; there exist a known $%
(x_{0,0},x_{0,2})\in \mathcal{Z}_{u_{1}}$, nonempty open intervals $\mathcal{%
I}_{1j}\subseteq \mathbb{R}$, $j=1,\ldots ,J$, and a nonempty open connected
set $\mathcal{X}_{1,2}^{\ast }$ such that $\mathcal{X}_{1}^{\ast
}:=\prod_{j=1}^{J}\mathcal{I}_{1j}\times \mathcal{X}_{1,2}^{\ast }\subseteq 
\mathcal{X}_{1}(x_{0,0},x_{0,2})$. Moreover, (i) for every $x_{1,2}\in 
\mathcal{X}_{1,2}^{\ast }$ and $j=1,\ldots ,J$, $x_{1,1,j}\mapsto
u_{1,j}(x_{1,1,j},x_{1,2})$ is nonconstant on $\mathcal{I}_{1j}$; (ii) for
every $(x_{0},x_{2})\in \mathcal{Z}_{u_{1}}$, $u_{1}(\mathcal{X}%
_{1}(x_{0},x_{2}))=u_{1}(\mathcal{X}_{1}^{\ast }\cap \mathcal{X}%
_{1}(x_{0},x_{2}))$.
\end{assumption}

\begin{theorem}
\label{Th: RI ARUM u1 eps neq 0} Suppose Assumptions~\ref{ass: Affine ind}
and~\ref{ass:idependentXWZande} hold and $J\geq 2$. Write $%
x_{1}=(x_{1,1},x_{1,2})$, where $x_{1,1}=(x_{1,1,1},\ldots ,x_{1,1,J})$, and
suppose 
\begin{equation*}
u_{1,j}(x_{1})=u_{1,j}(x_{1,1,j},x_{1,2}),\qquad j=1,\ldots ,J,
\end{equation*}%
with each $u_{1,j}$ continuously differentiable in $x_{1,1,j}$ on
neighbourhoods of the relevant supports. Suppose Assumption~\ref{Ass: RI
ARUM u1 support} holds. Then $u_{1}$ is identified on $\mathcal{X}_{1,%
\mathrm{reg}}$ under the following normalisations: For some known $%
x_{1,0}=(x_{1,1,0},x_{1,2,0})\in\mathcal{X}_{1}^{\ast}$, 
\begin{equation*}
u_{1}(x_{1,1,0},x_{1,2,0})=0,\text{ \ \ }\nabla
_{x_{1,1,1}}u_{1,1}(x_{1,1,0},x_{1,2,0})=1.
\end{equation*}
\end{theorem}

Alternative support conditions for this result are given in Appendix \ref%
{sec: support conds}.

Here, observe that, under the stated support conditions, we are able to
achieve the same level of identification of $u_{1}$ as in the case of $%
\varepsilon =0$. In fact, the above result is generally stronger than the
one in Theorem \ref{Th: RI ARUM u1 lambda eps = 0}: The added variation due
to $\varepsilon \neq 0$ increases the set of values of $x_{1}$ for which we
can identify $u_{1}(x_{1})$. For example, if $\varepsilon _{0}|\left(
\varepsilon _{1},\varepsilon _{2}\right) $ has full support on $\mathbb{R}%
^{J}$, then the condition (\ref{eq: Pi u1 heterogeneity regular z}) is
satisfied for any $(x_{0},x_{2})$ in the support of $(X_{0},X_{2})$ and so $%
\mathcal{X}_{1,\mathrm{reg}}$ covers the interiors of the conditional
supports of $X_{1}$. In contrast, Theorem \ref{Th: RI ARUM u1 lambda eps = 0}
will generally only deliver identification on a subset of the support of $%
X_{1}$.

We next identify the information-cost function $\lambda $ directly from the
original CCP mapping $\Phi (x_{0},x_{1},x_{2})$, without requiring prior
identification of $u_{1}$. Part (iii) of Theorem \ref{Thm: P one-to-one}
motivates using the stacked mapping 
\begin{equation*}
\tilde{\Phi}(x_{2},z):=\left( 
\begin{array}{c}
\Phi (x_{0},x_{1},x_{2}) \\ 
\func{vec}\bigl(\nabla _{x_{1,1}}\Phi (x_{0},x_{1},x_{2})\bigr)%
\end{array}%
\right) ,\qquad z:=(x_{0},x_{1}),
\end{equation*}%
for identification. The corresponding structural mapping takes the form%
\begin{equation}
\Psi (\lambda ,z):=\left( 
\begin{array}{c}
\Pi (u_{0}(x_{0}),u_{1}(x_{1}),\lambda ) \\ 
\func{vec}\bigl(\nabla _{u_{1}}\Pi (u_{0}(x_{0}),u_{1}(x_{1}),\lambda
)\nabla _{x_{1,1}}u_{1}(x_{1})\bigr)%
\end{array}%
\right) .  \label{eq: Lambda def lambda}
\end{equation}%
We have not been able to verify that this mapping satisfies the conditions
of the general results of Section \ref{sec: ID composite injectivity} and so
instead develop an identification argument that relies on a global full rank
condition on $\lambda $. This argument requires us to impose the restriction
that $x_{2}=\left( x_{2,1},x_{2,2}\right) $ and%
\begin{equation}
\lambda (x_{2})=\lambda _{0}(x_{2})+\lambda _{1}(x_{2,2})>0,
\label{eq: lambda restriction}
\end{equation}%
where $\lambda _{0}(x_{2})$ is assumed known and satisfies a global rank
condition. See the theorem below. To state the assumptions of this theorem,
let $\mathcal{Z}_{\lambda }$ be the set of points of the support of $%
(X_{0},X_{1})$ for which there exists a nonempty open set $\mathcal{X}%
_{2}(x_{0},x_{1})$ contained in the conditional support of $%
X_{2}|(X_{0},X_{1})=(x_{0},x_{1})$ such that 
\begin{equation*}
x_{1,1}\in \func{int}\func{supp}(X_{1,1}\mid
X_{0}=x_{0},X_{1,2}=x_{1,2},X_{2}=x_{2}),\qquad x_{2}\in \mathcal{X}%
_{2}(x_{0},x_{1}).
\end{equation*}%
For each $(x_{0},x_{1})\in \mathcal{Z}_{\lambda }$, define 
\begin{eqnarray*}
\mathcal{X}_{2,\mathrm{reg}}(x_{0},x_{1}) &:&=\left\{ x_{2}\in \mathcal{X}%
_{2}(x_{0},x_{1}):\nabla _{x_{2,1}}\tilde{\Phi}(x_{2},(x_{0},x_{1}))\neq
0\right\} , \\
\mathcal{X}_{2,2,\mathrm{reg}} &:&=\func{proj}_{x_{2,2}}%
\bigcup_{(x_{0},x_{1})\in \mathcal{Z}_{\lambda }}\mathcal{X}_{2,\mathrm{reg}%
}(x_{0},x_{1}).
\end{eqnarray*}%
The following theorem identifies $\lambda _{1}$ in (\ref{eq: lambda
restriction}):

\begin{theorem}
\label{Th: RI ARUM lambda eps neq 0} Suppose Assumptions \ref{ass: Affine
ind} and \ref{ass:idependentXWZande} hold together with (\ref{eq: lambda
restriction}). Suppose $u_{1}$ is continuously differentiable in $x_{1,1}$
on neighbourhoods of the covariate values specified above. Moreover, suppose 
$\varepsilon _{2}|(\varepsilon _{0},\varepsilon _{1})$ has a density $%
f_{\varepsilon _{2}|\varepsilon _{0},\varepsilon _{1}}(e_{2}|e_{0},e_{1})$
on $[0,\infty )$ that is absolutely continuous in $e_{2}$, vanishes at $%
e_{2}=0$, and satisfies $E[\int_{0}^{\infty }|\nabla _{e_{2}}f_{\varepsilon
_{2}|\varepsilon _{0},\varepsilon _{1}}(e_{2}|\varepsilon _{0},\varepsilon
_{1})|de_{2}]<\infty $. Finally, suppose that $\lambda _{0}$ and $\lambda
_{1}$ are continuously differentiable, with $\nabla _{x_{2,1}}\lambda
_{0}(x_{2})\neq 0$ on $\bigcup_{(x_{0},x_{1})\in \mathcal{Z}_{\lambda }}%
\mathcal{X}_{2,\mathrm{reg}}(x_{0},x_{1})$. Then $\nabla _{x_{2,2}}\lambda
_{1}$ is identified on the open set $\mathcal{X}_{2,2,\mathrm{reg}}$, and $%
\lambda _{1}$ is identified on any connected open subset $\mathcal{X}%
_{2,2}^{\ast }\subseteq \mathcal{X}_{2,2,\mathrm{reg}}$ under the
normalisation $\lambda _{1}(x_{2,2}^{0})=0$ for some known $x_{2,2}^{0}\in 
\mathcal{X}_{2,2}^{\ast }$.
\end{theorem}

If $\mathcal{X}_{2,2,\mathrm{reg}}$ has several connected components, a
separate location normalisation is required on each component to identify $%
\lambda _{1}$ there, unless other restrictions determine the relative
constants.

\subsubsection{Identification of $u_{0}$}

We now analyse which features of $u_{0}$ can be identified. As in the case
of $\varepsilon =0$, we consider the two scenarios of $\pi _{V_{1}}$ being
known or unknown in turn.

\paragraph{Prior is partially known and correct}

We here pursue a strategy similar to the one that yielded Theorem \ref{Th:
RI ARUM u0 eps=0 pi known}. With $u_{1}$ identified on $\mathcal{X}_{1,%
\mathrm{reg}}$, cf. Theorem \ref{Th: RI ARUM u1 eps neq 0}, we can treat $%
U_{1}:=u_{1}\left( X_{1}\right) $ as observed for $X_{1}\in \mathcal{X}%
_{1}(X_{0},X_{2})$ and $\left( X_{0},X_{2}\right) \in \mathcal{Z}_{u_{1}}$,
where $\mathcal{Z}_{u_{1}}$ is defined in (\ref{eq: Pi u1 heterogeneity
regular z}). In particular, 
\begin{equation*}
\Phi _{\pi ,j}(x_{0},u_{1},x_{2}):=\Pr \left(
Y=j|X_{0}=x_{0},U_{1}=u_{1},X_{2}=x_{2}\right) ,\text{ \ \ }j=1,...,J,
\end{equation*}%
are identified for all $u_{1}\in u_{1}\left( \mathcal{X}_{1}(x_{0},x_{2})%
\right) $ and $\left( x_{0},x_{2}\right) \in \mathcal{Z}_{u_{1}}$ and
satisfy $\Phi _{\pi }(x_{0},v_{1},x_{2})=\Pi (u_{0}(x_{0}),v_{1},\lambda
(x_{2}))$. As in Theorem \ref{Th: RI ARUM u0 eps=0 pi known}, we now wish to
identify an integrated version of $\Phi _{\pi }$ that can be expressed as a
functional of $P^{0\ast }$. To this end, we first assume that the DM\ is
aware of the structure of the data--generating process $V_{1}=U_{1}+%
\varepsilon _{1}$ with $U_{1}\bot \varepsilon _{1}$, and so her prior takes
the form $\pi _{V_{1}}=\pi _{U_{1}}\oplus \pi _{\varepsilon _{1}}$, where $%
\oplus $ denotes convolution of probability measures and $\pi _{U_{1}}$ is
the DM's prior on $U_{1}=u_{1}(X_{1})$. Next, we impose the assumption that $%
\varepsilon _{1}\bot \left( \varepsilon _{0},\varepsilon _{2}\right) $ so
that $F_{\varepsilon }(e_{0},e_{1},e_{2})=F_{\varepsilon
_{1}}(e_{1})F_{\varepsilon _{0},\varepsilon _{2}}(e_{0},e_{2})$. Finally, we
impose 
\begin{equation}
\pi _{\varepsilon _{1}}=F_{\varepsilon _{1}}\text{ and }\pi _{U_{1}}\text{
is known to the Econometrician.}  \label{eq: prior part known}
\end{equation}%
This last assumption states that the DM has correct beliefs about $%
\varepsilon _{1}$ but potentially incorrect beliefs about $U_{1}$, and this
prior is known to the Econometrician. It includes as a special case the
scenario where the DM has fully correct beliefs and this is known to the
Econometrician since in this case $\pi _{U_{1}}=F_{U_{1}}$, where $F_{U_{1}}$
is identified.

For any $(x_{0},x_{2})\in \mathcal{Z}_{\pi }$, where 
\begin{equation*}
\mathcal{Z}_{\pi }:=\left\{ (x_{0},x_{2})\in \mathcal{Z}_{u_{1}}:\func{supp}%
(\pi _{U_{1}})\subseteq u_{1}(\mathcal{X}_{1}((x_{0},x_{2})))\right\} ,
\end{equation*}%
we can then compute/identify the following integrated version of $\Phi _{\pi
}$,%
\begin{equation}
\Phi _{\pi }^{0}(x_{0},x_{2}):=\int_{\mathcal{U}_{1}}\Phi _{\pi
}(x_{0},u_{1},x_{2})d\pi _{U_{1}}\left( u_{1}\right) .
\label{eq: Phi^0_pi def}
\end{equation}%
which is identified and can be expressed as an integral over $P^{0\ast }$.
Under the stated assumptions, $\Phi _{\pi }^{0}$ satisfies $\Phi _{\pi
}^{0}(x_{0},x_{2})=\Pi _{\pi }^{0}(u_{0}\left( x_{0}\right) ,\lambda
(x_{2})) $. Let $\mathcal{E}_{0,2}$ denote the joint support of $%
(\varepsilon _{0},\varepsilon _{2})$, which need not be independent. Then%
\begin{eqnarray*}
\Pi _{\pi }^{0}(u_{0},\lambda ) &=&\int_{\mathcal{U}_{1}}\Pi
(u_{0},u_{1},\lambda )d\pi _{U_{1}}\left( u_{1}\right) \\
&=&\int_{\mathcal{E}_{0,2}}\left[ \int_{\mathcal{U}_{1}}\int_{\mathcal{E}%
_{1}}P^{\ast }(u_{0}+e_{0},u_{1}+e_{1},\lambda +e_{2})dF_{\varepsilon
_{1}}(e_{1})d\pi _{U_{1}}\left( u_{1}\right) \right] dF_{\varepsilon
_{0},\varepsilon _{2}}(e_{0},e_{2}) \\
&=&\int_{\mathcal{E}_{0,2}}P^{0\ast }(u_{0}+e_{0},\lambda
+e_{2})dF_{\varepsilon _{0},\varepsilon _{2}}(e_{0},e_{2}),
\end{eqnarray*}%
where $P^{0\ast }=\int P^{\ast }d\pi _{V_{1}}$ is the vector of optimal
integrated CCPs. Importantly, $u_{0}\mapsto \Pi _{\pi }^{0}$ will inherit
many of the attractive properties of $u_{0}\mapsto P^{0\ast }$ and so we can
follow a strategy similar to the one used to show identification of $u_{1}$
in Theorem \ref{Th: RI ARUM u1 eps neq 0}. With%
\begin{equation*}
\mathcal{U}_{0,\mathrm{reg}}\left( \lambda \right) =\left\{ u_{0}\in \mathbb{%
R}^{J}:F_{\varepsilon }\bigl(\mathcal{E}_{\mathrm{reg}}(u_{0},\lambda )\bigr)%
>0\right\} ,
\end{equation*}%
where $\mathcal{E}_{\mathrm{reg}}(u_{0},\lambda )$ was defined in (\ref{eq:
E reg Pi definition}), we first obtain the following result regarding $%
u_{0}\mapsto \Pi _{\pi }^{0}$:

\begin{lemma}
\label{Lem: Pi0 u0 injectivity} Suppose Assumption \ref{ass: Affine ind} and
(\ref{eq: prior part known}) hold and $\varepsilon _{1}\bot (\varepsilon
_{0},\varepsilon _{2})$. Then, for any $\lambda >0$, $\Pi _{\pi }^{0}(\cdot
,\lambda )$ is continuous and monotone on $\mathbb{R}^{J}$, and strictly
monotone, hence injective, on $\mathcal{U}_{0,\mathrm{reg}}(\lambda )$.
\end{lemma}

While $\Pi _{\pi }^{0}(\cdot ,\lambda )$ is injective, we have been unable
to establish that the off-diagonal elements of its Jacobian are nonzero. We
therefore apply Theorem \ref{Th: Pi ID 2} to the composite representation 
\begin{equation*}
\Phi _{\pi }^{0}(x_{0},x_{2})=\Psi (u_{0}(x_{0}),x_{2}),\qquad \Psi
(v_{0},x_{2}):=\Pi _{\pi }^{0}(v_{0},\lambda (x_{2})),
\end{equation*}%
with $w=x_{0}$ and $z=x_{2}$. Since \eqref{eq: Phi^0_pi def} is identified
whenever $(x_{0},x_{2})\in \mathcal{Z}_{\pi }$, define 
\begin{eqnarray*}
\mathcal{X}_{0,\pi }(x_{2}) &:&=\left\{ x_{0}\in \func{int}\func{supp}%
(X_{0}\mid X_{2}=x_{2}):(x_{0},x_{2})\in \mathcal{Z}_{\pi }\right\} , \\
\mathcal{X}_{0,\pi ,\mathrm{reg}}(x_{2}) &:&=\left\{ x_{0}\in \mathcal{X}%
_{0,\pi }(x_{2}):u_{0}(x_{0})\in \mathcal{U}_{0,\mathrm{reg}}(\lambda
(x_{2}))\right\} ,
\end{eqnarray*}%
\begin{equation*}
\mathcal{X}_{2,\pi }:=\left\{ x_{2}:\mathcal{X}_{0,\pi ,\mathrm{reg}%
}(x_{2})\neq \varnothing \right\} ,\qquad \mathcal{U}_{0}(x_{2}):=u_{0}(%
\mathcal{X}_{0,\pi ,\mathrm{reg}}(x_{2})).
\end{equation*}%
For $x_{2}\in \mathcal{X}_{2,\pi }$, $\Pi _{\pi }^{0}(\cdot ,\lambda
(x_{2})) $ is injective on $\mathcal{U}_{0,\mathrm{reg}}(\lambda (x_{2}))$
by Lemma \ref{Lem: Pi0 u0 injectivity}.

\begin{assumption}
\label{Ass: RI ARUM u0 support eps neq 0} The function $u_{0,j}$ satisfies 
\begin{equation*}
u_{0,j}(x_{0,1,j},x_{0,2})=u_{0j}^{(0)}(x_{0,1,j},x_{0,2})+u_{0j}^{(1)}(x_{0,1,j})+u_{0j}^{(2)}(x_{0,2}),\qquad j=1,\ldots ,J,
\end{equation*}%
where $u_{0j}^{(0)}$ is known, $u_{0j}^{(1)}$ and $u_{0j}^{(2)}$ are
unknown, and all three components are continuous on their relevant domains.
There exist a known $x_{2,0}\in \mathcal{X}_{2,\pi }$, nonempty open
intervals $\mathcal{I}_{0j}\subseteq \mathbb{R}$, $j=1,\ldots ,J$, and a
nonempty open connected set $\mathcal{X}_{0,2}^{\ast }$ such that $\mathcal{X%
}_{0}^{\ast }:=\prod_{j=1}^{J}\mathcal{I}_{0j}\times \mathcal{X}_{0,2}^{\ast
}\subseteq \mathcal{X}_{0,\pi ,\mathrm{reg}}(x_{2,0})$. For a known $%
x_{0,0}=(x_{0,1,0},x_{0,2,0})\in \mathcal{X}_{0}^{\ast }$, suppose that: (i)
for every $j$ and $x_{0,1,j}\in \mathcal{I}_{0j}$, $x_{0,2}\mapsto
u_{0,j}(x_{0,1,j},x_{0,2})$ is nonconstant on $\mathcal{X}_{0,2}^{\ast }$;
(ii) for every $j$, $u_{0j}^{(1)}$ is differentiable at $x_{0,1,j,0}$, and $%
u_{0j}^{(0)}(\cdot ,x_{0,2})$ is differentiable there for every $x_{0,2}\in 
\mathcal{X}_{0,2}^{\ast }$, with 
\begin{equation*}
1+\nabla _{x_{0,1,j}}u_{0j}^{(0)}(x_{0,1,j,0},x_{0,2})-\nabla
_{x_{0,1,j}}u_{0j}^{(0)}(x_{0,1,j,0},x_{0,2,0})\neq 0,\qquad x_{0,2}\in 
\mathcal{X}_{0,2}^{\ast };
\end{equation*}%
(iii) for every $x_{2}\in \mathcal{X}_{2,\pi }$, $\mathcal{U}%
_{0}(x_{2})=u_{0}(\mathcal{X}_{0}^{\ast }\cap \mathcal{X}_{0,\pi}(x_{2}))$.
\end{assumption}

\begin{theorem}
\label{Thm: RI ARUM u0 eps neq 0} Suppose $u_{1}$ and $\lambda$ are
identified, $\pi_{U_{1}}$ is known, and the conditions of Lemma \ref{Lem:
Pi0 u0 injectivity} hold. Under Assumption \ref{Ass: RI ARUM u0 support eps
neq 0}, $u_{0}$ is identified on $\bigcup_{x_{2}\in\mathcal{X}_{2,\pi}}%
\mathcal{X}_{0,\pi,\mathrm{reg}}(x_{2})$ under the following normalisations:
for some known $x_{0,0}=(x_{0,1,0},x_{0,2,0})\in\mathcal{X}_{0}^{\ast}$, 
\begin{equation*}
u_{0}(x_{0,0})=0,\qquad \nabla_{x_{0,1,j}}u_{0,j}(x_{0,0})=1,\quad
j=1,\ldots,J.
\end{equation*}
Moreover, $\Pi_{\pi}^{0}(\cdot,\lambda(x_{2}))$ is identified on $\mathcal{U}%
_{0}(x_{2})$ for every $x_{2}\in\mathcal{X}_{2,\pi}$.
\end{theorem}

\paragraph{Prior is unknown}

In the case of the prior being completely unknown to the Econometrician and
unrestricted, we have not been able to show identification of $u_{0}$. One
could have hoped that, in this setting, the integrated version of $\Phi $
w.r.t. $F_{U_{1}}$ would still enjoy nice properties akin to those of $\Phi
_{\pi }^{0}(x_{0},x_{2})=\Pi _{\pi }^{0}(u_{0}\left( x_{0}\right) ,\lambda
(x_{2}))$ defined above. However,%
\begin{eqnarray*}
\Phi _{F}^{0}(x_{0},x_{2}) &=&\int_{\mathcal{U}_{1}}\Phi
(x_{0},u_{1},x_{2})dF_{U_{1}}\left( u_{1}\right) =\int_{\mathcal{U}_{1}}\Pi
(u_{0}\left( x_{0}\right) ,u_{1},\lambda (x_{2}))dF_{U_{1}}\left(
u_{1}\right) \\
&=&:\Pi _{F}^{0}\left( u_{0}\left( x_{0}\right) ,\lambda (x_{2})\right)
\end{eqnarray*}%
where $u_{0}\mapsto \Pi _{F}^{0}\left( u_{0},\lambda \right) $ will in
general not be injective, nor will it have a symmetric Jacobian with
non--zero off-diagonal elements. As such, point identification of $%
u_{0}\left( x_{0}\right) $ seems very unlikely.

Alternatively, one could have hoped to use the following identity for
identification, $\Phi (x_{0},u_{1},x_{2})=\Pi (u_{0}\left( x_{0}\right)
,u_{1},\lambda (x_{2}))$. However, we have also been unable to show that $%
u_{0}\mapsto \Pi (u_{0},u_{1},\lambda )$ is injective, even if $v_{0}\mapsto
P^{\ast }(v_{0},v_{1},\lambda )$ is. In fact, we are not even able to show
that $\nabla _{u_{0}}\Pi (u_{0},u_{1},\lambda )$ is non--singular. We will
not rule out that either, or both, of these properties actually hold, but we
have not been able to verify them. As such, the set of tools used to obtain
our previous identification results is not available to us and we are unable
to provide any identification results for $u_{0}$ in general.

However, if the DM's prior is not too badly misspecified, then we can
recover the identification result of Theorem \ref{Thm: RI ARUM u0 eps neq 0}%
. This idea is similar to the one of \citet{chenhansenhansen2024robust} who
also investigate identification of structural models under bounds on the
deviation of subjective beliefs from the physical law. However, in contrast
to their general set--identification results, we here show that point
identification can in fact be restored if the deviation is "moderate". The
key insight is that if $\pi _{V_{1}}$ is not too far away from $F_{V_{1}}$,
then $u_{0}\mapsto \Pi _{F}^{0}\left( u_{0},\lambda \right) $ will inherit
many of the desirable features of $u_{0}\mapsto \Pi _{\pi }^{0}\left(
u_{0},\lambda \right) $. In particular, $u_{0}\mapsto \Pi _{F}^{0}\left(
u_{0},\lambda \right) $ will be injective and $\nabla _{u_{0}}\Pi
_{F}^{0}\left( u_{0},\lambda \right) $ will be non--singular. With%
\begin{eqnarray*}
L_{f} &:&=\int_{\mathbb{R}^{J}\times \lbrack 0,\infty )}\left\Vert \nabla
_{e_{0}}f_{\varepsilon _{0},\varepsilon _{2}}(e_{0},e_{2})\right\Vert
\,de_{0}\,de_{2}, \\
c_{\mathcal{U}}(\lambda ) &:&=\inf_{u_{0}\in \mathcal{U}}\lambda _{\min
}\left( \nabla _{u_{0}}\Pi _{\pi }^{0}(u_{0},\lambda )\right) , \\
d_{\mathrm{TV}}(\pi _{V_{1}},F_{V_{1}}) &:&=\sup_{A}\left\vert \pi
_{V_{1}}(A)-F_{V_{1}}(A)\right\vert ,
\end{eqnarray*}
where we note that the eigenvalue in the definition of $c_{\mathcal{U}%
}(\lambda )$is well defined because $\nabla _{u_{0}}\Pi _{\pi }^{0}$ is
symmetric, we obtain the following:

\begin{lemma}
\label{Lem: PiF u0 injectivity} Suppose Assumption \ref{ass: Affine ind}
holds, $\varepsilon _{1}\bot (\varepsilon _{0},\varepsilon _{2})$, and $%
\varepsilon _{2}\geq 0$ almost surely, and the marginal distribution of $%
(\varepsilon _{0},\varepsilon _{2})$ has a density $f_{\varepsilon
_{0},\varepsilon _{2}}$ which is continuously differentiable w.r.t. $e_{0}$
with $L_{f}<\infty $. For any convex set $\mathcal{U}\subseteq \mathcal{U}%
_{0,\mathrm{reg}}(\lambda )$ with $c_{\mathcal{U}}(\lambda )>0$ and $%
2L_{f}\,d_{\mathrm{TV}}(\pi _{V_{1}},F_{V_{1}})<c_{\mathcal{U}}(\lambda )$, $%
\Pi _{F}^{0}(\cdot ,\lambda ):\mathcal{U}\mapsto \mathbb{R}^{J}$ is
continuously differentiable with non--singular Jacobian and is strictly
monotone, hence injective.
\end{lemma}

The condition $c_{\mathcal{U}}(\lambda )>0$ is not implied by boundedness of 
$\mathcal{U}$ alone, since it may fail as $u_{0}$ approaches the boundary of 
$\mathcal{U}_{0,\mathrm{reg}}(\lambda )$. If $\overline{\mathcal{U}}$ is
compact and contained in $\mathcal{U}_{0,\mathrm{reg}}(\lambda )$,
continuity of $\nabla _{u_{0}}\Pi _{\pi }^{0}$ and Lemma \ref{Lem: Pi0 u0
injectivity} imply $c_{\mathcal{U}}(\lambda )>0$. No differentiability or
integrable Jacobian bound for $v_{0}\mapsto P^{\ast }(v_{0},v_{1},\lambda
+e_{2})$ is required. One can now proceed as in the proof of Theorem \ref%
{Thm: RI ARUM u0 eps neq 0} and show identification of $u_{0}$ under the
conditions of this theorem together with the ones of the lemma above. We
will refrain from stating the result here. Compared to the results of %
\citet{chenhansenhansen2024robust}, above lemma provides a quantitative
robustness guarantee for point identification of $u_{0}$ rather than bounds
on an identified set for this function.

\subsubsection{(Non-)identification of the prior}

We have shown that, under suitable assumptions, the functions $u_{0}$, $%
u_{1} $ and $\lambda $ are identified even when $\varepsilon \neq 0$.
Consequently, the composite mapping $\Pi $ is identified on the
corresponding index regions. However, this does not seem to allow us to
identify the DM's prior $\pi _{V_{1}}$ without further restrictions. Recall
that%
\begin{equation*}
\Pi (u_{0},u_{1},\lambda )=\int_{\mathcal{E}}P_{\pi _{V_{1}}}^{\ast
}(u_{0}+e_{0},u_{1}+e_{1},\lambda +e_{2})dF_{\varepsilon
}(e_{0},e_{1},e_{2}),
\end{equation*}%
where we now index the DM's optimal CCP by $\pi _{V_{1}}$. While $\Pi $ is
known to us, the right--hand side contains two unknowns, $\pi $ and $%
F_{\varepsilon }$. Unless very strong restrictions are imposed on $\pi $ and 
$F_{\varepsilon }$, we do not see much hope to achieve identification of the
prior. We end our identification quest on this somewhat sad note.

\section{Conclusion}

Based on the discrete choice rational inattention model of %
\citet{matvejka2015rational}, we develop a flexible class of econometric
models for rationally inattentive agents. We provide a comprehensive
identification analysis of this class that shows what can and cannot be
identified under different scenarios.

\clearpage
\bibliographystyle{apalike}
\bibliography{references}

@article{matvejka2015rational,
  title={Rational inattention to discrete choices: A new foundation for the multinomial logit model},
  author={Mat{\v{e}}jka, Filip and McKay, Alisdair},
  journal={American Economic Review},
  volume={105},
  number={1},
  pages={272--98},
  year={2015}
}

@article{allen2019identification,
  title={Identification with additively separable heterogeneity},
  author={Allen, Roy and Rehbeck, John},
  journal={Econometrica},
  volume={87},
  number={3},
  pages={1021--1054},
  year={2019},
  publisher={Wiley Online Library}
}

@article{caplin2019rational,
  title={Rational inattention, optimal consideration sets, and stochastic choice},
  author={Caplin, Andrew and Dean, Mark and Leahy, John},
  journal={The Review of Economic Studies},
  volume={86},
  number={3},
  pages={1061--1094},
  year={2019},
  publisher={Oxford University Press}
}

@article{fosgerau2020discrete,
  title={Discrete choice and rational inattention: A general equivalence result},
  author={Fosgerau, Mogens and Melo, Emerson and De Palma, Andre and Shum, Matthew},
  journal={International economic review},
  volume={61},
  number={4},
  pages={1569--1589},
  year={2020},
  publisher={Wiley Online Library}
}

@article{mcfadden1973conditional,
  title={Conditional logit analysis of qualitative choice behavior},
  author={McFadden, Daniel and others},
  year={1973},
  publisher={Institute of Urban and Regional Development, University of California~…}
}

@article{mcfadden1981econometric,
  title={Econometric models of probabilistic choice},
  author={McFadden, Daniel},
  journal={Structural analysis of discrete data with econometric applications},
  volume={198272},
  year={1981},
  publisher={MIT Press}
}

@article{joo2023rational,
  title={Rational inattention as an empirical framework for discrete choice and consumer-welfare evaluation},
  author={Joo, Joonhwi},
  journal={Journal of Marketing Research},
  volume={60},
  number={2},
  pages={278--298},
  year={2023},
  publisher={SAGE Publications Sage CA: Los Angeles, CA}
}

@article{caplin2022rationally,
  title={Rationally inattentive behavior: Characterizing and generalizing Shannon entropy},
  author={Caplin, Andrew and Dean, Mark and Leahy, John},
  journal={Journal of Political Economy},
  volume={130},
  number={6},
  pages={1676--1715},
  year={2022},
  publisher={The University of Chicago Press Chicago, IL}
}

@article{brown2024endogenous,
  title={Endogenous Information and Simplifying Insurance Choice},
  author={Brown, Zach Y. and Jeon, Jihye},
  journal={Econometrica},
  volume={92},
  number={3},
  pages={881--911},
  year={2024},
  doi={10.3982/ECTA18555}
}

@article{liao2024identification,
  title={Identification of a rational inattention discrete choice model},
  author={Liao, Moyu},
  journal={Journal of Econometrics},
  volume={240},
  number={1},
  pages={105670},
  year={2024},
  publisher={Elsevier}
}

@article{sorensen2022mcfadden,
  title = {How {McFadden} met {Rockafellar} and learned to do more with less},
  author = {S{\o}rensen, Jesper R.-V. and Fosgerau, Mogens},
  journal = {Journal of Mathematical Economics},
  volume = {100},
  pages = {102629},
  year = {2022},
  publisher = {Elsevier}
}

@article{matzkin1993nonparametric,
  title={Nonparametric identification and estimation of polychotomous choice models},
  author={Matzkin, Rosa L},
  journal={Journal of Econometrics},
  volume={58},
  number={1-2},
  pages={137--168},
  year={1993},
  publisher={Elsevier}
}

@article{Berry2014identification,
 ISSN = {00129682, 14680262},
 URL = {http://www.jstor.org/stable/24029293},
 author = {Steven T. Berry and Philip A. Haile},
 journal = {Econometrica},
 number = {5},
 pages = {1749--1797},
 publisher = {The Econometric Society},
 title = {IDENTIFICATION IN DIFFERENTIATED PRODUCTS MARKETS USING MARKET LEVEL DATA},
 urldate = {2025-06-12},
 volume = {82},
 year = {2014}
}

@article{natan2025choice,
  title={Choice frictions in large assortments},
  author={Natan, Olivia R},
  journal={Marketing Science},
  volume={44},
  number={3},
  pages={593--625},
  year={2025},
  publisher={INFORMS}
}

@article{bertoli2020rational,
  title={Rational inattention and migration decisions},
  author={Bertoli, Simone and Moraga, Jes{\'u}s Fern{\'a}ndez-Huertas and Guichard, Lucas},
  journal={Journal of International Economics},
  volume={126},
  pages={103364},
  year={2020},
  publisher={Elsevier}
}

@Article{Fosgerau2020,
  author    = {Fosgerau, Mogens and Kristensen, Dennis},
  journal   = {The Econometrics Journal},
  title     = {Identification of a class of index models: A topological approach},
  year      = {2020},
  issn      = {1368-423X},
  month     = June,
  number    = {1},
  pages     = {121--133},
  volume    = {24},
  doi       = {10.1093/ectj/utaa016},
  publisher = {Oxford University Press (OUP)},
}

@article{chenhansenhansen2024robust,
  author  = {Chen, Xiaohong and Hansen, Lars Peter and Hansen, Peter G.},
  title   = {Robust inference for moment condition models without rational expectations},
  journal = {Journal of Econometrics},
  year    = {2024},
  volume  = {243},
  number  = {1--2},
  pages   = {105653},
  doi     = {10.1016/j.jeconom.2023.105653}
}

@article{anhuxiao2021dynamic,
  author  = {An, Yonghong and Hu, Yingyao and Xiao, Ruli},
  title   = {Dynamic decisions under subjective expectations: A structural analysis},
  journal = {Journal of Econometrics},
  year    = {2021},
  volume  = {222},
  number  = {1},
  pages   = {645--675},
  doi     = {10.1016/j.jeconom.2020.04.046}
}

@article{foxkimryanbajari2012,
  author  = {Fox, Jeremy T. and Kim, Kyoo il and Ryan, Stephen P. and Bajari, Patrick},
  title   = {The random coefficients logit model is identified},
  journal = {Journal of Econometrics},
  year    = {2012},
  volume  = {166},
  number  = {2},
  pages   = {204--212},
  doi     = {10.1016/j.jeconom.2011.09.002}
}

@book{widder1941laplace,
  author    = {Widder, David Vernon},
  title     = {The Laplace Transform},
  publisher = {Princeton University Press},
  address   = {Princeton, NJ},
  year      = {1941}
}

@unpublished{wang2026semiparametric,
  author = {Wang, Hanqi},
  title  = {Estimating a Rational Inattention Discrete Choice Model: A Semiparametric Approach},
  year   = {2026},
  month  = nov,
  note   = {Job market paper, Vancouver School of Economics, University of British Columbia, November 3, 2026}
}

@Article{Zhu2026BLP,
  author  = {Zhu, Ruolei},
  journal = {manuscript},
  title   = {Demand Estimation for Rationally Inattentive Consumers},
  year    = {2026},
}

@article{ichimuraThompson1998,
  author = {Ichimura, Hidehiko and Thompson, T. Scott},
  title = {Maximum Likelihood Estimation of a Binary Choice Model with Random Coefficients of Unknown Distribution},
  journal = {Journal of Econometrics},
  year = {1998}, volume = {86}, number = {2}, pages = {269--295}
}

@article{gautierKitamura2013,
  author = {Gautier, Eric and Kitamura, Yuichi},
  title = {Nonparametric Estimation in Random Coefficients Binary Choice Models},
  journal = {Econometrica}, year = {2013}, volume = {81}, number = {2}, pages = {581--607}
}

@techreport{dunkerHoderleinKaido2013,
  author = {Dunker, Fabian and Hoderlein, Stefan and Kaido, Hiroaki},
  title = {Random Coefficients in Static Games of Complete Information},
  institution = {Centre for Microdata Methods and Practice (cemmap)},
  type = {Working paper}, year = {2013}
}

@article{foxGandhi2016,
  author = {Fox, Jeremy T. and Gandhi, Amit},
  title = {Nonparametric Identification and Estimation of Random Coefficients in Multinomial Choice Models},
  journal = {RAND Journal of Economics}, year = {2016}, volume = {47}, number = {1}, pages = {118--139}
}

@incollection{matzkin2007heterogeneous,
  author = {Matzkin, Rosa L.},
  title = {Heterogeneous Choice},
  booktitle = {Advances in Economics and Econometrics: Theory and Applications, Ninth World Congress},
  editor = {Blundell, Richard and Newey, Whitney K. and Persson, Torsten},
  publisher = {Cambridge University Press}, year = {2007}
}

@techreport{berryHaile2009,
  author = {Berry, Steven T. and Haile, Philip A.},
  title = {Nonparametric Identification of Multinomial Choice Demand Models with Heterogeneous Consumers},
  institution = {Cowles Foundation, Yale University},
  type = {Discussion Paper}, year = {2009}
}

@article{brieschChintaguntaMatzkin2010,
  author = {Briesch, Richard and Chintagunta, Pradeep and Matzkin, Rosa L.},
  title = {Nonparametric Discrete Choice Models with Unobserved Heterogeneity},
  journal = {Journal of Business \& Economic Statistics},
  year = {2010}, volume = {28}, number = {2}, pages = {291--307}
}

@article{matzkin2012limited,
  author = {Matzkin, Rosa L.},
  title = {Identification in Nonparametric Limited Dependent Variable Models with Simultaneity and Unobserved Heterogeneity},
  journal = {Journal of Econometrics}, year = {2012}, volume = {166}, number = {1}, pages = {106--115}
}

@article{matzkin1991monotone,
  author = {Matzkin, Rosa L.},
  title = {Semiparametric Estimation of Monotone and Concave Utility Functions for Polychotomous Choice Models},
  journal = {Econometrica}, year = {1991}, pages = {1315--1327}
}

@article{matzkin1992binary,
  author = {Matzkin, Rosa L.},
  title = {Nonparametric and Distribution-Free Estimation of the Binary Threshold Crossing and the Binary Choice Models},
  journal = {Econometrica}, year = {1992}, volume = {60}, pages = {239--270}
}

@incollection{matzkin1994restrictions,
  author = {Matzkin, Rosa L.},
  title = {Restrictions of Economic Theory in Nonparametric Methods},
  booktitle = {Handbook of Econometrics},
  editor = {Engle, Robert F. and McFadden, Daniel L.},
  volume = {4}, publisher = {Elsevier}, year = {1994}, pages = {2523--2558}
}

@incollection{matzkin2007nonparametric,
  author = {Matzkin, Rosa L.},
  title = {Nonparametric Identification},
  booktitle = {Handbook of Econometrics},
  editor = {Heckman, James J. and Leamer, Edward E.},
  volume = {6B}, publisher = {Elsevier}, year = {2007}
}

@article{matzkin2019constructive,
  author = {Matzkin, Rosa L.},
  title = {Constructive Identification in Some Nonseparable Discrete Choice Models},
  journal = {Journal of Econometrics},
  volume = {211},
  number = {1},
  pages = {83--103},
  year = {2019},
  doi = {10.1016/j.jeconom.2018.12.007}
}

@article{berryHaile2024micro,
  author = {Berry, Steven T. and Haile, Philip A.},
  title = {Nonparametric Identification of Differentiated Products Demand Using Micro Data},
  journal = {Econometrica},
  year = {2024},
  volume = {92},
  number = {4},
  pages = {1135--1162},
  doi = {10.3982/ECTA20731}
}

@article{fox2021note,
  author = {Fox, Jeremy T.},
  title = {A Note on Nonparametric Identification of Distributions of Random Coefficients in Multinomial Choice Models},
  journal = {Annals of Economics and Statistics},
  year = {2021},
  number = {142},
  pages = {305--310},
  doi = {10.15609/annaeconstat2009.142.0305}
}

@article{allenRehbeck2025latent,
  author = {Allen, Roy and Rehbeck, John},
  title = {Identification of Random Coefficient Latent Utility Models},
  journal = {Quantitative Economics},
  year = {2025},
  volume = {16},
  number = {4},
  pages = {1223--1265},
  doi = {10.3982/QE1809}
}

@Article{Zhu2026MLE,
  author  = {Zhu, Ruolei},
  journal = {manuscript},
  title   = {Estimation of Rational Inattention–Additive Random Utility Models},
  year    = {2026},
}

@article{caplin2020competitive,
  title={Rational Inattention, Competitive Supply, and Psychometrics},
  author={Caplin, Andrew and Csaba, D{\'a}niel and Leahy, John and Nov, Oded},
  journal={The Quarterly Journal of Economics},
  volume={135},
  number={3},
  pages={1681--1724},
  year={2020},
  doi={10.1093/qje/qjaa011}
}

@article{turlo2025discrete,
  title={Discrete Choice in Marketing through the Lens of Rational Inattention},
  author={Turlo, Sergey and Fina, Matteo and Kasinger, Johannes and Laghaie, Arash and Otter, Thomas},
  journal={Quantitative Marketing and Economics},
  volume={23},
  number={1},
  pages={45--104},
  year={2025},
  doi={10.1007/s11129-025-09292-9}
}
\clearpage

\appendix

\section{Proofs of Section \protect\ref{sec: DMs problem} \label%
{appendix:DMCCP}}

\begin{proof}[Proof of Lemma~\protect\ref{lemma:DMCCP}]
For each fixed $v_{0}$, this is the standard Shannon rational inattention
problem with prior $\pi _{V|V_{0}}(\cdot |v_{0})$. The result therefore
follows from Theorem 1 of \citet{matvejka2015rational}; see in particular
their Equation (13), which gives the generalized multinomial logit form of
the state-contingent choice probabilities. Their Lemma 1 and Corollary 1
establish the preceding reduction from signal rules to state-contingent
choice probabilities. Applying these results conditional on $V_{0}=v_{0}$
yields \eqref{eq:DMCCP}.
\end{proof}

\bigskip

\begin{proof}[Proof of Lemma~\protect\ref{lemma:optimalp0}]
The reduced maximisation problem is Lemma 2 of \citet{matvejka2015rational},
applied conditional on $V_{0}=v_{0}$. Substituting \eqref{eq:DMCCP} into the
state-contingent choice problem yields the reduced maximisation problem %
\eqref{eq: P0 charac} with objective \eqref{eq: Q0 def}. The Kuhn--Tucker
conditions for this concave problem include inequalities for alternatives
assigned zero unconditional probability; compare Corollary 2 of %
\citet{matvejka2015rational} and \citet{caplin2019rational}.

Existence of a solution to the Shannon problem and the reduced formulation
are established in \citet{matvejka2015rational}; see their Lemma 2 and the
discussion following Corollary 2.
\end{proof}

\bigskip

\begin{proof}[Proof of Lemma~\protect\ref{lemma:outsidenormalization}]
Let $Q^{0}(P^{0};v_{0},\lambda )$ be the objective function corresponding to
the prior $\pi $. Replace $\pi $ with a new prior $\tilde{\pi}$ that changes 
$V_{1,j}$ to $V_{1,j}+C$, where $C$ is integrable, and let $\tilde{Q}%
^{0}(P^{0};v_{0},\lambda )$ be the objective function with this new prior.
It satisfies%
\begin{eqnarray*}
\tilde{Q}^{0}(P^{0};v_{0},\lambda ) &=&\lambda E\left[ \ln \left( \sum_{j\in 
\mathcal{J}}P_{j}^{0}(v_{0})\exp \left( (v_{0,j}+V_{1,j}+C)/\lambda \right)
\right) \right] \\
&=&E\left[ C\right] +\lambda E\left[ \ln \left( \sum_{j\in \mathcal{J}%
}P_{j}^{0}(v_{0})\exp \left( (v_{0,j}+V_{1,j})/\lambda \right) \right) %
\right] \\
&=&E\left[ C\right] +Q^{0}(P^{0};v_{0},\lambda ),
\end{eqnarray*}%
and so $\arg \max_{P^{0}}\tilde{Q}^{0}(P^{0};v_{0},\lambda )=\arg
\max_{P^{0}}Q^{0}(P^{0};v_{0},\lambda )$.
\end{proof}

\bigskip

\bigskip

\begin{proof}[Proof of Lemma~\protect\ref{lemma:Q0_concavity}]
Write 
\begin{equation*}
\sigma (v_{0},v_{1};P^{0})=1+\sum_{j=1}^{J}P_{j}^{0}\left\{ \exp
((v_{0,j}+v_{1,j})/\lambda )-1\right\} ,
\end{equation*}%
so that $Q^{0}(P^{0};v_{0},\lambda )=\lambda \int_{\mathcal{V}_{1}}\log
\sigma (v_{0},v_{1};P^{0})d\pi _{V_{1}}(v_{1})$. For each $v_{1}$, $\sigma
(v_{0},v_{1};P^{0})$ is affine in $P^{0}$ and strictly positive on $\Delta (%
\mathcal{J})$. Since the logarithm is concave, $P^{0}\mapsto \log \sigma
(v_{0},v_{1};P^{0})$ is concave, and integration preserves concavity.

Next, consider $P^{0}\in \func{int}\Delta (\mathcal{J})$. Since $%
P_{0}^{0}=1-\sum_{j=1}^{J}P_{j}^{0}>0$, $\sigma (v_{0},v_{1};P^{0})\geq
P_{0}^{0}$ and 
\begin{equation*}
\log \sigma (v_{0},v_{1};P^{0})\leq \frac{1}{\lambda }\max \left\{
0,\max_{1\leq j\leq J} (v_{0,j}+v_{1,j})\right\} .
\end{equation*}%
Under Assumption \ref{ass: Affine ind}, $E_{\pi _{V_{1}}} [\|V_{1}\|]<\infty 
$, so $Q^{0}(P^{0};v_{0},\lambda )$ is finite in the interior.

For any $a\in \mathbb{R}^{J}$, differentiation yields 
\begin{equation*}
\begin{split}
& a^{\prime }\nabla _{P^{0}P^{0}}^{2}Q^{0}(P^{0};v_{0},\lambda )a \\
& \quad =-\lambda \int_{\mathcal{V}_{1}}\frac{\left[ \sum_{j=1}^{J}a_{j}%
\left\{ \exp ((v_{0,j}+v_{1,j})/\lambda )-1\right\} \right] ^{2}}{\sigma
(v_{0},v_{1};P^{0})^{2}}d\pi _{V_{1}}(v_{1})\leq 0.
\end{split}%
\end{equation*}%
Differentiation under the integral is justified because, in a neighbourhood
of any interior $P^{0}$, the integrand in the above expression is uniformly
bounded, using the positive lower bounds on $P^{0}$. Equality in the
quadratic form for some $a\neq 0$ would imply 
\begin{equation*}
\sum_{j=1}^{J}a_{j}\exp (v_{0,j}/\lambda )\exp (V_{1,j}/\lambda
)-\sum_{j=1}^{J}a_{j}=0\quad \pi _{V_{1}}\text{-a.s.}
\end{equation*}%
By Assumption \ref{ass: Affine ind}, the random variables $\exp
(V_{1,1}/\lambda ),\ldots ,\exp (V_{1,J}/\lambda )$ are affinely
independent. Consequently, $a=0$, a contradiction. The Hessian is therefore
negative definite throughout the interior, establishing strict concavity.

Finally, for fixed interior $P^{0}$, 
\begin{equation*}
\nabla_{v_{0,j}}\left[ \lambda \log \sigma (v_{0},v_{1};P^{0})\right] =\frac{%
P_{j}^{0}\exp ((v_{0,j}+v_{1,j})/\lambda )}{\sigma (v_{0},v_{1};P^{0})}%
=:p_{j}(v_{0},v_{1};P^{0}).
\end{equation*}%
Since the right--hand side is bounded between 0 and 1, differentiation under
the integral is justified by dominated convergence. Similarly, for $%
j,k=1,\ldots ,J$, 
\begin{equation*}
\nabla_{v_{0,j}v_{0,k}}^{2}Q^{0}=\frac{1}{\lambda }\int_{\mathcal{V}_{1}}%
\left[ \mathbf{1}\{j=k\}p_{j}-p_{j}p_{k}\right] d\pi _{V_{1}}(v_{1}).
\end{equation*}%
The integrand is uniformly bounded and continuous in $v_{0}$. Dominated
convergence establishes existence and continuity of the second derivatives,
completing the proof.
\end{proof}

\bigskip

\begin{proof}[Proof of Theorem~\protect\ref{Thm: P0 one-to-one}]
Since $\lambda >0$ is kept fixed, for parts (i)--(iii), we suppress its
dependence in the following. For any given $P^{0}\in \Delta (\mathcal{J})$,
the mapping $v_{0}\mapsto Q^{0}(P^{0};v_{0},\lambda )$ is convex because its
integrand is a log-sum-exp function of $v_{0}$. Hence $Q_{\max }^{0}$ is
convex as the pointwise maximum of convex functions. Since the maximiser $%
P^{0\ast }(v_{0})$ is unique, Danskin's theorem and Lemma~\ref%
{lemma:Q0_concavity} give 
\begin{equation}
\nabla _{v_{0}}Q_{\max }^{0}(v_{0},\lambda )=\int_{\mathcal{V}_{1}}P^{\ast
}(v_{1}|v_{0})d\pi (v_{1})=P^{0\ast }(v_{0}).  \label{eq: gradient W}
\end{equation}

We next establish strict convexity of $W(v_{0})$ along regular line segments
in $\mathcal{V}_{0,\mathrm{reg}}$. Suppose that there exists distinct $v_{0},%
\tilde{v}_{0}\in \mathcal{V}_{0,\mathrm{reg}}$ and the line segment joining
them is contained in $\mathcal{V}_{0,\mathrm{reg}}$. For $t\in (0,1)$, let $%
v_{0}^{t}=(1-t)v_{0}+t\tilde{v}_{0}$ and $P_{t}^{0\ast }=P^{0\ast
}(v_{0}^{t})$. Since $P_{t}^{0\ast }\in \func{int}\Delta (\mathcal{J})$, the
mapping $v_{0}\mapsto Q^{0}(P_{t}^{0\ast };v_{0})$ is strictly convex: for
every nonzero $a\in \mathbb{R}^{J}$, the Hessian of its integrand satisfies 
\begin{equation*}
a^{\top }\nabla _{v_{0}v_{0}}^{2}\{\lambda \ln e(P_{t}^{0\ast
},v_{0},v_{1})\}a=\frac{1}{\lambda }\left\{
\sum_{j=1}^{J}p_{j}a_{j}^{2}-\left( \sum_{j=1}^{J}p_{j}a_{j}\right)
^{2}\right\} \geq \frac{p_{0}}{\lambda }\sum_{j=1}^{J}p_{j}a_{j}^{2}>0,
\end{equation*}%
where $p_{j}>0$, $j=0,\ldots ,J$, are the corresponding log-sum-exp weights
and $a_{0}=0$. Hence 
\begin{eqnarray*}
Q_{\max }^{0}(v_{0}^{t}) &=&Q^{0}(P_{t}^{0\ast
};v_{0}^{t})<(1-t)Q^{0}(P_{t}^{0\ast };v_{0})+tQ^{0}(P_{t}^{0\ast };\tilde{v}%
_{0}) \\
&\leq &(1-t)W(v_{0})+tW(\tilde{v}_{0}).
\end{eqnarray*}

Next, we use above to show that $\mathcal{V}_{0,\mathrm{reg}}$ is non-empty
and open and that $P^{0\ast }(v_{0})$ is one-to-one on $\mathcal{V}_{0,%
\mathrm{reg}}$. That is, for any $\bar{P}^{0}\in \func{int}\Delta (\mathcal{J%
})$, there exists a unique $\bar{v}_{0}\in \mathcal{V}_{0,\mathrm{reg}}$ so
that $P^{0\ast }(v_{0})=\bar{P}^{0}$. This is done by convex duality. Recall
the definition of the convex conjugate of $Q_{\max }^{0}(v_{0},\lambda )$, $%
Q_{\max }^{0\ast }(P_{0})=\sup_{v_{0}\in \mathbb{R}^{J}}\left\{
(P^{0})^{\prime }v_{0}-Q_{\max }^{0}(v_{0},\lambda )\right\} $. For any $%
\bar{P}^{0}\in \func{int}\Delta (\mathcal{J})$, we claim that there exists a
unique $\bar{v}_{0}\in \arg \max_{v_{0}\in \mathbb{R}^{J}}\left\{
(P^{0})^{\prime }v_{0}-W(v_{0})\right\} $. The first order condition to this
problem then yields $\nabla Q_{\max }^{0}(\bar{v}_{0})=\bar{P}^{0}$, which
combined with (\ref{eq: gradient W}) yields $\bar{v}_{0}\in \mathcal{V}_{0,%
\mathrm{reg}}$.

Since $Q_{\max }^{0}(v_{0},\lambda )$ is continuous, to show its supremum is
attained at some $\bar{v}_{0}$, it suffices to demonstrate that $%
v_{0}\mapsto (\bar{P}^{0})^{\prime }v_{0}-Q_{\max }^{0}(v_{0},\lambda )$
tends to $-\infty $ as $\Vert v_{0}\Vert _{\infty }\rightarrow \infty $.
Choosing $P^{0}$ that assigns probability one to the outside option or to
alternative $j$ gives 
\begin{equation*}
Q_{\max }^{0}(v_{0},\lambda )\geq \max \left\{ 0,\max_{1\leq j\leq J}\left[
v_{0,j}+E[V_{1,j}]\right] \right\} \geq \bar{v}_{0}-\bar{V}_{1},
\end{equation*}%
where $\bar{v}_{0}=\max_{1\leq j\leq J}(v_{0,j})_{+}$ and $\bar{V}%
_{1}=\max_{1\leq j\leq J}|E[V_{1,j}]|$. Thus, with $\bar{P}_{\min
}^{0}=\min_{0\leq j\leq J}\bar{P}_{j}^{0}>0$, 
\begin{eqnarray*}
(\bar{P}^{0})^{\prime }v_{0}-Q_{\max }^{0}(v_{0},\lambda ) &\leq &(\bar{P}%
^{0})^{\prime }v_{0}-\bar{v}_{0}+\bar{V}_{1}\leq -\sum_{j=1}^{J}\bar{P}%
_{j}^{0}(v_{0,j})_{-}-\bar{P}_{0}^{0}\bar{v}_{0}+\bar{V}_{1} \\
&\leq &\bar{V}_{1}-\bar{P}_{\min }^{0}\Vert v_{0}\Vert _{\infty }\rightarrow
-\infty .
\end{eqnarray*}

To see that $\bar{v}_{0}$ is unique, suppose that $\tilde{v}_{0}$ is another
distinct maximiser. Concavity of $v_{0}\mapsto (\bar{P}^{0})^{\prime
}v_{0}-Q_{\max }^{0}(v_{0},\lambda )$ would then imply that every $%
v_{0}^{t}=(1-t)v_{0}+t\tilde{v}_{0}$ is also a maximiser. Hence its
first-order condition gives 
\begin{equation*}
P^{0\ast }(v_{0}^{t})=\nabla Q_{\max }^{0}(v_{0}^{t})=\bar{P}^{0}\in \func{%
int}\Delta (\mathcal{J}),\qquad t\in \lbrack 0,1],
\end{equation*}%
so the entire segment is contained in $\mathcal{V}_{0,\mathrm{reg}}$. But $W$
is strictly convex along this segment, and therefore $v_{0}\mapsto (\bar{P}%
^{0})^{\prime }v_{0}-W(v_{0})$ is strictly concave there, contradicting that
both endpoints are maximisers. Thus $P^{0\ast }$ is one-to-one on $\mathcal{V%
}_{0,\mathrm{reg}}$. Equivalently, convex conjugacy and uniqueness of the
maximiser give 
\begin{equation}
\nabla Q_{\max }^{0\ast }(P^{0})=[P^{0\ast }]^{-1}(P^{0}),\qquad P^{0}\in 
\func{int}\Delta (\mathcal{J}).  \label{eq: conjugate inverse P0}
\end{equation}

Since $P^{0\ast }=\nabla _{P^{0}}Q_{\max }^{0}$ is continuous and $\func{int}%
\Delta (\mathcal{J})$ is open, $\mathcal{V}_{0,\mathrm{reg}}=(P^{0\ast
})^{-1}(\func{int}\Delta (\mathcal{J}))$ is non--empty and open.

Next, we characterise the open set. For any $v_{0}\in \mathcal{V}_{0,\mathrm{%
reg}}$, define 
\begin{equation*}
\nu _{0,j}=\frac{P_{j}^{0\ast }(v_{0})}{P_{0}^{0\ast }(v_{0})}\exp
(v_{0,j}/\lambda ),\qquad j=1,\ldots ,J.
\end{equation*}%
The conditional CCPs in \eqref{eq:DMCCP} then have denominator $%
1+\sum_{j=1}^{J}\nu _{0,j}\exp (V_{1,j}/\lambda )$ and the integrated CCPs
satisfy $P_{j}^{0\ast }(v_{0})=p_{j}^{0}(\nu _{0},\lambda )$, $j=0,\ldots ,J$%
. Solving the expression for $\nu _{0,j}$ yields \eqref{eq: V0reg exact}.
Conversely, for any $\nu _{0}\in (0,\infty )^{J}$, define $v_{0}$ by the
right-hand side of \eqref{eq: V0reg exact} and set $P^{0}=(p_{j}^{0}(\nu
_{0},\lambda ))_{j=1}^{J}$. Then $P^{0}\in \func{int}\Delta (\mathcal{J})$
and 
\begin{equation*}
\sigma (P^{0},v_{0},v_{1})=p_{0}^{0}(\nu _{0},\lambda )\left[
1+\sum_{j=1}^{J}\nu _{0,j}\exp (v_{1,j}/\lambda )\right] .
\end{equation*}%
Since $p_{0}^{0}(\nu _{0},\lambda )+\sum_{j=1}^{J}p_{j}^{0}(\nu _{0},\lambda
)=1$, the definitions of $p_{j}^{0}(\nu _{0},\lambda )$ imply 
\begin{equation*}
\int_{\mathcal{V}_{1}}\frac{1}{\sigma (P^{0},v_{0},v_{1})}d\pi
_{V_{1}}(v_{1})=1,\qquad \int_{\mathcal{V}_{1}}\frac{\exp
((v_{0,j}+v_{1,j})/\lambda )}{\sigma (P^{0},v_{0},v_{1})}d\pi
_{V_{1}}(v_{1})=1.
\end{equation*}%
Thus $\nabla _{P^{0}}Q^{0}(P^{0};v_{0},\lambda )=0$ and strict concavity
implies $P^{0\ast }(v_{0})=P^{0}$. This proves \eqref{eq: V0reg exact}. The
mapping on its right-hand side is $C^{1}$, since differentiation of $%
p_{j}^{0}(\nu _{0},\lambda )$ with respect to $\nu _{0}$ is justified
locally by bounded domination. It is bijective, with inverse 
\begin{equation*}
\nu _{0,j}=\frac{P_{j}^{0\ast }(v_{0})}{P_{0}^{0\ast }(v_{0})}\exp
(v_{0,j}/\lambda ),\qquad j=1,\ldots ,J.
\end{equation*}%
The inverse is $C^{1}$ by the implicit-function theorem applied to the
interior first-order conditions, using Lemma~\ref{lemma:Q0_concavity}. Thus
the parametrisation is a $C^{1}$ diffeomorphism. Since $(0,\infty )^{J}$ is
convex, its image under this diffeomorphism is path-connected and
contractible; in particular, it is connected.

To show (iii), let $v_{0}\in \mathcal{V}_{0,\mathrm{reg}}$ and evaluate all
derivatives below at $(P^{0\ast }(v_{0}),v_{0})$. The interior first-order
condition is $\nabla _{P^{0}}Q^{0\ast }:=\nabla _{P^{0}}Q^{0}(P^{0\ast
}(v_{0});v_{0})=0$. By Lemma~\ref{lemma:Q0_concavity}, $\nabla
_{P^{0}P^{0}}^{2}Q^{0\ast }\prec 0$. The implicit-function theorem therefore
gives%
\begin{equation*}
\nabla _{v_{0}}P^{0\ast }(v_{0})=-\left[ \nabla _{P^{0}P^{0}}^{2}Q^{0\ast }%
\right] ^{-1}\nabla _{P^{0}v_{0}}^{2}Q^{0\ast }.
\end{equation*}%
Differentiating the envelope identity \eqref{eq: gradient W} gives 
\begin{equation*}
\nabla _{v_{0}}P^{0\ast }(v_{0})=\nabla _{v_{0}v_{0}}^{2}Q_{\max
}^{0}(v_{0})=\nabla _{v_{0}v_{0}}^{2}Q^{0\ast }-\nabla
_{v_{0}P^{0}}^{2}Q^{0\ast }\left[ \nabla _{P^{0}P^{0}}^{2}Q^{0\ast }\right]
^{-1}\nabla _{P^{0}v_{0}}^{2}Q^{0\ast }.
\end{equation*}%
Since $\nabla _{P^{0}P^{0}}^{2}Q^{0\ast }\prec 0$, $\nabla
_{v_{0}P^{0}}^{2}Q^{0\ast }\left[ \nabla _{P^{0}P^{0}}^{2}Q^{0\ast }\right]
^{-1}\nabla _{P^{0}v_{0}}^{2}Q^{0\ast }\preccurlyeq 0$. Moreover, since $%
P^{0\ast }(v_{0})$ is interior, the Hessian calculation above gives $\nabla
_{v_{0}v_{0}}^{2}Q^{0\ast }\succ 0$. It follows that $\nabla
_{v_{0}}P^{0\ast }(v_{0})=\nabla _{v_{0}v_{0}}^{2}Q_{\max
}^{0}(v_{0},\lambda )\succ 0$. Together with bijectivity, the
inverse-function theorem implies that $P^{0\ast }:\mathcal{V}_{0,\mathrm{reg}%
}\mapsto \func{int}\Delta (\mathcal{J})$ is a monotonic $C^{1}$
diffeomorphism, with inverse given by \eqref{eq: conjugate
inverse P0}.

For part (iv), observe that the interior first-order condition $\nabla
_{P^{0}}Q^{0\ast }=0$ is continuously differentiable in $\lambda $.
Differentiation under the integral with respect to $\lambda $ follows from $%
E_{\pi _{V_{1}}}\Vert V_{1}\Vert <\infty $, locally around each interior
optimum and $\lambda >0$. Since $\nabla _{P^{0}P^{0}}^{2}Q^{0\ast }\prec 0$,
the implicit-function theorem gives \eqref{eq: P0 lambda derivative}.
\end{proof}

\bigskip

\begin{proof}[Proof of Theorem~\protect\ref{Thm: P one-to-one}]
For fixed $(v_{0},\lambda )$, Lemma \ref{lemma:DMCCP} expresses $P^{\ast }$
as a multinomial-logit function of $v_{1}$, with weights $P^{0\ast
}(v_{0},\lambda )$. Differentiation gives \eqref{eq: P v1 Jacobian}. For any 
$a\in \mathbb{R}^{J}$, 
\begin{equation*}
a^{\prime }\Sigma (P^{\ast })a=\sum_{j=1}^{J}P_{j}^{\ast }a_{j}^{2}-\left(
\sum_{j=1}^{J}P_{j}^{\ast }a_{j}\right) ^{2}\geq 0.
\end{equation*}%
This is the variance of a random variable taking values $0,a_{1},\ldots
,a_{J}$ with respective probabilities $P_{0}^{\ast },\ldots ,P_{J}^{\ast }$.

When $v_{0}\in \mathcal{V}_{0,\mathrm{reg}}(\lambda )$, all $J+1$
probabilities are positive. The variance is then strictly positive for every 
$a\neq 0$. This proves positive definiteness; integration along line
segments gives injectivity. Formula \eqref{eq: P v1 cross derivative}
follows directly from \eqref{eq: P v1 Jacobian}, completing (i).

For (ii) the conditional log odds implied by \eqref{eq:DMCCP} satisfy 
\begin{equation}
\log \frac{P_{j}^{\ast }(v_{0},v_{1},\lambda )}{P_{0}^{\ast
}(v_{0},v_{1},\lambda )}=\frac{v_{0,j}+v_{1,j}}{\lambda }+\log \frac{%
P_{j}^{0\ast }(v_{0},\lambda )}{P_{0}^{0\ast }(v_{0},\lambda )},\qquad
j=1,\ldots ,J.  \label{eq: conditional log odds v0}
\end{equation}%
Differentiating these log odds with respect to $v_{0}$ and using the
multinomial-logit Jacobian gives \eqref{eq: P v0 Jacobian}. On $\mathcal{V}%
_{0,\mathrm{reg}}(\lambda )$, both $\Sigma (P^{\ast })$ and $\Sigma
(P^{0\ast })$ are positive definite. Moreover, $\nabla _{v_{0}}P^{0\ast }$
is symmetric positive semidefinite by Theorem \ref{Thm: P0 one-to-one}. Thus 
$\Sigma (P^{0\ast })^{-1}\nabla _{v_{0}}P^{0\ast }$ is similar to a
symmetric positive semidefinite matrix. Consequently the bracket in 
\eqref{eq: P v0
Jacobian} is nonsingular and the Jacobian has rank $J$. The product in %
\eqref{eq: P v0 Jacobian} need not be symmetric.

To show (ii), let $v_{1}\in \mathbb{R}^{J}$ be given and suppose that $%
P^{\ast }(v_{1}|v_{0})=P^{\ast }(v_{1}|\tilde{v}_{0})$ for some $v_{0},%
\tilde{v}_{0}\in \mathcal{V}_{0,\mathrm{reg}}$. Subtracting 
\eqref{eq: conditional log odds
v0} evaluated at $v_{0}$ from the same expression evaluated at $\tilde{v}%
_{0} $ yields 
\begin{equation}
\frac{v_{0,j}-\tilde{v}_{0,j}}{\lambda }+\log \left( \frac{P_{j}^{0\ast
}(v_{0})}{P_{0}^{0\ast }(v_{0})}\right) -\log \left( \frac{P_{j}^{0\ast }(%
\tilde{v}_{0})}{P_{0}^{0\ast }(\widetilde{v}_{0})}\right) =0.
\label{eq: conditional equal log odds v0}
\end{equation}%
Multiplying \eqref{eq: conditional equal log odds v0} by $\Delta
P_{j}^{0\ast }=P_{j}^{0\ast }(v_{0})-P_{j}^{0\ast }(\tilde{v}_{0})$ and
summing over $j=1,\ldots ,J$ gives 
\begin{equation}
0=\frac{1}{\lambda }\left\{ P^{0\ast }(v_{0})-P^{0\ast }(\tilde{v}%
_{0})\right\} ^{\prime }(v_{0}-\tilde{v}_{0})+\sum_{j=0}^{J}\Delta
P_{j}^{0\ast }\left\{ \log P_{j}^{0\ast }(v_{0})-\ln P_{j}^{0\ast }(\tilde{v}%
_{0})\right\}  \label{eq: conditional monotonicity v0}
\end{equation}%
By Theorem \ref{Thm: P0 one-to-one}, $P^{0\ast }=\nabla _{v_{0}}Q_{\max
}^{0} $, and convexity of $Q_{\max }^{0}$ implies that the first term in %
\eqref{eq: conditional monotonicity v0} is nonnegative. Since $v_{0},\tilde{v%
}_{0}\in \mathcal{V}_{0,\mathrm{reg}}$, all probabilities in the second term
are strictly positive. As $\log (\cdot )$ is strictly increasing, this term
is also nonnegative, and it is strictly positive whenever $P^{0\ast
}(v_{0})\neq P^{0\ast }(\tilde{v}_{0})$. If $v_{0}\neq \tilde{v}_{0}$,
injectivity of $P^{0\ast }$ from Theorem \ref{Thm: P0 one-to-one} implies
precisely this inequality, contradicting 
\eqref{eq: conditional monotonicity
v0}. Hence $v_{0}=\tilde{v}_{0}$, and the map $v_{0}\mapsto P^{\ast
}(v_{1}|v_{0})$ is one-to-one on $\mathcal{V}_{0,\mathrm{reg}}$.

For (iii), taking traces in \eqref{eq: P v1 Jacobian} gives 
\eqref{eq: P
stacked lambda identification}, whose denominator is strictly positive by
(i). Subtracting \eqref{eq: conditional log odds v0} at $v_{1}$ and $\tilde{v%
}_{1}$ gives \eqref{eq: P lambda log odds difference}.
\end{proof}

\section{Proofs of Section~\protect\ref{sec: ID composite injectivity} \label%
{sec: Proofs ID composite}}

\setcounter{equation}{0} \renewcommand{\theequation}{B.\arabic{equation}}

\begin{proof}[Proof of Theorem~\protect\ref{Th: Pi ID 1}]
The proof proceeds as in \citet{Fosgerau2020}. Let $z\in \mathcal{Z}_{0}$ be
given. For $w_{2},\tilde{w}_{2}\in \mathcal{W}_{0,2}(z)$, suppose that 
\begin{equation*}
\left\{ \mathcal{V}_{0,0}(w_{2},z)+\nu_{2}(w_{2})\right\} \cap \left\{ 
\mathcal{V}_{0,0}(\tilde{w}_{2},z)+\nu_{2}(\tilde{w}_{2})\right\} \neq
\varnothing .
\end{equation*}%
Then there exist $w_{1}\in \mathcal{W}_{0,1}(w_{2},z)$ and $\tilde{w}_{1}\in 
\mathcal{W}_{0,1}(\tilde{w}_{2},z)$ such that 
\begin{equation*}
\nu_{0}(w_{1},w_{2})+\nu_{2}(w_{2})=\nu_{0}(\tilde{w}_{1},\tilde{w}%
_{2})+\nu_{2}(\tilde{w}_{2})\in \mathcal{V}_{0}(z).
\end{equation*}%
Equivalently, by injectivity of $\Psi (\cdot ,z)$ on $\mathcal{V}_{0}(z)$,
these pairs are characterized by $\Phi (w_{1},w_{2},z)=\Phi (\tilde{w}_{1},%
\tilde{w}_{2},z)$. Since $\nu_{0}$ is known, such a match identifies 
\begin{equation}
\nu_{2}(w_{2})-\nu_{2}(\tilde{w}_{2})=\nu_{0}(\tilde{w}_{1},\tilde{w}%
_{2})-\nu_{0}(w_{1},w_{2}).  \label{eq: u2 pair difference}
\end{equation}%
Thus, whenever the two displayed local index regions overlap, the difference
between the corresponding values of $\nu_{2}$ is identified.

Now consider arbitrary $w_{2},\tilde{w}_{2}\in \mathcal{W}_{0,2}(z)$. Since
the sets $\mathcal{V}_{0,0}(w_{2}^{\prime },z)+\nu_{2}(w_{2}^{\prime })$, $%
w_{2}^{\prime }\in \mathcal{W}_{0,2}(z)$, are open and cover the connected
set $\mathcal{V}_{0}(z)$, there exists a finite sequence $%
w_{2}^{(0)}=w_{2},\ w_{2}^{(1)},\ldots ,\ w_{2}^{(M)}=\tilde{w}_{2}$ such
that, for each $m=0,\ldots ,M-1$, the local index regions associated with $%
w_{2}^{(m)}$ and $w_{2}^{(m+1)}$ overlap. Applying 
\eqref{eq: u2 pair
difference} to each consecutive pair identifies $\nu_{2}(w_{2}^{(m)})-%
\nu_{2}(w_{2}^{(m+1)})$, $m=0,\ldots ,M-1$. Therefore 
\begin{equation*}
\nu_{2}(w_{2})-\nu_{2}(\tilde{w}_{2})=\sum_{m=0}^{M-1}\left\{
\nu_{2}(w_{2}^{(m)})-\nu_{2}(w_{2}^{(m+1)})\right\}
\end{equation*}%
is identified. Hence, for each $z$, $\nu_{2}$ is identified on $\mathcal{W}%
_{0,2}(z)$ up to one additive constant. More explicitly, choose an arbitrary 
$w_{2,z}\in \mathcal{W}_{0,2}(z)$ and define the identified function $%
\nu_{0,z}(w_{2}):=\nu_{2}(w_{2})-\nu_{2}(w_{2,z})$. Then 
\begin{equation}
\nu_{2}(w_{2})=\nu_{0,z}(w_{2})+b(z)  \label{eq: u2 within z rep}
\end{equation}%
where only $b(z)$ remains unidentified.

We next identify these constants relative to one another. Suppose that $%
z_{1},z_{2}\in \mathcal{Z}_{0}$ satisfy the overlap condition in part (ii),
and let $w_{2}\in \mathcal{W}_{0,2}(z_{1})\cap \mathcal{W}_{0,2}(z_{2})$ be
the common value supplied by that condition. Since $\nu_{2}$ does not depend
on $z$, equation \eqref{eq: u2 within z rep} gives $%
\nu_{0,z_{1}}(w_{2})+b(z_{1})=\nu_{0,z_{2}}(w_{2})+b(z_{2})$. Consequently, 
\begin{equation}
b(z_{1})-b(z_{2})=\nu_{0,z_{2}}(w_{2})-\nu_{0,z_{1}}(w_{2}),
\label{eq: b z difference}
\end{equation}%
and the right-hand side is identified.

It remains to show that these pairwise differences determine all $b(z)$ up
to one common additive constant. Starting from $z_{0}$ in the normalisation,
let $\mathcal{Z}^{\ast }$ denote the set of all $z\in \mathcal{Z}_{0}$ that
can be reached from $z_{0}$ through a finite sequence $%
z_{0}=z^{(0)},z^{(1)},\ldots ,z^{(M)}=z$ such that every consecutive pair
satisfies the overlap condition in part (ii). If $\mathcal{Z}^{\ast }\neq 
\mathcal{Z}_{0}$, then $\mathcal{Z}^{\ast }$ and $\mathcal{Z}_{0}\setminus 
\mathcal{Z}^{\ast }$ form a partition of $\mathcal{Z}_{0}$ into two nonempty
sets. Part (ii) would then supply an overlapping pair with one element in
each set, contradicting the definition of $\mathcal{Z}^{\ast }$. Hence $%
\mathcal{Z}^{\ast }=\mathcal{Z}_{0}$.

For any $z\in \mathcal{Z}_{0}$, choose such a finite sequence. Applying %
\eqref{eq: b z difference} to its consecutive pairs identifies 
\begin{equation*}
b(z)-b(z_{0})=\sum_{m=0}^{M-1}\left\{ b(z^{(m+1)})-b(z^{(m)})\right\} .
\end{equation*}%
Thus all $b(z)$ are identified relative to $b(z_{0})$. Finally, $%
\nu_{2}(w_{0,2})=0$ and \eqref{eq: u2 within z rep} at $z=z_{0}$ imply $%
b(z_{0})=-\nu_{0,z_{0}}(w_{0,2})$, so $b(z_{0})$, and hence every $b(z)$, is
identified. It follows that $\nu_{2}$ is identified on $\bigcup_{z\in 
\mathcal{Z}_{0}}\mathcal{W}_{0,2}(z)$. Finally, for every $%
\nu=\nu_{0}(w_{1},w_{2})+\nu_{2}(w_{2})\in \mathcal{V}_{0}(z)$, $\Psi
(\nu,z)=\Phi (w_{1},w_{2},z)$, which identifies $\Psi (\cdot ,z)$ on $%
\mathcal{V}_{0}(z)$.
\end{proof}

\bigskip

\begin{proof}[Proof of Theorem~\protect\ref{Th: Pi ID joint rank}]
Let $(\tilde{\nu},\tilde{\Psi})$ be an observationally equivalent
representation. For $\nu \in \nu (\mathcal{W}^{\ast }\left( z_{0}\right) )$,
choose $w\in \mathcal{W}^{\ast }\left( z_{0}\right) $ with $\nu (w)=\nu $
and put $h(\nu )=\tilde{\nu}(w)$. By observational equivalence and
injectivity of $\tilde{\Psi}(\cdot ,z_{0})$ and $\Psi (\cdot ,z_{0})$, $h$
is well defined and injective. The reference-support condition gives $\Psi
(\nu ,z)=\tilde{\Psi}(h(\nu ),z)$, $z\in \mathcal{Z}^{\ast }$. The reference
images for both representations are open. By invariance of domain and the
injectivity of the outer mappings, $h$ and $h^{-1}$ are continuous. At each $%
\nu $, choose $z\in\mathcal{Z}^{\ast }$ such that $\nabla _{\tilde{\nu}}%
\tilde{\Psi}(h(\nu),z)$ is nonsingular. The inverse function theorem and $%
\Psi(\nu,z)=\tilde{\Psi}(h(\nu),z)$ give locally $h=\tilde{\Psi}%
(\cdot,z)^{-1}\circ\Psi(\cdot,z)$. Thus $h$ is continuously differentiable.
Applying the same argument to $h^{-1}$ with a possibly different $z\in%
\mathcal{Z}^{\ast }$ gives nonsingularity of $Dh(\nu)$. Differentiating $%
\Psi (\nu ,z)=\tilde{\Psi}(h(\nu ),z)$ gives $\nabla _{\nu }\Psi (\nu
,z)[Dh(\nu )]^{-1}=\nabla _{\nu }\tilde{\Psi}(h(\nu ),z)$, $z\in \mathcal{Z}%
^{\ast }$, where the right-hand side is symmetric. The matrices in %
\eqref{eq: Pi joint rank sufficient condition} are symmetric and linearly
independent. Since $\dim\mathbb{S}^{J}=J(J+1)/2$, they form a basis of $%
\mathbb{S}^{J}$. Put $H=[Dh(\nu )]^{-1}$. Then $\nabla _{\nu }\Psi (\nu
,z)H\in\mathbb{S}^{J}$ for every $z\in\mathcal{Z}^{\ast }$, and hence $AH\in%
\mathbb{S}^{J}$ for every $A\in\mathbb{S}^{J}$. Taking $A=I_{J}$ yields $%
H=H^{\prime }$, so $AH=HA$ for every $A\in\mathbb{S}^{J}$. Taking $%
A=e_{j}e_{j}^{\prime }$, $j=1,\ldots,J$, gives $H=\func{diag}%
(h_{1},\ldots,h_{J})$. Taking $A=e_{j}e_{k}^{\prime }+e_{k}e_{j}^{\prime }$, 
$j\neq k$, yields $h_{j}=h_{k}$. Thus $H=hI_{J}$, and consequently $Dh(\nu
)=a(\nu )I_{J}$ with $a(\nu )\neq 0$. Since $J\geq 2$ and $\nu (\mathcal{W}%
^{\ast }\left( z_{0}\right) )$ is open and connected, $a(\nu )$ is constant
and $h(\nu )=a\nu +b$ on $\nu (\mathcal{W}^{\ast }\left( z_{0}\right) )$.
Consequently, $\tilde{\nu}(w)=a\nu (w)+b$, $w\in \mathcal{W}^{\ast }(z_{0})$%
. The normalisations \eqref{eq: Pi joint rank normalizations} imply $a=1$
and $b=0$.

Since Assumption \ref{Ass: Pi reference support} holds for both
representations and $\tilde{\nu}=\nu$ on $\mathcal{W}^{\ast }(z_{0})$, its
coverage condition gives, for each $z\in\mathcal{Z}_{0}$, 
\begin{equation*}
\tilde{\mathcal{V}}_{0}(z)=\tilde{\nu}(\mathcal{W}^{\ast }(z_{0})\cap 
\mathcal{W}(z))=\nu(\mathcal{W}^{\ast }(z_{0})\cap\mathcal{W}(z)) =\mathcal{V%
}_{0}(z).
\end{equation*}

For any $w\in \mathcal{W}_{0}(z)$, Assumption \ref{Ass: Pi reference support}
supplies $w^{\ast }\in \mathcal{W}^{\ast }(z_{0})\cap\mathcal{W}(z)$ with $%
\nu(w)=\nu(w^{\ast })$. Since $\nu(w)\in\mathcal{V}_{0}(z)$, $w^{\ast }\in%
\mathcal{W}_{0}(z)$ by \eqref{eq: U_0 def}. Observational equivalence and
injectivity of $\tilde{\Psi}(\cdot ,z)$ give $\tilde{\nu}(w)=\tilde{\nu}%
(w^{\ast })=\nu (w^{\ast })=\nu (w)$. Identification of $\Psi (\cdot ,z)$ on 
$\mathcal{V}_{0}(z)$ follows.
\end{proof}

\bigskip

\begin{proof}[Proof of Theorem~\protect\ref{Th: Pi ID 2}]
Let $(\tilde{\nu},\tilde{\Psi})$ be an observationally equivalent
representation. Injectivity of $\tilde{\Psi}(\cdot,z_0)$ implies that, for $%
w,\bar w\in\mathcal{W}^*$, $\nu(w)=\nu(\bar w)$ entails $\tilde{\nu}(w)=%
\tilde{\nu}(\bar w)$. Since $\nu_{0j}$ is known, 
\begin{equation}
\Delta_j(w_{1,j},w_2):=\tilde{\nu}_j(w_{1,j},w_2)-\nu_j(w_{1,j},w_2)
=a_j(w_{1,j})+b_j(w_2),\qquad j=1,\ldots,J,
\label{eq: ASV known interaction difference}
\end{equation}%
where $a_j=\tilde{\nu}_{1j}-\nu_{1j}$ and $b_j=\tilde{\nu}_{2j}-\nu_{2j}$.
The normalisations and the known interaction imply, for every $w_2\in%
\mathcal{W}_2^*$, 
\begin{equation}
\nabla_{w_{1,j}}\tilde{\nu}_j(w_{0,1,j},w_2)
=\nabla_{w_{1,j}}\nu_j(w_{0,1,j},w_2)
=1+\nabla_{w_{1,j}}\nu_{0j}(w_{0,1,j},w_2)
-\nabla_{w_{1,j}}\nu_{0j}(w_{0,1,j},w_{0,2})\neq0.
\label{eq: ASV reference derivative equality}
\end{equation}

Fix $\hat{w}_{2}\in \mathcal{W}_{2}^{\ast }$. By 
\eqref{eq: ASV reference
derivative equality}, for each $k$ there exist two points of $\mathcal{I}%
_{k} $ on opposite sides of $w_{0,1,k}$ whose index values at $\hat{w}_{2}$
strictly bracket $\nu _{k}(w_{0,1,k},\hat{w}_{2})$. Continuity therefore
gives a connected open neighbourhood $\mathcal{W}_{2,\mathrm{loc}}$ of $\hat{%
w}_{2}$ and open intervals $\mathcal{V}_{k,\mathrm{loc}}$ such that 
\begin{equation*}
\nu _{k}(w_{0,1,k},w_{2})\in \mathcal{V}_{k,\mathrm{loc}}\subseteq \nu _{k}(%
\mathcal{I}_{k},w_{2}),\qquad w_{2}\in \mathcal{W}_{2,\mathrm{loc}},\quad
k=1,\ldots ,J.
\end{equation*}%
For $w_{2},\bar{w}_{2}\in \mathcal{W}_{2,\mathrm{loc}}$ and any $v_{k}\in 
\mathcal{V}_{k,\mathrm{loc}}$, choose $w_{1,k},\bar{w}_{1,k}\in \mathcal{I}%
_{k}$ such that $\nu _{k}(w_{1,k},w_{2})=\nu _{k}(\bar{w}_{1,k},\bar{w}%
_{2})=v_{k}$ for all $k$. By injectivity, $\tilde{\nu}_{k}(w_{1,k},w_{2})=%
\tilde{\nu}_{k}(\bar{w}_{1,k},\bar{w}_{2})$, so for each $j$ there is a
well-defined $h_{j}$ on $\mathcal{V}_{j,\mathrm{loc}}$ satisfying $\tilde{\nu%
}_{j}=h_{j}\circ \nu _{j}$ there. Taking difference quotients at $w_{0,1,j}$%
, using the nonzero derivatives in 
\eqref{eq: ASV reference derivative
equality} and continuity to attain all nearby index values, gives 
\begin{equation*}
h_{j}^{\prime }(\nu _{j}(w_{0,1,j},w_{2}))=\frac{\nabla _{w_{1,j}}\tilde{\nu}%
_{j}(w_{0,1,j},w_{2})}{\nabla _{w_{1,j}}\nu _{j}(w_{0,1,j},w_{2})}=1,\qquad
w_{2}\in \mathcal{W}_{2,\mathrm{loc}}.
\end{equation*}%
Since $\nu _{j}(w_{0,1,j},\mathcal{W}_{2,\mathrm{loc}})$ is an interval, $%
\Delta _{j}(w_{0,1,j},w_{2})$ is constant on $\mathcal{W}_{2,\mathrm{loc}}$.
Connectedness of $\mathcal{W}_{2}^{\ast }$ and the normalisation at $w_{0}$
yield $\Delta _{j}(w_{0,1,j},w_{2})=0$, $w_{2}\in \mathcal{W}_{2}^{\ast }$.
Hence \eqref{eq: ASV known interaction difference} becomes 
\begin{equation}
\Delta _{j}(w_{1,j},w_{2})=a_{j}(w_{1,j})-a_{j}(w_{0,1,j}),\qquad
(w_{1,j},w_{2})\in \mathcal{I}_{j}\times \mathcal{W}_{2}^{\ast }.
\label{eq: ASV known interaction reduced difference}
\end{equation}

Fix $w_{1,j}\in \mathcal{I}_{j}$. By nonconstancy and connectedness, $%
w_{2}\mapsto \nu _{j}(w_{1,j},w_{2})$ is not locally constant at some $\hat{w%
}_{2}\in \mathcal{W}_{2}^{\ast }$. Choose an open ball $\mathcal{W}_{2,%
\mathrm{loc}}$ about $\hat{w}_{2}$ such that the index images $\nu _{k}(%
\mathcal{I}_{k},w_{2})$, $k\neq j$, contain common nonempty open intervals
for all $w_{2}\in \mathcal{W}_{2,\mathrm{loc}}$; this follows from %
\eqref{eq: ASV reference derivative equality} as above. Choose $%
w_{2}^{\prime },w_{2}^{\prime \prime }\in \mathcal{W}_{2,\mathrm{loc}}$ with 
$\nu _{j}(w_{1,j},w_{2}^{\prime })<\nu _{j}(w_{1,j},w_{2}^{\prime \prime })$
and set $v=\tfrac{1}{2}[\nu _{j}(w_{1,j},w_{2}^{\prime })+\nu
_{j}(w_{1,j},w_{2}^{\prime \prime })]$. For every $\bar{w}_{1,j}$
sufficiently close to $w_{1,j}$, continuity gives $\nu _{j}(\bar{w}%
_{1,j},w_{2}^{\prime })<v<\nu _{j}(\bar{w}_{1,j},w_{2}^{\prime \prime })$.
The intermediate value theorem on the ball therefore supplies $w_{2},\bar{w}%
_{2}\in \mathcal{W}_{2,\mathrm{loc}}$ with $\nu _{j}(w_{1,j},w_{2})=v=\nu
_{j}(\bar{w}_{1,j},\bar{w}_{2})$. Matching the remaining coordinates in
their common index intervals gives two points of $\mathcal{W}^{\ast }\left(
z_{0}\right) $ with identical full index vectors. Injectivity and 
\eqref{eq: ASV known
interaction reduced difference} therefore give $a_{j}(w_{1,j})=a_{j}(\bar{w}%
_{1,j})$ and so $a_{j}$ is locally constant on the connected interval $%
\mathcal{I}_{j}$, therefore constant. It follows that $\tilde{\nu}=\nu $ on $%
\mathcal{W}^{\ast }(z_{0})$.

Since Assumption \ref{Ass: Pi reference support} holds for both
representations and $\tilde{\nu}=\nu$ on $\mathcal{W}^{\ast }(z_{0})$, its
coverage condition gives, for each $z\in\mathcal{Z}_{0}$, 
\begin{equation*}
\tilde{\mathcal{V}}_{0}(z)=\tilde{\nu}(\mathcal{W}^{\ast }(z_{0})\cap 
\mathcal{W}(z))=\nu(\mathcal{W}^{\ast }(z_{0})\cap\mathcal{W}(z)) =\mathcal{V%
}_{0}(z).
\end{equation*}

For $w\in \mathcal{W}_{0}(z)$, Assumption \ref{Ass: Pi reference support}
supplies $\bar{w}\in\mathcal{W}^{\ast }(z_{0})\cap\mathcal{W}(z)$ with $%
\nu(w)=\nu(\bar{w})$. Hence $\bar{w}\in\mathcal{W}_{0}(z)$ by 
\eqref{eq: U_0
def}. Injectivity of $\tilde{\Psi}(\cdot ,z)$ implies $\tilde{\nu}(w)=\tilde{%
\nu}(\bar{w})=\nu (w)$. Identification of $\Psi (\cdot ,z)$ on $\mathcal{V}%
_{0}(z)$ follows.
\end{proof}

\bigskip

\bigskip

\begin{proof}[Proof of Theorem~\protect\ref{Th: Pi ID 3}]
Let $(\tilde{\nu},\tilde{\Psi})$ be an observationally equivalent
representation. For any given $w_{2}\in \mathcal{W}_{2}^{\ast }$ and for
every edge $\{j,k\}$ of $\mathcal{G}_{\Psi }(w_{2})$, the chain rule and
symmetry of both outer Jacobians imply 
\begin{equation}
\nabla _{w_{1,j}}\nu _{j}(w_{1,j},w_{2})\nabla _{w_{1,k}}\tilde{\nu}%
_{k}(w_{1,k},w_{2})=\nabla _{w_{1,k}}\nu _{k}(w_{1,k},w_{2})\nabla _{w_{1,j}}%
\tilde{\nu}_{j}(w_{1,j},w_{2}),  \label{eq: Pi3 matching ratio}
\end{equation}%
for $(w_{1,j},w_{1,k})\in \mathcal{I}_{j}\times \mathcal{I}_{k}$. Since $\nu
_{j}(\cdot ,w_{2})$ is nonconstant and continuously differentiable, choose $%
w_{1,j}\in \mathcal{I}_{j}$ with $\nabla _{w_{1,j}}\nu
_{j}(w_{1,j},w_{2})\neq 0$ for each $j$. Equation 
\eqref{eq:
Pi3 matching ratio} and graph connectedness yield a common $a(w_{2})$ such
that $\nabla _{w_{1,j}}\tilde{\nu}_{j}(w_{1,j},w_{2})=a(w_{2})\nabla
_{w_{1,j}}\nu _{j}(w_{1,j},w_{2})$ for all $j$ and $w_{1,j}\in \mathcal{I}%
_{j}$. Integrating gives 
\begin{equation}
\tilde{\nu}(w_{1},w_{2})=a(w_{2})\nu (w_{1},w_{2})+b(w_{2}),\qquad w_{1}\in
\prod_{j=1}^{J}\mathcal{I}_{j}.  \label{eq: Pi3 matching affine local}
\end{equation}%
Injectivity implies $a(w_{2})\neq 0$. For each $w_{2}\in \mathcal{W}%
_{2}^{\ast }$, the image $\nu (\prod_{j}\mathcal{I}_{j},w_{2})$ contains a
nonempty open rectangle, since each $\nu _{j}(\cdot ,w_{2})$ is continuous
and nonconstant. By joint continuity of $\nu $ on $\mathcal{W}^{\ast }(z_{0})
$, there is a neighbourhood $\mathcal{N}(w_{2})\subseteq \mathcal{W}%
_{2}^{\ast }$ such that $\nu \Bigl(\prod_{j}\mathcal{I}_{j},w_{2}\Bigr)\cap
\nu \Bigl(\prod_{j}\mathcal{I}_{j},\bar{w}_{2}\Bigr)$ contains two distinct
points for every $\bar{w}_{2}\in \mathcal{N}(w_{2})$. Observational
equivalence, injectivity of $\tilde{\Psi}(\cdot ,z_{0})$, and \eqref{eq: Pi3
matching affine local} at these two index values yield $a(w_{2})=a(\bar{w}%
_{2})$ and $b(w_{2})=b(\bar{w}_{2})$. Since $\mathcal{W}_{2}^{\ast }$ is
connected, $a(w_{2})=a$ and $b(w_{2})=b$. The normalisations \eqref{eq: Pi3
matching normalization} give $a=1$ and $b=0$, hence $\tilde{\nu}=\nu $ on $%
\mathcal{W}^{\ast }(z_{0})$. Since Assumption \ref{Ass: Pi reference support}
holds for both representations and $\tilde{\nu}=\nu $ on $\mathcal{W}^{\ast
}(z_{0})$, its coverage condition gives, for each $z\in \mathcal{Z}_{0}$, 
\begin{equation*}
\tilde{\mathcal{V}}_{0}(z)=\tilde{\nu}(\mathcal{W}^{\ast }(z_{0})\cap 
\mathcal{W}(z))=\nu (\mathcal{W}^{\ast }(z_{0})\cap \mathcal{W}(z))=\mathcal{%
V}_{0}(z).
\end{equation*}

For $w\in\mathcal{W}_{0}(z)$, Assumption \ref{Ass: Pi reference support}
supplies $w^{\ast }\in\mathcal{W}^{\ast }(z_{0})\cap\mathcal{W}(z)$ with $%
\nu(w)=\nu(w^{\ast })$. Hence $w^{\ast }\in\mathcal{W}_{0}(z)$ by 
\eqref{eq:
U_0 def}. Thus injectivity of $\tilde{\Psi}(\cdot ,z)$ gives $\tilde{\nu}(w)=%
\tilde{\nu}(w^{\ast })=\nu (w)$. Identification of $\Psi (\cdot ,z)$ on $%
\mathcal{V}_{0}(z)$ follows.
\end{proof}

\section{Proofs of Section~\protect\ref{Sec: ID RI-ARUM} \label{sec: Proofs
ID RI-ARUM}}

\setcounter{equation}{0} \renewcommand{\theequation}{C.\arabic{equation}}

\begin{proof}[Proof of Theorem~\protect\ref{Th: RI ARUM u1 lambda eps = 0}]
Apply Theorem \ref{Th: Pi ID 3} with $w=x_{1}$, $z=(x_{0},x_{2})$, $%
\nu=u_{1} $ and $\Psi(v_{1},z)=P^{\ast}(u_{0}(x_{0}),v_{1}, \lambda(x_{2}))$%
. By Theorem \ref{Thm: P one-to-one}, for $z\in\mathcal{Z}_{0}$, $%
\Psi(\cdot,z)$ is injective, with 
\begin{equation*}
\nabla_{v_{1}}\Psi(v_{1},z)=\frac{1}{\lambda(x_{2})}\Sigma(P^{\ast})
\succ0,\qquad \nabla_{v_{1,k}}\Psi_{j}(v_{1},z) =-\frac{P_{j}^{\ast}P_{k}^{%
\ast}}{\lambda(x_{2})}<0,\quad j\neq k.
\end{equation*}
Thus the interaction graph is complete on $\mathcal{X}_{1}^{\ast}$.
Assumption \ref{Ass: RI ARUM u1 lambda support} verifies the reference
support and coverage conditions of Assumption \ref{Ass: Pi reference support}%
. The alternative-specific restriction and nonconstancy conditions verify
the remaining conditions of Theorem \ref{Th: Pi ID 3}. The normalisations
therefore identify $u_{1}$ on $\mathcal{X}_{1,\mathrm{reg}}^{0}$.

For $x_{2}\in \func{proj}_{x_{2}}\mathcal{Z}_{0}^{0}$, choose $x_{0}$ such
that $(x_{0},x_{2})\in \mathcal{Z}_{0}^{0}$ and $x_{1},x_{1}^{\prime }\in 
\mathcal{X}_{1}(x_{0},x_{2})$ with $u_{1,j}(x_{1})\neq u_{1,j}(x_{1}^{\prime
})$ for some $j$. Equation \eqref{eq: P lambda log odds difference} gives 
\begin{equation*}
\lambda (x_{2})=\frac{u_{1,j}(x_{1})-u_{1,j}(x_{1}^{\prime })}{\log \frac{%
\Phi _{j}(x_{0},x_{1},x_{2})}{\Phi _{0}(x_{0},x_{1},x_{2})}-\log \frac{\Phi
_{j}(x_{0},x_{1}^{\prime },x_{2})}{\Phi _{0}(x_{0},x_{1}^{\prime },x_{2})}},
\end{equation*}%
which is identified.
\end{proof}

\bigskip

\begin{proof}[Proof of Theorem~\protect\ref{Th: RI ARUM u0 eps=0 pi unknown}]

With $w=x_{0}$ and $z=(x_{1},x_{2})$, Theorem \ref{Thm: P one-to-one}(ii)
implies that the outer map $v_{0}\mapsto \Psi (v_{0},z)=P^{\ast
}(v_{0},u_{1}(x_{1}),\lambda (x_{2}))$ is injective on $\mathcal{V}_{0,%
\mathrm{reg}}(\lambda (x_{2}))$, and hence on $\mathcal{U}_{0}(x_{1},x_{2})$%
. The composite representation $\Phi (x_{0},x_{1},x_{2})=\Psi
(u_{0}(x_{0}),x_{1},x_{2})$ therefore satisfies Assumption \ref{Ass: Pi
injective}. Since the coverage condition is imposed with $\mathcal{X}%
_{0}(x_{1},x_{2})$ on its right-hand side, it verifies the matching-coverage
condition of Theorem \ref{Th: Pi ID 2}. With \eqref{eq: u0 known interaction}
and Assumption~\ref{Ass: RI ARUM u0 support}, that theorem applies and
yields the conclusions.
\end{proof}

\bigskip

\begin{proof}[Proof of Theorem~\protect\ref{Th: RI ARUM prior ID}]
Let $(x_{1,\pi },x_{2,\pi })$ and $\mathcal{X}_{0,\pi }$ be as in the
theorem, and set $\lambda =\lambda (x_{2,\pi })$. By Theorem \ref{Th: RI
ARUM u1 lambda eps = 0}, $u_{1}(x_{1,\pi })$ and $\lambda $ are identified
on $\mathcal{X}_{1,\mathrm{reg}}^{0}$ and $\func{proj}_{x_{2}}\mathcal{Z}%
_{0}^{0}$, respectively. By Theorem \ref{Th: RI ARUM u0 eps=0 pi unknown}, $%
u_{0}$ is identified on $\bigcup_{(x_{1},x_{2})\in \mathcal{Z}_{u_{0}}^{0}}%
\mathcal{X}_{0,\mathrm{reg}}^{0}(x_{1},x_{2})$, which contains $\mathcal{X}%
_{0,\pi }$. Since $\mathcal{X}_{0,\pi }\subseteq \mathcal{X}_{0,\mathrm{reg}%
}^{0}(x_{1,\pi },x_{2,\pi })$, the CCPs $\Phi (x_{0},x_{1,\pi },x_{2,\pi })$
are identified and $u_{0}(x_{0})\in \mathcal{V}_{0,\mathrm{reg}}(\lambda )$
for every $x_{0}\in \mathcal{X}_{0,\pi }$. Thus \eqref{eq: prior P0 odds}
identifies $P^{0\ast }(u_{0}(x_{0}),\lambda )$ for every $x_{0}\in \mathcal{X%
}_{0,\pi }$. The set $\mathcal{V}_{0,\pi }(\lambda ):=\func{int}u_{0}(%
\mathcal{X}_{0,\pi })$ is nonempty and open by assumption and is contained
in $\mathcal{V}_{0,\mathrm{reg}}(\lambda )$. Hence $P^{0\ast }(v_{0},\lambda
)$ is identified for every $v_{0}\in \mathcal{V}_{0,\pi }(\lambda )$.

Next, for any $v_{0}\in \mathcal{V}_{0,\pi }\left( \lambda \right) $, the
identity $P_{0}^{0\ast }(v_{0},\lambda )=\int P_{0}^{\ast
}(v_{0},v_{1},\lambda )d\pi _{V_{1}}\left( v_{1}\right) $ together with %
\eqref{eq:DMCCP} yield 
\begin{equation}
\int_{\mathbb{R}^{J}}\frac{1}{1+\sum_{j=1}^{J}a_{j}(v_{0},\lambda )\exp
(v_{1,j}/\lambda )}d\pi _{V_{1}}(v_{1})=P_{0}^{0\ast }(v_{0},\lambda )
\label{eq: prior Stieltjes}
\end{equation}%
where $a_{j}(v_{0},\lambda ):=[P_{j}^{0\ast }(v_{0},\lambda )/P_{0}^{0\ast
}(v_{0},\lambda )]\exp (v_{0,j}/\lambda )$ is identified. Moreover, 
\begin{equation*}
\nabla _{v_{0}}\log a(v_{0},\lambda )=\lambda ^{-1}I_{J}+\Sigma (P^{0\ast
})^{-1}\nabla _{v_{0}}P^{0\ast }.
\end{equation*}%
Since $\Sigma (P^{0\ast })\succ 0$ and $\nabla _{v_{0}}P^{0\ast }$ is
symmetric positive semidefinite, this matrix is nonsingular and so the
inverse function theorem implies that $a(\mathcal{V}_{0,\pi }\left( \lambda
\right) ,\lambda )$ contains a nonempty open subset of $(0,\infty )^{J}$.

We now show that there exists a unique solution $\pi _{V_{1}}$ to 
\eqref{eq:
prior Stieltjes}. Let $\pi _{V_{1}}$ and $\tilde{\pi}_{V_{1}}$ both satisfy 
\eqref{eq: prior
Stieltjes} and define $\Delta \pi _{V_{1}}=\pi _{V_{1}}-\tilde{\pi}_{V_{1}}$%
. With $\Delta \tilde{\pi}$ denoting the transformed version of $\Delta \pi
_{V_{1}}$ under the bijection $v_{1}\mapsto (\exp (v_{1,1}/\lambda ),\ldots
,\exp (v_{1,J}/\lambda ))$, we have $S(a(v_{0},\lambda ))=0$, where 
\begin{equation*}
S(a)=\int_{\mathbb{R}^{J}}\frac{1}{1+\sum_{j=1}^{J}a_{j}\exp
(v_{1,j}/\lambda )}d\Delta \pi _{V_{1}}(v_{1})=\int_{(0,\infty )^{J}}\frac{%
d\Delta \tilde{\pi}(t)}{1+a^{\prime }t}.
\end{equation*}%
on a nonempty open subset of $(0,\infty )^{J}$. For complex $a$ with $\Re
a_{j}>0$, the integrand is holomorphic in $a$ and bounded in modulus by one.
Dominated holomorphic integration thus shows that $S$ is holomorphic on this
product of half-planes. The identity theorem, applied successively to each
coordinate, implies $S(a)=0$ for all $a\in (0,\infty )^{J}$.

For each $a\in (0,\infty )^{J}$, the measure $\Delta \tilde{\pi}_{a}$
obtained under the transformation $t\mapsto a^{\prime }t$ consequently
satisfies $\int_{(0,\infty )}1/\left( 1+rs\right) d\Delta \tilde{\pi}%
_{a}(s)=0$, $r>0$. Uniqueness of the one-dimensional Stieltjes transform
gives $\Delta \tilde{\pi}_{a}=0$; see \citet{widder1941laplace}. Therefore,
for every $a\in (0,\infty )^{J}$, $\int \exp (-a^{\prime }t)d\Delta \tilde{%
\pi}(t)=0$. Uniqueness of the multivariate Laplace transform (by successive
application of one-dimensional uniqueness; see \citet{widder1941laplace})
implies $\Delta \tilde{\pi}=0$. The exponential map is bijective, so $\pi
_{V_{1}}=\tilde{\pi}_{V_{1}}$.
\end{proof}

\bigskip

\begin{proof}[Proof of Lemma~\protect\ref{Lem: Pi u1 injectivity}]
By Theorem \ref{Thm: P one-to-one}(i), for every $e\in\mathcal{E}$ the
mapping $u_1\mapsto P^\ast(u_0+e_0,u_1+e_1,\lambda+e_2)$ is continuously
differentiable, with Jacobian $(\lambda+e_2)^{-1}\Sigma(P^\ast)$. Every
entry is bounded in absolute value by $(\lambda+e_2)^{-1}\leq \lambda^{-1}$.
Dominated differentiation gives \eqref{eq: Pi u1 Jacobian}. For any sequence 
$u_1^{(n)}\to u_1$, the conditional Jacobians converge pointwise and satisfy
the same bound. Dominated convergence therefore shows that $\nabla_{u_1}\Pi$
is continuous, proving that $\Pi$ is continuously differentiable with
respect to $u_1$.

For every $a\in \mathbb{R}^{J}$, positive semidefiniteness gives 
\begin{equation*}
a^{\prime }\nabla _{u_{1}}\Pi (u_{0},u_{1},\lambda )a=0\Leftrightarrow
a^{\prime }\nabla _{v_{1}}P^{\ast }(u_{0}+e_{0},u_{1}+e_{1},\lambda
+e_{2})a=0\quad F_{\varepsilon }\text{-a.s.}
\end{equation*}%
Hence neither does the null space of $\nabla _{u_{1}}\Pi $. Under 
\eqref{eq:
Pi u1 regular condition}, the conditional Jacobian is positive definite on a
set of positive probability, so $\nabla _{u_{1}}\Pi \succ 0$. Theorem \ref%
{Thm: P one-to-one}(i) also gives $\nabla _{u_{1,k}}\Pi _{j}<0$ for $j\neq k$%
. For $u_{1}\neq \bar{u}_{1}$, 
\begin{equation*}
(u_{1}-\bar{u}_{1})^{\prime }[\Pi (u_{0},u_{1},\lambda )-\Pi (u_{0},\bar{u}%
_{1},\lambda )]=\int_{0}^{1}(u_{1}-\bar{u}_{1})^{\prime }\nabla _{u_{1}}\Pi
(u_{0},\bar{u}_{1}+t(u_{1}-\bar{u}_{1}),\lambda )(u_{1}-\bar{u}_{1})\,dt>0.
\end{equation*}%
Thus $\Pi (u_{0},\cdot ,\lambda )$ is injective.
\end{proof}

\bigskip

\begin{proof}[Proof of Theorem~\protect\ref{Th: RI ARUM u1 eps neq 0}]
Apply Theorem \ref{Th: Pi ID 3} with $w=x_{1}$, $z=(x_{0},x_{2})$, $%
\nu=u_{1} $ and $\Psi(v_{1},z)=\Pi(u_{0}(x_{0}),v_{1},\lambda(x_{2}))$.
Lemma \ref{Lem: Pi u1 injectivity} and 
\eqref{eq: Pi u1 heterogeneity
regular z} imply that $\Psi(\cdot,z)$ is injective, with continuously
differentiable symmetric positive definite Jacobian and $\nabla_{v_{1,k}}%
\Psi_{j}<0$ for $j\neq k$. Hence the interaction graph is complete.
Assumption \ref{Ass: RI ARUM u1 support} verifies Assumption \ref{Ass: Pi
reference support} and the remaining variation condition. Theorem \ref{Th:
Pi ID 3} yields identification of $u_{1}$ on $\mathcal{X}_{1,\mathrm{reg}}$
under the stated normalisations.
\end{proof}

\bigskip

\begin{proof}[Proof of Theorem~\protect\ref{Th: RI ARUM lambda eps neq 0}]
By the chain rule, $\tilde{\Phi}(x_{2},z)=\Psi (\lambda (x_{2}),z)$, where $%
\Psi (\lambda ,z)$ was defined in (\ref{eq: Lambda def lambda}) and $%
z=(x_{0},x_{1})$. We first show that $\lambda \mapsto \Psi (\lambda ,z)$ is
continuously differentiable. By Lemma \ref{Lem: Pi u1 injectivity}, $\nabla
_{u_{1}}\Pi $ exists and is continuous. Writing $t=\lambda +e_{2}$ in the
conditional integral defining $\Pi $ and differentiating the translated
density in $L^{1}$ gives 
\begin{equation*}
\nabla _{\lambda }\Pi (u_{0},u_{1},\lambda )=-E_{\varepsilon
_{0},\varepsilon _{1}}\left[ \int_{0}^{\infty }P^{\ast }(u_{0}+\varepsilon
_{0},u_{1}+\varepsilon _{1},\lambda +e_{2})\nabla _{e_{2}}f_{\varepsilon
_{2}|\varepsilon _{0},\varepsilon _{1}}(e_{2}|\varepsilon _{0},\varepsilon
_{1})\,de_{2}\right] .
\end{equation*}%
Applying the same argument to \eqref{eq: Pi u1 Jacobian} yields 
\begin{equation*}
\nabla _{\lambda }\nabla _{u_{1}}\Pi (u_{0},u_{1},\lambda )=-E_{\varepsilon
_{0},\varepsilon _{1}}\left[ \int_{0}^{\infty }\frac{\func{diag}(P^{\ast
})-P^{\ast }P^{\ast \prime }}{\lambda +e_{2}}\nabla _{e_{2}}f_{\varepsilon
_{2}|\varepsilon _{0},\varepsilon _{1}}(e_{2}|\varepsilon _{0},\varepsilon
_{1})\,de_{2}\right] ,
\end{equation*}%
where $P^{\ast }=P^{\ast }(u_{0}+\varepsilon _{0},u_{1}+\varepsilon
_{1},\lambda +e_{2})$. Since $0\leq P_{j}^{\ast }\leq 1$ and $\lambda
+e_{2}\geq \lambda >0$, the assumed integrability of the density derivative
and dominated convergence establish existence and continuity of both
derivatives locally in $\lambda $. Because $\nabla _{x_{1,1}}u_{1}(x_{1})$
does not depend on $\lambda $, the mapping $\Psi (\cdot ,z)$ is continuously
differentiable.

The chain rule now gives 
\begin{eqnarray*}
\nabla _{x_{2,1}}\tilde{\Phi}(x_{2},z) &=&\nabla _{\lambda }\Psi (\lambda
(x_{2}),z)\nabla _{x_{2,1}}\lambda _{0}(x_{2}), \\
\nabla _{x_{2,2}}\tilde{\Phi}(x_{2},z) &=&\nabla _{\lambda }\Psi (\lambda
(x_{2}),z)\left[ \nabla _{x_{2,2}}\lambda _{0}(x_{2})+\nabla
_{x_{2,2}}\lambda _{1}(x_{2,2})\right] .
\end{eqnarray*}%
For $x_{2}\in \mathcal{X}_{2,\mathrm{reg}}(z)$, choose a coordinate $k$ such
that $\partial _{x_{2,1,k}}\lambda _{0}(x_{2})\neq 0$. Then 
\begin{equation*}
\nabla _{\lambda }\Psi (\lambda (x_{2}),z)=\frac{\partial _{x_{2,1,k}}\tilde{%
\Phi}(x_{2},z)}{\partial _{x_{2,1,k}}\lambda _{0}(x_{2})}\neq 0.
\end{equation*}%
Hence 
\begin{equation*}
\nabla _{x_{2,2}}\lambda _{1}(x_{2,2})=\frac{\nabla _{\lambda }\Psi (\lambda
(x_{2}),z)^{\prime }\nabla _{x_{2,2}}\tilde{\Phi}(x_{2},z)}{\nabla _{\lambda
}\Psi (\lambda (x_{2}),z)^{\prime }\nabla _{\lambda }\Psi (\lambda (x_{2}),z)%
}-\nabla _{x_{2,2}}\lambda _{0}(x_{2}).
\end{equation*}%
The right-hand side is identified at every $x_{2}\in \mathcal{X}_{2,\mathrm{%
reg}}(z)$, so $\nabla _{x_{2,2}}\lambda _{1}$ is identified on $\mathcal{X}%
_{2,2,\mathrm{reg}}$. On a connected open subset $\mathcal{X}_{2,2}^{\ast }$%
, any two continuously differentiable functions with the same gradient
differ by a constant, which is determined by the stated location
normalisation.
\end{proof}

\bigskip

\begin{proof}[Proof of Lemma~\protect\ref{Lem: Pi0 u0 injectivity}]
By Theorem \ref{Thm: P0 one-to-one}, $P^{0\ast }(\cdot ,t)=\nabla
_{v_{0}}Q_{\max }^{0}(\cdot ,t)$ for $t>0$, where $Q_{\max }^{0}(\cdot ,t)$
is convex. Thus, for $u_{0},\tilde{u}_{0}\in \mathbb{R}^{J}$, 
\begin{equation*}
(u_{0}-\tilde{u}_{0})^{\prime }[P^{0\ast }(u_{0}+e_{0},\lambda
+e_{2})-P^{0\ast }(\tilde{u}_{0}+e_{0},\lambda +e_{2})]\geq 0
\end{equation*}%
for every $(e_{0},e_{2})$. Integration establishes monotonicity of $\Pi
_{\pi }^{0}(\cdot ,\lambda )$. Continuity follows by dominated convergence,
since $P^{0\ast }$ is continuous and bounded.

Now suppose $u_{0}\neq \tilde{u}_{0}$ and $u_{0}\in \mathcal{U}_{0,\mathrm{%
reg}}(\lambda )$. On the event $\mathcal{E}_{\mathrm{reg}}(u_{0},\lambda )$
, the point $u_{0}+e_{0}$ belongs to the open set $\mathcal{V}_{0,\mathrm{reg%
}}(\lambda +e_{2})$, where $\nabla _{v_{0}}P^{0\ast }$ is positive definite
by Theorem \ref{Thm: P0 one-to-one}. Writing $h=u_{0}-\tilde{u}_{0}$, the
function $t\longmapsto h^{\prime }P^{0\ast }(\tilde{u}_{0}+e_{0}+th,\lambda
+e_{2})$ is nondecreasing on $[0,1]$ and strictly increasing on a
sufficiently short interval ending at $1$. Consequently, 
\begin{equation*}
h^{\prime }[P^{0\ast }(u_{0}+e_{0},\lambda +e_{2})-P^{0\ast }(\tilde{u}%
_{0}+e_{0},\lambda +e_{2})]>0
\end{equation*}%
on $\mathcal{E}_{\mathrm{reg}}(u_{0},\lambda )$, which has positive
probability. Integrating yields $(u_{0}-\tilde{u}_{0})^{\prime }[\Pi _{\pi
}^{0}(u_{0},\lambda )-\Pi _{\pi }^{0}(\tilde{u}_{0},\lambda )]>0$. This
proves strict monotonicity and injectivity on $\mathcal{U}_{0,\mathrm{reg}%
}(\lambda )$.
\end{proof}

\bigskip

\begin{proof}[Proof of Theorem~\protect\ref{Thm: RI ARUM u0 eps neq 0}]
For $x_{2}\in\mathcal{X}_{2,\pi}$ and $x_{0}\in\mathcal{X}_{0,\pi,\mathrm{reg%
}}(x_{2})$, $(x_{0},x_{2})\in\mathcal{Z}_{\pi}$ implies that $%
\Phi_{\pi}^{0}(x_{0},x_{2})$ is identified by \eqref{eq:
Phi^0_pi def}. Moreover, 
\begin{equation*}
\Phi_{\pi}^{0}(x_{0},x_{2})=\Psi(u_{0}(x_{0}),x_{2}),\qquad
\Psi(v_{0},x_{2})=\Pi_{\pi}^{0}(v_{0},\lambda(x_{2})).
\end{equation*}
Lemma \ref{Lem: Pi0 u0 injectivity} gives injectivity of $\Psi(\cdot,x_{2})$
on $\mathcal{U}_{0,\mathrm{reg}}(\lambda(x_{2}))$. With $w=x_{0}$, $z=x_{2}$
and $\nu=u_{0}$, the observable conditional support for the composite
mapping is $\mathcal{X}_{0,\pi}(x_{2})$, and Assumption \ref{Ass: RI ARUM u0
support eps neq 0} verifies the matching-coverage condition of Theorem \ref%
{Th: Pi ID 2}. The remaining restrictions of that theorem also hold, and its
conclusions follow.
\end{proof}

\bigskip

\begin{proof}[Proof of Lemma~\protect\ref{Lem: PiF u0 injectivity}]
Since $0\leq P_{j}^{\ast }\leq 1$, translation of $f_{\varepsilon
_{0},\varepsilon _{2}}$ in the $e_{0}$ coordinates, together with its
integrable weak derivatives, permits differentiation of the convolution
defining $\Pi _{F}^{0}$. In matrix form, 
\begin{equation*}
\nabla _{u_{0}}\Pi _{F}^{0}(u_{0},\lambda )=-\int_{\mathbb{R}^{J}\times
\lbrack 0,\infty )}\int_{\mathbb{R}^{J}}P^{\ast }(u_{0}+e_{0},v_{1},\lambda
+e_{2})\nabla _{e_{0}}f_{\varepsilon _{0},\varepsilon
_{2}}(e_{0},e_{2})^{\prime }dF_{V_{1}}(v_{1})de_{0}de_{2}.
\end{equation*}%
Continuity of the Jacobian follows from continuity of translations in $L^{1}$
applied to $\nabla _{e_{0}}f_{\varepsilon _{0},\varepsilon _{2}}$. The same
representation holds for $\Pi _{\pi }^{0}$, with $F_{V_{1}}$ replaced by $%
\pi _{V_{1}}$. For fixed $\lambda >0$, define the centred integrated surplus 
\begin{equation*}
\bar{Q}_{\pi }^{0}(u_{0},\lambda ):=\int [Q_{\max }^{0}(u_{0}+e_{0},\lambda
+e_{2})-Q_{\max }^{0}(e_{0},\lambda +e_{2})]dF_{\varepsilon _{0},\varepsilon
_{2}}(e_{0},e_{2}).
\end{equation*}%
By Theorem \ref{Thm: P0 one-to-one}, $Q_{\max }^{0}(\cdot ,t)$ is convex
with gradient $P^{0\ast }(\cdot ,t)$ for every $t>0$. Since this gradient is
a vector of choice probabilities, 
\begin{equation*}
|Q_{\max }^{0}(u_{0}+e_{0},t)-Q_{\max }^{0}(e_{0},t)|\leq \Vert u_{0}\Vert
_{\infty }
\end{equation*}%
for every $(e_{0},t)$. Thus $\bar{Q}_{\pi }^{0}$ is finite and convex in $%
u_{0}$. Bounded convergence applied to the difference quotients gives $%
\nabla _{u_{0}}\bar{Q}_{\pi }^{0}(u_{0},\lambda )=\Pi _{\pi
}^{0}(u_{0},\lambda )$. Hence the continuous Jacobian $\nabla _{u_{0}}\Pi
_{\pi }^{0}$ is symmetric positive semidefinite. For $u_{0}\in \mathcal{U}%
_{0,\mathrm{reg}}(\lambda )$, the nonnegative difference quotients and
Fatou's lemma imply that this Jacobian is positive definite by Theorem \ref%
{Thm: P0 one-to-one}. Since $\Vert P^{\ast }\Vert \leq 1$, the definition of
total variation yields 
\begin{eqnarray*}
&&\sup_{u_{0}\in \mathbb{R}^{J}}\left\Vert \nabla _{u_{0}}\Pi
_{F}^{0}(u_{0},\lambda )-\nabla _{u_{0}}\Pi _{\pi }^{0}(u_{0},\lambda
)\right\Vert  \\
&\leq &2d_{\mathrm{TV}}(\pi _{V_{1}},F_{V_{1}})\int_{\mathbb{R}^{J}\times
\lbrack 0,\infty )}\Vert \nabla _{e_{0}}f_{\varepsilon _{0},\varepsilon
_{2}}(e_{0},e_{2})\Vert de_{0}de_{2} \\
&=&2L_{f}\,d_{\mathrm{TV}}(\pi _{V_{1}},F_{V_{1}}).
\end{eqnarray*}%
Since $\nabla _{u_{0}}\Pi _{\pi }^{0}$ is symmetric, the symmetric part of $%
\nabla _{u_{0}}\Pi _{F}^{0}$ satisfies, for $u_{0}\in \mathcal{U}$, 
\begin{equation*}
\lambda _{\min }\left( \frac{\nabla _{u_{0}}\Pi _{F}^{0}(u_{0},\lambda
)+\nabla _{u_{0}}\Pi _{F}^{0}(u_{0},\lambda )^{\prime }}{2}\right) \geq c_{%
\mathcal{U}}(\lambda )-2L_{f}\,d_{\mathrm{TV}}(\pi _{V_{1}},F_{V_{1}})>0.
\end{equation*}%
Thus $\nabla _{u_{0}}\Pi _{F}^{0}$ is nonsingular. For distinct $u_{0},%
\tilde{u}_{0}\in \mathcal{U}$, convexity and the fundamental theorem of
calculus give%
\begin{equation*}
(u_{0}-\tilde{u}_{0})^{\prime }[\Pi _{F}^{0}(u_{0},\lambda )-\Pi _{F}^{0}(%
\tilde{u}_{0},\lambda )]=\int_{0}^{1}(u_{0}-\tilde{u}_{0})^{\prime }\nabla
_{u_{0}}\Pi _{F}^{0}(\tilde{u}_{0}+t(u_{0}-\tilde{u}_{0}),\lambda )(u_{0}-%
\tilde{u}_{0})dt>0.
\end{equation*}%
Hence $\Pi _{F}^{0}(\cdot ,\lambda )$ is strictly monotone, hence injective,
on $\mathcal{U}$.
\end{proof}

\section{Alternative support conditions\label{sec: support conds}}

We here provide an alternative set of support conditions under which the
results of Theorems \ref{Th: Pi ID joint rank} and \ref{Th: Pi ID 3} hold.
We first consider Theorem \ref{Th: Pi ID joint rank}. For every $z\in 
\mathcal{Z}_{0}$, let $\mathcal{W}_{2}^{\ast }(z)\subseteq \mathcal{W}%
_{0,2}(z)$ and, for $w_{2}\in \mathcal{W}_{2}^{\ast }(z)$, let $\mathcal{R}%
(w_{2},z)\subseteq \mathcal{W}_{0,1}(w_{2},z)$. Define 
\begin{equation*}
\mathcal{V}^{\ast }(w_{2},z):=\nu (\mathcal{R}(w_{2},z),w_{2}),\qquad 
\mathcal{V}^{\ast }(z):=\bigcup_{w_{2}\in \mathcal{W}_{2}^{\ast }(z)}%
\mathcal{V}^{\ast }(w_{2},z).
\end{equation*}

\begin{assumption}
\label{Ass: Pi joint rank alternative support} For every $z\in \mathcal{Z}%
_{0}$, there exist a nonempty set $\mathcal{W}_{2}^{\ast }(z)$ and nonempty
sets $\mathcal{R}(w_{2},z)$, $w_{2}\in \mathcal{W}_{2}^{\ast }(z)$, as
above, such that: (i) For each $z\in \mathcal{Z}_{0}$ and $w_{2}\in \mathcal{%
W}_{2}^{\ast }(z)$, $\mathcal{V}^{\ast }(w_{2},z)$ is open and connected,
and $\mathcal{R}(w_{2},z)\times \{w_{2}\}\subseteq \bigcap_{z^{\prime }\in 
\mathcal{Z}^{\ast }}\mathcal{W}_{0}(z^{\prime })$; (ii) for every $z\in 
\mathcal{Z}_{0}$ and every partition of $\mathcal{W}_{2}^{\ast }(z)$ into
two nonempty disjoint sets, there exist $w_{2},\bar{w}_{2}$, one in each
set, such that $\mathcal{V}^{\ast }(w_{2},z)\cap \mathcal{V}^{\ast }(\bar{w}%
_{2},z)$ contains two distinct points; (iii) for every partition of $%
\mathcal{Z}_{0}$ into two nonempty disjoint sets $\mathcal{Z}_{1},\mathcal{Z}%
_{2}$, there exist $z_{i}\in \mathcal{Z}_{i}$, $i=1,2$, and $w_{2}\in 
\mathcal{W}_{2}^{\ast }(z_{1})\cap \mathcal{W}_{2}^{\ast }(z_{2})$ such that 
$\mathcal{R}(w_{2},z_{1})\cap \mathcal{R}(w_{2},z_{2})$ contains points $%
w_{1},\bar{w}_{1}$ with $\nu (w_{1},w_{2})\neq \nu (\bar{w}_{1},w_{2})$;
(iv) $\mathcal{V}_{0}(z)=\mathcal{V}^{\ast }(z)$ for every $z\in \mathcal{Z}%
_{0}$.
\end{assumption}

Compared with Assumption \ref{Ass: Pi reference support}, this condition
replaces the single product reference set $\mathcal{W}^*(z_0)$ by slices $%
\mathcal{R}(w_2,z)\times\{w_2\}$ that may vary with both $w_2$ and $z$.
Thus, the conditional reference support for $w_1$ can shift with $w_2$ and $%
z $, while $\mathcal{W}_2^*(z)$ need not be open or connected; in
particular, discrete or disconnected supports for $w_2$ are allowed. Each
slice must nevertheless have an open, connected image $\mathcal{V}^*(w_2,z)$
lying in the regular region for every $z^{\prime }\in\mathcal{Z}^*$, and
conditions (ii)--(iii) require two-point overlaps that link the local affine
representations, first across $w_2$ and then across $z$. Condition (iv)
requires coverage of the attained regular index image by these local
reference images. In contrast, Assumption \ref{Ass: Pi reference support}
requires a common open, connected product reference set and coverage from
that set, but not an open reference image at every $(w_2,z)$ or the stated
pairwise overlap conditions. Neither support condition therefore implies the
other. For example, with $J\geq2$, $\nu_j(w_{1,j},w_2)=w_{1,j}+w_2$ and $%
\mathcal{W}_{0,1}(w_2,z)=\prod_{j=1}^J(-1-w_2,1-w_2)$, all reference slices
have the same image $(-1,1)^J$, whereas no nondegenerate Cartesian product
contained in this moving support attains that entire image. Conversely, the
common-reference condition can accommodate a value of $z$ for which the
regular conditional support of $w_1$ contains no slice with an open index
image, provided its attained index values are covered by the common
reference set; condition (i) above rules out such a configuration.

\begin{theorem}
\label{Th: Pi ID joint rank alternative support} Suppose Assumptions \ref%
{Ass: Pi injective} and \ref{Ass: Pi joint rank alternative support} hold, $%
J\geq 2$, and conditions (i)--(ii) of Theorem \ref{Th: Pi ID joint rank}
hold. Then $\nu$ is identified on $\bigcup_{z\in\mathcal{Z}_0}\mathcal{W}%
_0(z)$ under the following normalisations: for known $w_0,\bar w_0$ in the
same reference region $\mathcal{R}(w_2,z)\times\{w_2\}$ with $\nu_1(\bar
w_0)\neq\nu_1(w_0)$, 
\begin{equation*}
\nu(w_0)=0,\qquad\nu_1(\bar w_0)=1.
\end{equation*}
Moreover, $\Psi(\cdot,z)$ is identified on $\mathcal{V}_0(z)$ for every $z\in%
\mathcal{Z}_0$.
\end{theorem}

\begin{proof}[Proof of Theorem~\protect\ref{Th: Pi ID joint rank alternative
support}]
Let $(\tilde\nu,\tilde\Psi)$ be an observationally equivalent
representation. Fix $z\in\mathcal{Z}_0$ and $w_2\in\mathcal{W}_2^*(z)$. By
Assumption~\ref{Ass: Pi joint rank alternative support}(i), the reference
slice $\mathcal{R}(w_2,z)\times\{w_2\}$ belongs to $\mathcal{W}_0(z^{\prime
})$ for every $z^{\prime }\in\mathcal{Z}^*$, and $\mathcal{V}^*(w_2,z)$ is
open and connected. Thus, on this image the construction of the
transformation $h$ and the differentiation argument in the proof of Theorem~%
\ref{Th: Pi ID joint rank} apply. In particular, symmetry and the spanning
condition imply $[Dh(v)]^{-1}\in\func{span}\{I_J\}$. Since $J\geq2$ and $%
\mathcal{V}^*(w_2,z)$ is open and connected, $h(v)=a(z,w_2)v+b(z,w_2)$,
where $a(z,w_2)\neq0$. Hence 
\begin{equation*}
\tilde\nu(w_1,w_2)=a(z,w_2)\nu(w_1,w_2)+b(z,w_2), \qquad w_1\in\mathcal{R}%
(w_2,z).
\end{equation*}

The difference from the proof of Theorem~\ref{Th: Pi ID joint rank} is that
these affine coefficients may initially depend on $(z,w_2)$. For fixed $z$,
suppose $v\neq\bar v$ belong to $\mathcal{V}^*(w_2,z)\cap\mathcal{V}^*(\bar
w_2,z)$. Observational equivalence and injectivity of $\tilde\Psi(\cdot,z)$
give 
\begin{equation*}
[a(z,w_2)-a(z,\bar w_2)]v =b(z,\bar w_2)-b(z,w_2) =[a(z,w_2)-a(z,\bar
w_2)]\bar v.
\end{equation*}
Thus the affine coefficients coincide. The overlap condition in Assumption~%
\ref{Ass: Pi joint rank alternative support}(ii) propagates this equality
across $\mathcal{W}_2^*(z)$, so the coefficients depend only on $z$; write
them as $a(z)$ and $b(z)$. For $z_1,z_2$ and the two common covariate points 
$w,\bar w$ supplied by condition (iii), the two affine representations give 
\begin{equation*}
[a(z_1)-a(z_2)][\nu(w)-\nu(\bar w)]=0.
\end{equation*}
Because $\nu(w)\neq\nu(\bar w)$, we have $a(z_1)=a(z_2)$ and then $%
b(z_1)=b(z_2)$. Condition (iii) propagates this equality across $\mathcal{Z}%
_0$. The stated normalisations imply that the common coefficients are $a=1$
and $b=0$. Hence $\tilde\nu=\nu$ on every reference region.

Finally, condition (iv) supplies, for each $w\in\mathcal{W}_0(z)$, a
reference point $w^*$ at the same $z$ with $\nu(w^*)=\nu(w)$. Injectivity of 
$\tilde\Psi(\cdot,z)$ gives $\tilde\nu(w)=\tilde\nu(w^*)=\nu(w)$. Thus $\nu$
is identified on the stated union of regular supports, and observational
equivalence identifies $\Psi(\cdot,z)$ on $\mathcal{V}_0(z)$.
\end{proof}

We next consider Theorem \ref{Th: Pi ID 3}. The reference regions are now
open rectangles, and the symmetry and graph conditions are imposed on each
such region. For $z\in\mathcal{Z}_0$ and $w_2\in\mathcal{W}_2^*(z)$, let $%
\mathcal{G}_\Psi(w_2,z)$ be the undirected graph with vertices $1,\ldots,J$
and an edge $\{j,k\}$ whenever 
\begin{equation*}
\nabla_{\nu_k}\Psi_j(\nu(w_1,w_2),z)\neq0\qquad\text{for every } w_1\in%
\mathcal{R}(w_2,z).
\end{equation*}

\begin{assumption}
\label{Ass: Pi 3 alternative support} For every $z\in \mathcal{Z}_{0}$,
there exist a nonempty set $\mathcal{W}_{2}^{\ast }(z)\subseteq \mathcal{W}%
_{0,2}(z)$ and, for every $w_{2}\in \mathcal{W}_{2}^{\ast }(z)$, a nonempty
open rectangle $\mathcal{R}(w_{2},z)=\prod_{j=1}^{J}\mathcal{I}%
_{j}(w_{2},z)\subseteq \mathcal{W}_{0,1}(w_{2},z)$. With $\mathcal{V}^{\ast
}(w_{2},z):=\nu (\mathcal{R}(w_{2},z),w_{2})$ and $\mathcal{V}^{\ast
}(z):=\bigcup_{w_{2}\in \mathcal{W}_{2}^{\ast }(z)}\mathcal{V}^{\ast
}(w_{2},z)$, the following hold: (i) For every $z\in \mathcal{Z}_{0}$ and
every partition of $\mathcal{W}_{2}^{\ast }(z)$ into two nonempty disjoint
sets, there exist $w_{2},\bar{w}_{2}$, one in each set, such that $\mathcal{V%
}^{\ast }(w_{2},z)\cap \mathcal{V}^{\ast }(\bar{w}_{2},z)$ contains two
distinct points; (ii) for every partition of $\mathcal{Z}_{0}$ into two
nonempty disjoint sets $\mathcal{Z}_{1},\mathcal{Z}_{2}$, there exist $%
z_{i}\in \mathcal{Z}_{i}$, $i=1,2$, and $w_{2}\in \mathcal{W}_{2}^{\ast
}(z_{1})\cap \mathcal{W}_{2}^{\ast }(z_{2})$ such that $\mathcal{R}%
(w_{2},z_{1})\cap \mathcal{R}(w_{2},z_{2})$ contains points $w_{1},\bar{w}%
_{1}$ with $\nu (w_{1},w_{2})\neq \nu (\bar{w}_{1},w_{2})$; (iii) $\mathcal{V%
}_{0}(z)=\mathcal{V}^{\ast }(z)$ for every $z\in \mathcal{Z}_{0}$.
\end{assumption}

The comparison of above with Assumption \ref{Ass: Pi reference support} is
similar to the following Assumption \ref{Ass: Pi joint rank alternative
support} and so is left out.

\begin{theorem}
\label{Th: Pi ID 3 alternative support} Suppose $J\geq2$, Assumptions \ref%
{Ass: Pi injective} and \ref{Ass: Pi 3 alternative support} hold, and %
\eqref{eq: ASV common structure} holds. Suppose $\nu$ is continuous on the
reference regions and each $\nu_j$ is continuously differentiable in $%
w_{1,j} $ on neighbourhoods of the reference regions, and $\Psi(\cdot,z)$ is
continuously differentiable with symmetric Jacobian $\nabla_\nu\Psi(\nu,z)$
on $\mathcal{V}^*(w_2,z)$ for every $z\in\mathcal{Z}_0$ and $w_2\in\mathcal{W%
}_2^*(z)$. Suppose further that, for every such $(w_2,z)$: (i) each $%
\nu_j(\cdot,w_2)$ is nonconstant on $\mathcal{I}_j(w_2,z)$; (ii) the graph $%
\mathcal{G}_\Psi(w_2,z)$ is connected. Then $\nu$ is identified on $%
\bigcup_{z\in\mathcal{Z}_0}\mathcal{W}_0(z)$ under the following
normalisations: for a known $w_0=(w_{0,1},w_{0,2})$ in a reference region
with $\nabla_{w_{1,1}}\nu_1(w_0)\neq0$, 
\begin{equation*}
\nu(w_0)=0,\qquad\nabla_{w_{1,1}}\nu_1(w_0)=1.
\end{equation*}
Moreover, $\Psi(\cdot,z)$ is identified on $\mathcal{V}_0(z)$ for every $z\in%
\mathcal{Z}_0$.
\end{theorem}

\begin{proof}[Proof of Theorem~\protect\ref{Th: Pi ID 3 alternative support}]

Let $(\tilde\nu,\tilde\Psi)$ be an observationally equivalent
representation. Fix $z\in\mathcal{Z}_0$ and $w_2\in\mathcal{W}_2^*(z)$. For
each edge $\{j,k\}$ of $\mathcal{G}_\Psi(w_2,z)$, the chain rule, the
alternative-specific structure \eqref{eq: ASV common structure}, and
symmetry of both outer Jacobians imply, exactly as in the proof of Theorem~%
\ref{Th: Pi ID 3}, 
\begin{equation*}
\partial_{w_{1,j}}\nu_j(w_{1,j},w_2)\,
\partial_{w_{1,k}}\tilde\nu_k(w_{1,k},w_2) =
\partial_{w_{1,k}}\nu_k(w_{1,k},w_2)\,
\partial_{w_{1,j}}\tilde\nu_j(w_{1,j},w_2).
\end{equation*}
Since each $\nu_j(\cdot,w_2)$ is nonconstant, there is a point in $\mathcal{I%
}_j(w_2,z)$ where its derivative is nonzero. The connected graph therefore
forces a common derivative ratio $a(z,w_2)$ on the whole reference
rectangle. Integration gives 
\begin{equation*}
\tilde\nu(w_1,w_2)=a(z,w_2)\nu(w_1,w_2)+b(z,w_2), \qquad w_1\in\mathcal{R}%
(w_2,z).
\end{equation*}
Injectivity and nonconstancy give $a(z,w_2)\neq0$.

Unlike the common-reference proof of Theorem~\ref{Th: Pi ID 3}, these affine
coefficients initially depend on both $w_2$ and $z$. The remainder is the
overlap argument in the preceding proof: Assumption~\ref{Ass: Pi 3
alternative support}(i) makes the coefficients constant across $w_2$ at each 
$z$, and (ii) links them across $z$. The stated normalisations give $a=1$
and $b=0$. Finally, the coverage condition (iii), together with injectivity
of $\tilde\Psi(\cdot,z)$, extends $\tilde\nu=\nu$ from the reference
rectangles to every $\mathcal{W}_0(z)$ and identifies $\Psi(\cdot,z)$ on $%
\mathcal{V}_0(z)$, as in the preceding proof.
\end{proof}

\bigskip

Next, we apply these results to the two applications of Theorem \ref{Th: Pi
ID 3} to identification of $u_{1}$ in Section \ref{Sec: ID RI-ARUM}. These
conditions correspond to Assumption \ref{Ass: Pi 3 alternative support}. In
both applications, we take the CCPs to be known on $\mathcal{X}_{1}(z)$ and
on the reference slices at each $z$.

\begin{assumption}
\label{Ass: RI ARUM u1 lambda alternative support} Suppose $\mathcal{Z}%
_{0}^{0}\neq \varnothing$. For each $z\in\mathcal{Z}_{0}^{0}$, there exists
a nonempty set $\mathcal{X}_{1,2}^{\ast}(z)$ of common-covariate values,
and, for each $x_{1,2}\in\mathcal{X}_{1,2}^{\ast}(z)$, a nonempty open
rectangle $\mathcal{R}_{1}(x_{1,2},z)=\prod_{j=1}^{J}\mathcal{I}%
_{1j}(x_{1,2},z)$. Write 
\begin{equation*}
\mathcal{U}_{1}^{\ast}(x_{1,2},z):= u_{1}(\mathcal{R}%
_{1}(x_{1,2},z),x_{1,2}),\qquad \mathcal{U}_{1}^{\ast}(z):=\bigcup_{x_{1,2}%
\in\mathcal{X}_{1,2}^{\ast}(z)} \mathcal{U}_{1}^{\ast}(x_{1,2},z).
\end{equation*}
Suppose: (i) $\mathcal{R}_{1}(x_{1,2},z)\subseteq \func{int}\func{supp}%
(X_{1,1}\mid X_{1,2}=x_{1,2},(X_{0},X_{2})=z)$ and $x_{1,1,j}\mapsto
u_{1,j}(x_{1,1,j},x_{1,2})$ is nonconstant on $\mathcal{I}_{1j}(x_{1,2},z)$
for every $j$; (ii) for every $z\in\mathcal{Z}_{0}^{0}$ and every partition
of $\mathcal{X}_{1,2}^{\ast}(z)$ into two nonempty disjoint sets, there
exist $x_{1,2},\bar{x}_{1,2}$ in different sets such that $\mathcal{U}%
_{1}^{\ast}(x_{1,2},z)\cap\mathcal{U}_{1}^{\ast}(\bar{x}_{1,2},z)$ contains
two distinct points; (iii) for every partition of $\mathcal{Z}_{0}^{0}$ into
two nonempty disjoint sets $\mathcal{Z}_{1},\mathcal{Z}_{2}$, there exist $%
z_{i}\in\mathcal{Z}_{i}$, $i=1,2$, and $x_{1,2}\in\mathcal{X}%
_{1,2}^{\ast}(z_{1})\cap\mathcal{X}_{1,2}^{\ast}(z_{2})$ such that $\mathcal{%
R}_{1}(x_{1,2},z_{1})\cap \mathcal{R}_{1}(x_{1,2},z_{2})$ contains points $%
x_{1,1},\bar{x}_{1,1}$ with $u_{1}(x_{1,1},x_{1,2})\neq u_{1}(\bar{x}%
_{1,1},x_{1,2})$; (iv) $u_{1}(\mathcal{X}_{1}(z))=\mathcal{U}_{1}^{\ast}(z)$
for every $z\in\mathcal{Z}_{0}^{0}$.
\end{assumption}

\begin{corollary}
\label{Cor: RI ARUM u1 lambda alternative support} The conclusion of Theorem %
\ref{Th: RI ARUM u1 lambda eps = 0} holds if Assumption \ref{Ass: RI ARUM u1
lambda support} is replaced by Assumption \ref{Ass: RI ARUM u1 lambda
alternative support}, with the same normalisations imposed at a known $%
x_{1,0}=(x_{1,1,0},x_{1,2,0})$ belonging to one of the reference rectangles $%
\mathcal{R}_{1}(x_{1,2,0},z_{0})\times\{x_{1,2,0}\}$.
\end{corollary}

\begin{proof}
Apply Theorem \ref{Th: Pi ID 3 alternative support} with $w=x_{1}$, $%
z=(x_{0},x_{2})$, $\nu=u_{1}$ and $\Psi(v_{1},z)=P^{%
\ast}(u_{0}(x_{0}),v_{1},\lambda(x_{2}))$. The argument identifying $\lambda$
in the proof of Theorem \ref{Th: RI ARUM u1 lambda eps = 0} is unchanged.
\end{proof}

\begin{assumption}
\label{Ass: RI ARUM u1 alternative support} Suppose $\mathcal{Z}_{u_{1}}\neq
\varnothing$. For each $z\in\mathcal{Z}_{u_{1}}$, there exists a nonempty
set $\mathcal{X}_{1,2}^{\ast}(z)$ and, for each $x_{1,2}\in\mathcal{X}%
_{1,2}^{\ast}(z)$, a nonempty open rectangle $\mathcal{R}_{1}(x_{1,2},z)=%
\prod_{j=1}^{J}\mathcal{I}_{1j}(x_{1,2},z)$. Write 
\begin{equation*}
\mathcal{U}_{1}^{\ast}(x_{1,2},z):=u_{1}(\mathcal{R}%
_{1}(x_{1,2},z),x_{1,2}),\qquad \mathcal{U}_{1}^{\ast}(z):=\bigcup_{x_{1,2}%
\in\mathcal{X}_{1,2}^{\ast}(z)} \mathcal{U}_{1}^{\ast}(x_{1,2},z).
\end{equation*}
Suppose: (i) $\mathcal{R}_{1}(x_{1,2},z)\subseteq \func{int}\func{supp}%
(X_{1,1}\mid X_{1,2}=x_{1,2},(X_{0},X_{2})=z)$ and $x_{1,1,j}\mapsto
u_{1,j}(x_{1,1,j},x_{1,2})$ is nonconstant on $\mathcal{I}_{1j}(x_{1,2},z)$
for every $j$; (ii) for every $z\in\mathcal{Z}_{u_{1}}$ and every partition
of $\mathcal{X}_{1,2}^{\ast}(z)$ into two nonempty disjoint sets, there
exist $x_{1,2},\bar{x}_{1,2}$ in different sets such that $\mathcal{U}%
_{1}^{\ast}(x_{1,2},z)\cap\mathcal{U}_{1}^{\ast}(\bar{x}_{1,2},z)$ contains
two distinct points; (iii) for every partition of $\mathcal{Z}_{u_{1}}$ into
two nonempty disjoint sets $\mathcal{Z}_{1},\mathcal{Z}_{2}$, there exist $%
z_{i}\in\mathcal{Z}_{i}$, $i=1,2$, and $x_{1,2}\in\mathcal{X}%
_{1,2}^{\ast}(z_{1})\cap\mathcal{X}_{1,2}^{\ast}(z_{2})$ such that $\mathcal{%
R}_{1}(x_{1,2},z_{1})\cap \mathcal{R}_{1}(x_{1,2},z_{2})$ contains points $%
x_{1,1},\bar{x}_{1,1}$ with $u_{1}(x_{1,1},x_{1,2})\neq u_{1}(\bar{x}%
_{1,1},x_{1,2})$; (iv) $u_{1}(\mathcal{X}_{1}(z))=\mathcal{U}_{1}^{\ast}(z)$
for every $z\in\mathcal{Z}_{u_{1}}$.
\end{assumption}

\begin{corollary}
\label{Cor: RI ARUM u1 alternative support} The conclusion of Theorem \ref%
{Th: RI ARUM u1 eps neq 0} holds if Assumption \ref{Ass: RI ARUM u1 support}
is replaced by Assumption \ref{Ass: RI ARUM u1 alternative support}, with
the same normalisations imposed at a known $x_{1,0}=(x_{1,1,0},x_{1,2,0})$
belonging to one of the reference rectangles $\mathcal{R}%
_{1}(x_{1,2,0},z_{0})\times\{x_{1,2,0}\}$.
\end{corollary}

\begin{proof}
Apply Theorem \ref{Th: Pi ID 3 alternative support} with $w=x_{1}$, $%
z=(x_{0},x_{2})$, $\nu=u_{1}$ and $\Psi(v_{1},z)=\Pi(u_{0}(x_{0}),v_{1},%
\lambda(x_{2}))$. The injectivity, symmetry and graph conditions follow from
Lemma \ref{Lem: Pi u1 injectivity}.
\end{proof}

\end{document}